\documentclass[longauth]{aa}  

\usepackage{euclid}
\usepackage{natbib}
\usepackage{enumitem}
\usepackage{graphicx}
\usepackage{float}
\usepackage{placeins}
\usepackage{subfig}
\usepackage{multicol}
\usepackage{supertabular}
\usepackage{multirow}
\usepackage{comment}
\usepackage{adjustbox}

\usepackage{pdflscape}
\usepackage[varg]{txfonts}

\usepackage{oplotsymbl}

\usepackage{xcolor,colortbl}

 \usepackage{stackengine}
\setstackEOL{\\}
\newcounter{tmpctr}

\newcommand\FancyRoman[1]{%
  \setcounter{tmpctr}{#1}%
  \setbox0=\hbox{\kern.2pt\textsf{\Roman{tmpctr}}}%
  \setstackgap{S}{-.5pt}%
  \Shortstack{\rule{\dimexpr\wd0+.1ex}{.7pt}\\\copy0\\
              \rule{\dimexpr\wd0+.1ex}{.7pt}}%
}

\newcommand{\ECTile}{\texttt{ECTile}\xspace}

\def\bbeta{\ensuremath{\hspace{0.0555em}\beta}}  

\def\ddeg#1.#2{{#1\fdg #2}} 

\DeclareMathOperator{\asinh}{asinh}

\def\dalpha{\ensuremath{\delta_\text{AA}}\xspace}
\def\dSAA{\ensuremath{\delta_\text{SAA}}\xspace}

\def\Xsc{$X_{\rm SC}$\xspace}
\def\Ysc{$Y_{\rm SC}$\xspace}
\def\Zsc{$Z_{\rm SC}$\xspace}

\def\rlen{\ensuremath{\text{len}}}
\def\rcap{\ensuremath{\text{cap}}}
\def\rthreads{\ensuremath{n_\text{thread}}}

\def\stageone{{stage-1}\xspace}
\def\stagetwo{{stage-2}\xspace}

\def\fmark#1%
  {\frame{\centerline{\parbox{.4\textwidth}{\strut\tt FIGURE: #1\strut}}}}

\usepackage[pdfencoding=auto,psdextra]{hyperref}
\hypersetup{
    colorlinks=true,
    linkcolor=blue,
    filecolor=magenta,      
    urlcolor=blue,
    citecolor=blue,
    pdftitle = {Euclid Preparation I. The Euclid Wide Survey},
    pdfauthor = {R. Scaramella, et. al.},
    pdfsubject = {Version 0.8}
}
\makeatletter
\renewcommand*\aa@pageof{, page \thepage{} of \pageref*{LastPage}}
\makeatother

\newcommand{\ergscmsq}{\,\mathrm{erg}\,\mathrm{s}^{-1}\,\mathrm{cm}^{-2}}

\newcommand{\ergscmsqA}{\mathrm{erg}\;\mathrm{s}^{-1}\,\mathrm{cm}^{-2}\,\text{\AA}^{-1}}

\begin{document} 

\newlength{\twocolspan}
\setlength{\twocolspan}{0.99\textwidth}
\newlength{\onecolspan}
\setlength{\onecolspan}{0.49\textwidth}


 
 \title{\Euclid preparation: I. The Euclid Wide Survey}

%
   \subtitle{}
%
 
\author{\normalsize Euclid Collaboration: R.~Scaramella$^{1,2}$\thanks{\email{roberto.scaramella@inaf.it}}, J.~Amiaux$^{3}$, Y.~Mellier$^{4,5}$, C.~Burigana$^{6,7,8}$, C.S.~Carvalho$^{9}$, J.-C.~Cuillandre$^{3}$, A.~Da Silva$^{10,11}$, A.~Derosa$^{12}$, J.~Dinis$^{10,11}$, E.~Maiorano$^{12}$
% ...
}

\institute{$^{1}$ INFN-Sezione di Roma, Piazzale Aldo Moro, 2 - c/o Dipartimento di Fisica, Edificio G. Marconi, I-00185 Roma, Italy\\
$^{2}$ INAF-Osservatorio Astronomico di Roma, Via Frascati 33, I-00078 Monteporzio Catone, Italy\\
$^{3}$ AIM, CEA, CNRS, Universit\'{e} Paris-Saclay, Universit\'{e} de Paris, F-91191 Gif-sur-Yvette, France\\
$^{4}$ CEA Saclay, DFR/IRFU, Service d'Astrophysique, Bat. 709, 91191 Gif-sur-Yvette, France\\
$^{5}$ Institut d'Astrophysique de Paris, 98bis Boulevard Arago, F-75014, Paris, France\\
$^{6}$ INFN-Bologna, Via Irnerio 46, I-40126 Bologna, Italy\\
$^{7}$ Dipartimento di Fisica e Scienze della Terra, Universit\'a degli Studi di Ferrara, Via Giuseppe Saragat 1, I-44122 Ferrara, Italy\\
$^{8}$ INAF, Istituto di Radioastronomia, Via Piero Gobetti 101, I-40129 Bologna, Italy\\
$^{9}$ Instituto de Astrof\'isica e Ci\^encias do Espa\c{c}o, Faculdade de Ci\^encias, Universidade de Lisboa, Tapada da Ajuda, PT-1349-018 Lisboa, Portugal\\
$^{10}$ Departamento de F\'isica, Faculdade de Ci\^encias, Universidade de Lisboa, Edif\'icio C8, Campo Grande, PT1749-016 Lisboa, Portugal\\
$^{11}$ Instituto de Astrof\'isica e Ci\^encias do Espa\c{c}o, Faculdade de Ci\^encias, Universidade de Lisboa, Campo Grande, PT-1749-016 Lisboa, Portugal\\
$^{12}$ INAF-Osservatorio di Astrofisica e Scienza dello Spazio di Bologna, Via Piero Gobetti 93/3, I-40129 Bologna, Italy\\
$^{13}$ IFPU, Institute for Fundamental Physics of the Universe, via Beirut 2, 34151 Trieste, Italy\\
$^{14}$ INAF-Osservatorio Astronomico di Trieste, Via G. B. Tiepolo 11, I-34131 Trieste, Italy\\
$^{15}$ European Space Agency/ESTEC, Keplerlaan 1, 2201 AZ Noordwijk, The Netherlands\\
$^{16}$ ESAC/ESA, Camino Bajo del Castillo, s/n., Urb. Villafranca del Castillo, 28692 Villanueva de la Ca\~nada, Madrid, Spain\\
$^{17}$ FRACTAL S.L.N.E., calle Tulip\'an 2, Portal 13 1A, 28231, Las Rozas de Madrid, Spain\\
$^{18}$ European Space Agency/ESOC, Robert-Bosch-Str. 5, D-64293 Darmstadt, Germany\\
$^{19}$ Max-Planck-Institut f\"ur Astronomie, K\"onigstuhl 17, D-69117 Heidelberg, Germany\\
$^{20}$ Mullard Space Science Laboratory, University College London, Holmbury St Mary, Dorking, Surrey RH5 6NT, UK\\
$^{21}$ Univ Lyon, Univ Claude Bernard Lyon 1, CNRS/IN2P3, IP2I Lyon, UMR 5822, F-69622, Villeurbanne, France\\
$^{22}$ INAF-IASF Milano, Via Alfonso Corti 12, I-20133 Milano, Italy\\
$^{23}$ Aix-Marseille Univ, CNRS/IN2P3, CPPM, Marseille, France\\
$^{24}$ INAF-Osservatorio Astronomico di Brera, Via Brera 28, I-20122 Milano, Italy\\
$^{25}$ Dipartimento di Fisica "Aldo Pontremoli", Universit\'a degli Studi di Milano, Via Celoria 16, I-20133 Milano, Italy\\
$^{26}$ INFN-Sezione di Milano, Via Celoria 16, I-20133 Milano, Italy\\
$^{27}$ Leiden Observatory, Leiden University, Niels Bohrweg 2, 2333 CA Leiden, The Netherlands\\
$^{28}$ INFN-Sezione di Bologna, Viale Berti Pichat 6/2, I-40127 Bologna, Italy\\
$^{29}$ Department of Physics, Oxford University, Keble Road, Oxford OX1 3RH, UK\\
$^{30}$ Argelander-Institut f\"ur Astronomie, Universit\"at Bonn, Auf dem H\"ugel 71, 53121 Bonn, Germany\\
$^{31}$ Perimeter Institute for Theoretical Physics, Waterloo, Ontario N2L 2Y5, Canada\\
$^{32}$ Department of Physics and Astronomy, University of Waterloo, Waterloo, Ontario N2L 3G1, Canada\\
$^{33}$ Centre for Astrophysics, University of Waterloo, Waterloo, Ontario N2L 3G1, Canada\\
$^{34}$ Carnegie Observatories, Pasadena, CA 91101, USA\\
$^{35}$ Universit\'e Paris-Saclay, CNRS, Institut d'astrophysique spatiale, 91405, Orsay, France\\
$^{36}$ Institute of Cosmology and Gravitation, University of Portsmouth, Portsmouth PO1 3FX, UK\\
$^{37}$ INAF-Osservatorio Astronomico di Padova, Via dell'Osservatorio 5, I-35122 Padova, Italy\\
$^{38}$ Universit\"ats-Sternwarte M\"unchen, Fakult\"at f\"ur Physik, Ludwig-Maximilians-Universit\"at M\"unchen, Scheinerstrasse 1, 81679 M\"unchen, Germany\\
$^{39}$ Max Planck Institute for Extraterrestrial Physics, Giessenbachstr. 1, D-85748 Garching, Germany\\
$^{40}$ INAF-Osservatorio Astrofisico di Torino, Via Osservatorio 20, I-10025 Pino Torinese (TO), Italy\\
$^{41}$ INFN-Sezione di Roma Tre, Via della Vasca Navale 84, I-00146, Roma, Italy\\
$^{42}$ Department of Mathematics and Physics, Roma Tre University, Via della Vasca Navale 84, I-00146 Rome, Italy\\
$^{43}$ Institut de Recherche en Astrophysique et Plan\'etologie (IRAP), Universit\'e de Toulouse, CNRS, UPS, CNES, 14 Av. Edouard Belin, F-31400 Toulouse, France\\
$^{44}$ INAF-Osservatorio Astronomico di Capodimonte, Via Moiariello 16, I-80131 Napoli, Italy\\
$^{45}$ Department of Astronomy \& Astrophysics, University of Toronto, 50 St George Street, Toronto, Ontario M5S 3H4, Canada\\
$^{46}$ Institut de F\'{i}sica d’Altes Energies (IFAE), The Barcelona Institute of Science and Technology, Campus UAB, 08193 Bellaterra (Barcelona), Spain\\
$^{47}$ Institute of Space Sciences (ICE, CSIC), Campus UAB, Carrer de Can Magrans, s/n, 08193 Barcelona, Spain\\
$^{48}$ Institut d’Estudis Espacials de Catalunya (IEEC), Carrer Gran Capit\'a 2-4, 08034 Barcelona, Spain\\
$^{49}$ Department of Physics "E. Pancini", University Federico II, Via Cinthia 6, I-80126, Napoli, Italy\\
$^{50}$ INFN section of Naples, Via Cinthia 6, I-80126, Napoli, Italy\\
$^{51}$ Dipartimento di Fisica e Astronomia “Augusto Righi” - Alma Mater Studiorum Università di Bologna, via Piero Gobetti 93/2, I-40129 Bologna, Italy\\
$^{52}$ INAF-Osservatorio Astrofisico di Arcetri, Largo E. Fermi 5, I-50125, Firenze, Italy\\
$^{53}$ Institut national de physique nucl\'eaire et de physique des particules, 3 rue Michel-Ange, 75794 Paris C\'edex 16, France\\
$^{54}$ Centre National d'Etudes Spatiales, Toulouse, France\\
$^{55}$ Institute for Astronomy, University of Edinburgh, Royal Observatory, Blackford Hill, Edinburgh EH9 3HJ, UK\\
$^{56}$ Jodrell Bank Centre for Astrophysics, School of Physics and Astronomy, University of Manchester, Oxford Road, Manchester M13 9PL, UK\\
$^{57}$ European Space Agency/ESRIN, Largo Galileo Galilei 1, 00044 Frascati, Roma, Italy\\
$^{58}$ Aix-Marseille Univ, CNRS, CNES, LAM, Marseille, France\\
$^{59}$ Institute of Physics, Laboratory of Astrophysics, Ecole Polytechnique F\'{e}d\'{e}rale de Lausanne (EPFL), Observatoire de Sauverny, 1290 Versoix, Switzerland\\
$^{60}$ Department of Astronomy, University of Geneva, ch. d\'Ecogia 16, CH-1290 Versoix, Switzerland\\
$^{61}$ INFN-Padova, Via Marzolo 8, I-35131 Padova, Italy\\
$^{62}$ Istituto Nazionale di Astrofisica (INAF) - Osservatorio di Astrofisica e Scienza dello Spazio (OAS), Via Gobetti 93/3, I-40127 Bologna, Italy\\
$^{63}$ Istituto Nazionale di Fisica Nucleare, Sezione di Bologna, Via Irnerio 46, I-40126 Bologna, Italy\\
$^{64}$ Institute of Theoretical Astrophysics, University of Oslo, P.O. Box 1029 Blindern, N-0315 Oslo, Norway\\
$^{65}$ Jet Propulsion Laboratory, California Institute of Technology, 4800 Oak Grove Drive, Pasadena, CA, 91109, USA\\
$^{66}$ von Hoerner \& Sulger GmbH, Schlo{\ss}Platz 8, D-68723 Schwetzingen, Germany\\
$^{67}$ University of Lyon, UCB Lyon 1, CNRS/IN2P3, IUF, IP2I Lyon, France\\
$^{68}$ Universit\'e de Gen\`eve, D\'epartement de Physique Th\'eorique and Centre for Astroparticle Physics, 24 quai Ernest-Ansermet, CH-1211 Gen\`eve 4, Switzerland\\
$^{69}$ Department of Physics and Helsinki Institute of Physics, Gustaf H\"allstr\"omin katu 2, 00014 University of Helsinki, Finland\\
$^{70}$ NOVA optical infrared instrumentation group at ASTRON, Oude Hoogeveensedijk 4, 7991PD, Dwingeloo, The Netherlands\\
$^{71}$ Institute for Computational Cosmology, Department of Physics, Durham University, South Road, Durham, DH1 3LE, UK\\
$^{72}$ Universit\'e C\^{o}te d'Azur, Observatoire de la C\^{o}te d'Azur, CNRS, Laboratoire Lagrange, Bd de l'Observatoire, CS 34229, 06304 Nice cedex 4, France\\
$^{73}$ University of Applied Sciences and Arts of Northwestern Switzerland, School of Engineering, 5210 Windisch, Switzerland\\
$^{74}$ Observatoire de Sauverny, Ecole Polytechnique F\'ed\'erale de Lau- sanne, CH-1290 Versoix, Switzerland\\
$^{75}$ Department of Physics and Astronomy, University of Aarhus, Ny Munkegade 120, DK–8000 Aarhus C, Denmark\\
$^{76}$ Institute of Space Science, Bucharest, Ro-077125, Romania\\
$^{77}$ Departamento de Astrof\'{i}sica, Universidad de La Laguna, E-38206, La Laguna, Tenerife, Spain\\
$^{78}$ Instituto de Astrof\'{i}sica de Canarias, Calle V\'{i}a L\`actea s/n, 38204, San Crist\`obal de la Laguna, Tenerife, Spain\\
$^{79}$ Dipartimento di Fisica e Astronomia “G.Galilei", Universit\'a di Padova, Via Marzolo 8, I-35131 Padova, Italy\\
$^{80}$ Centre for Electronic Imaging, Open University, Walton Hall, Milton Keynes, MK7~6AA, UK\\
$^{81}$ Centro de Investigaciones Energ\'eticas, Medioambientales y Tecnol\'ogicas (CIEMAT), Avenida Complutense 40, 28040 Madrid, Spain\\
$^{82}$ Infrared Processing and Analysis Center, California Institute of Technology, Pasadena, CA 91125, USA\\
$^{83}$ Universidad Polit\'ecnica de Cartagena, Departamento de Electr\'onica y Tecnolog\'ia de Computadoras, 30202 Cartagena, Spain\\
$^{84}$ INAF-IASF Bologna, Via Piero Gobetti 101, I-40129 Bologna, Italy\\
$^{85}$ Kapteyn Astronomical Institute, University of Groningen, PO Box 800, 9700 AV Groningen, The Netherlands\\
$^{86}$ Universit\'e de Paris, CNRS, Astroparticule et Cosmologie, F-75013 Paris, France\\
$^{87}$ INFN-Sezione di Torino, Via P. Giuria 1, I-10125 Torino, Italy\\
$^{88}$ Dipartimento di Fisica, Universit\'a degli Studi di Torino, Via P. Giuria 1, I-10125 Torino, Italy\\
$^{89}$ School of Physics and Astronomy, Cardiff University, The Parade, Cardiff, CF24 3AA, UK\\
$^{90}$ Department of Physics \& Astronomy, University of Sussex, Brighton BN1 9QH, UK\\
$^{91}$ Space Science Data Center, Italian Space Agency, via del Politecnico snc, 00133 Roma, Italy\\
$^{92}$ SISSA, International School for Advanced Studies, Via Bonomea 265, I-34136 Trieste TS, Italy\\
$^{93}$ INFN, Sezione di Trieste, Via Valerio 2, I-34127 Trieste TS, Italy\\
$^{94}$ Universidad de la Laguna, E-38206, San Crist\'{o}bal de La Laguna, Tenerife, Spain\\
$^{95}$ Dipartimento di Fisica - Sezione di Astronomia, Universit\'a di Trieste, Via Tiepolo 11, I-34131 Trieste, Italy\\
$^{96}$ Dept. of Physics, IIT Hyderabad, Kandi, Telangana 502285, India\\
$^{97}$ INAF-Istituto di Astrofisica e Planetologia Spaziali, via del Fosso del Cavaliere, 100, I-00100 Roma, Italy\\
$^{98}$ School of Physics, HH Wills Physics Laboratory, University of Bristol, Tyndall Avenue, Bristol, BS8 1TL, UK\\
$^{99}$ Research Program in Systems Oncology, Faculty of Medicine, University of Helsinki, Helsinki, Finland\\
$^{100}$ Department of Physics, P.O. Box 64, 00014 University of Helsinki, Finland\\
$^{101}$ Department of Physics, Lancaster University, Lancaster, LA1 4YB, UK\\
$^{102}$ Helsinki Institute of Physics, Gustaf H{\"a}llstr{\"o}min katu 2, University of Helsinki, Helsinki, Finland\\
$^{103}$ Centre de Calcul de l'IN2P3, 21 avenue Pierre de Coubertin F-69627 Villeurbanne Cedex, France\\
$^{104}$ Instituto de F\'isica Te\'orica UAM-CSIC, Campus de Cantoblanco, E-28049 Madrid, Spain\\
$^{105}$ Institut f\"ur Theoretische Physik, University of Heidelberg, Philosophenweg 16, 69120 Heidelberg, Germany\\
$^{106}$ Zentrum f\"ur Astronomie, Universit\"at Heidelberg, Philosophenweg 12, D- 69120 Heidelberg, Germany\\
$^{107}$ Sorbonne Universit{\'e}s, UPMC Univ Paris 6 et CNRS, UMR 7095, Institut d'Astrophysique de Paris, 98 bis bd Arago, 75014 Paris, France\\
$^{108}$ ICC\&CEA, Department of Physics, Durham University, South Road, DH1 3LE, UK\\
$^{109}$ INFN, Sezione di Lecce, Via per Arnesano, CP-193, I-73100, Lecce, Italy\\
$^{110}$ Department of Mathematics and Physics E. De Giorgi, University of Salento, Via per Arnesano, CP-I93, I-73100, Lecce, Italy\\
$^{111}$ Institute for Computational Science, University of Zurich, Winterthurerstrasse 190, 8057 Zurich, Switzerland\\
$^{112}$ Departamento de F\'isica, FCFM, Universidad de Chile, Blanco Encalada 2008, Santiago, Chile\\
$^{113}$ Department of Physics, P.O.Box 35 (YFL), 40014 University of Jyv\"askyl\"a, Finland\\
$^{114}$ Department of Physics and Astronomy, University College London, Gower Street, London WC1E 6BT, UK\\
}

\date{Received August XXXX, 2021; accepted XXXX, 2021}

\titlerunning{The Euclid Wide Survey}
\authorrunning{R. Scaramella et al.}

%

 %

\abstract{ %
\Euclid is a mission of the European Space Agency, designed to constrain the properties of dark energy and gravity via weak gravitational lensing and galaxy clustering. It will carry out a wide area imaging and spectroscopy survey (EWS) in visible and near infrared bands, covering approximately $15\,000\,\mathrm{deg}^2$ of
extragalactic sky on six years. The wide-field telescope and instruments are optimized for pristine PSF and reduced straylight, producing very crisp images.

This paper presents the building of the Euclid reference survey: the sequence of pointings of EWS, Deep fields and Auxiliary fields for 
calibrations, and spacecraft movements followed by Euclid as it operates in a step-and-stare mode from its orbit around the Lagrange point L2.

Each EWS pointing has four dithered frames; we simulate the dither pattern at pixel level to analyse the effective coverage. We use up-to-date models for the sky background to define the Euclid region-of-interest (RoI). The building of the reference survey is highly constrained from calibration cadences, spacecraft constraints, and
background levels; synergies with ground-based coverage are also considered. Via purposely-built software, we first generate a schedule for the Auxiliary and Deep fields observations. On a second stage, the RoI is tiled and scheduled with EWS transit observations, with an algorithm optimized to prioritize best sky areas, produce a compact coverage, and ensure thermal stability.

The reference survey RSD\_2021A is the optimized result of a modern survey design. It fulfills all constraints and is a good proxy for the final solution. The wide survey covers $\approx 14\,500\,\mathrm{deg}^2$. The limiting AB magnitudes ($5\sigma$ point-like source) achieved in its footprint are estimated to be 26.2 (visible) and 24.5 (near infrared); for spectroscopy, the H$_\alpha$ line flux limit is $2\times 10^{-16}$\,erg\,cm$^{-2}$\,s$^{-1}$ at $1600\,\mathrm{nm}$; and for diffuse emission the surface brightness limits are 29.8 (visible) and 28.4 (near infrared) mag arcsec$^{-2}$. 
}


\maketitle

\section{Introduction}\label{sec:intro}
    
Observations of distant type Ia supernovae \citep[e.g.][]{Riess1998AJ_116_1009R,Perlmutter1999ApJ_517_565P} together with those of the cosmic microwave background \citep[CMB; e.g.][]{2000Natur404_955D,Boomerang2002_ApJ564_559D,MAXIMA2000ApJ545L_5H, DASI2002ApJ568_46P,WMAP9yrs_2013_ApJS_208_19H,Planck2020A&A_641A_1P,Planck2020A&A_641A_6P} suggest that the spatial curvature of the Universe is close to zero. Despite the large contribution of dark matter (DM), however, the total matter density is still much lower than the critical matter density. As a consequence, a  non-zero value for the cosmological constant $\Lambda$ is usually introduced to complete the cosmological model. Although a cosmological constant can describe the data, it is not generally appealing  \citep{Weinberg1989RvMP61_1W}, and alternative solutions have been investigated, such as an evolving quantum field (dark energy, or DE), and a modification to general relativity on cosmological scales. We refer to \cite{Amendola18} for an extensive review of theoretical models.

To learn more about the nature of DE and DM, we need to quantify their impact on cosmological observations. In particular, we need to determine $H(z)$, the expansion history of the Universe as a function of redshift $z$, using geometrical tests, and measure the growth of large-scale structures through gravitational instability. The latter can be captured using the time derivative of the matter density contrast $\delta \equiv \delta\rho/\rho$, 
\begin{equation}
    \frac{{\rm d}\ln (\delta)}{{\rm d}\ln(a)} \, \simeq \, \Omega_{\rm m}^\gamma,
\end{equation}
where $a\equiv 1/(1+z)$ is the cosmic scale factor, and $\Omega_{\rm m}$ is the mean density divided by the critical density. For the canonical $\Lambda$ cold dark matter ($\Lambda$CDM) model in linear theory, $\gamma\simeq 0.55$, whereas it differs for other models of DE \citep{Amendola18}. We note that the growth of cosmic structures is also influenced by the DM characteristics. 

The exact nature of dark energy can be tested via its equation-of-state, $w=p/(\rho c^2)$, which directly influences the expansion history. In the general case, $w$ is a function of the scale factor, $w=w(a)$, and simplifies to $w=-1$ in case of a cosmological constant. Using a truncated Taylor expansion, one can write $w=w_0+w_a \,(1-a)$ and seek constraints on the possible values in the $w_0-w_a$ plane. 
Ideally, we should have a model that describes the redshift dependence of $w(a)$, and predicts how the growth of structure is affected by a modification of gravity. Nonetheless, $\gamma$, $w_0$ and $w_a$ provide convenient generic parameterizations that can be used to compare the expected performance of various cosmological probes. When Fisher matrix techniques are used \citep{euclid_coll_istf_2020}, the confidence areas in two-dimensional parameter spaces are ellipses. The inverse of the area of the 2-$\sigma$ ellipse in the $w_0-w_a$ plane, after marginalisation over all other cosmological and nuisance parameters, defines the DE figure of merit \citep[FoM;][]{Albrecht06,RedBook}. Hence, the larger is the FoM, the better (more informative) is the experiment.

Two of the best cosmological probes are galaxy clustering (GC) and weak gravitational lensing (WL), especially once combined in the so-called $3\times2$pt statistics. These are the two-point correlation of galaxies positions (GC uses galaxies as test particles in the expanding space-time to map the mass density contrast $\delta$ over time), the shear two-point correlations \citep[WL exploits the cumulative distortion effect of the tidal gravitational fields along the line of sight on the shapes of the galaxy images; see e.g.][for a review]{Kilbinger2015},
and the cross-correlation of the lens positions with the shear of the source galaxies, known as galaxy-galaxy lensing. Despite tremendous progress in GC and WL experiments in recent years \citep[e.g.][]{BOSS17, DES21, eBOSS20, Hildebrandt20}, much larger cosmological volumes need to be surveyed. 

Even though a clear ``target precision'' is lacking,  \cite{RedBook} argues that a ${\rm FoM} \geq 400$ provides constraints on $w(a)$ and $\gamma$ that can test key aspects of our current cosmological model. This target ${\rm FoM} \geq 400$ has driven the design of \Euclid, a medium class mission of the European Space Agency (ESA) that combines GC and WL. 

 To meet its primary science goal, \Euclid \citep{RedBook, Racca16} has to observe a large fraction of the extra-galactic sky both with multiband imaging and slitless spectroscopy. The sky area and mean number density of galaxies are specified by the scientific requirements of the GC and WL experiments \citep{RedBook, Rassat08, Cropper13, Massey13}:

\begin{itemize}
	\item  a $15\,000\,\deg^2$ survey of the extra-galactic sky, jointly for WL and GC to be completed in six years with all the necessary calibrations;
	\item an average galaxy number density of $30 \, {\rm arcmin}^{-2}$ that are useful for WL in the optical imaging data;
	\item an average galaxy number density of $1700\,\deg^{-2}$ with reliable redshifts from the  $\mathrm{H}_\alpha$ emission line spectroscopic data, useful for GC.\footnote{This is a revision of the \cite{RedBook} value after the removal in 2014 of blue-grism exposures in the wide survey.} 
\end{itemize}	
Moreover, we want to minimise systematic residuals in the error budget by maximising the uniformity in the coverage of the observed sky \citep{RedBook,Scaramell15}. 

The resulting survey is the Euclid Wide Survey (EWS) that will cover $15\,000\,\deg^2$ to a minimum depth of $m_{\rm AB}=24.5$\,mag in the visible band with a signal-to-noise ratio (SNR) of $10$ for sources extended as the $z \sim 1$ galaxies \citep[details in Sect.~\ref{sec:roi_SNR} and][]{RedBook,VIS16}. In the near-infrared \emph{Y, J} and \emph{H} bands, a depth of $m_{\rm AB}=24.0$\,mag will be reached with a minimum SNR of $5$ for point sources \citep{RedBook}. This is sufficient to complement ground-based multi-band observations that will be used to determine photometric redshifts (photo-$z$s) for the WL sources  \citep{Desprez2020}. Using slitless spectroscopy, \Euclid will detect line emission with a sensitivity of $f_{{\rm H}\alpha} \geq 2 \times 10^{-16}\,\ergscmsq $; and a SNR of 3.5 for a typical source of size $0\farcs5$ \citep{NISP14,NISP16}. Space-based observations provide excellent and consistent image quality at visible wavelengths for WL, and sufficient depth in the NIR bands, both unattainable from the ground for this type of survey. 
\begin{figure}[htb]
 \begin{centering}
 \includegraphics[width=\onecolspan]{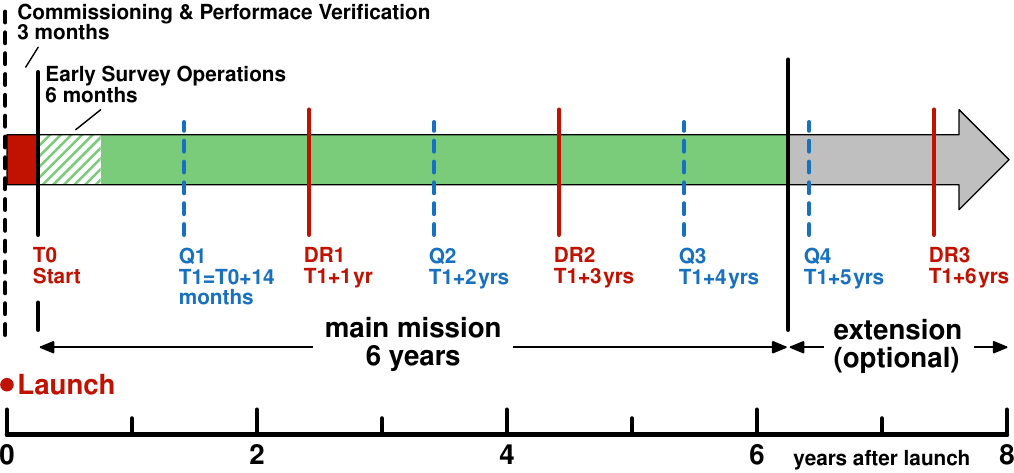}
 \caption{The \Euclid  timeline. The EWS has to be carried out within the six year mission baseline and will start 3 months after the launch, following a commissioning (1 month) and a Performance Verification period (2 months). An extension of operations beyond the six years is possible and will be decided in due time. A faster replanning is allowed during the Early Operation Phase of six months. The three main Data Releases (DR\#) are shown. The plan is to have $2500\,\deg^2$ made public in DR1, to grow to $7500$ in DR2 and be complete at DR3 for $15\,000\,\deg^2$. In addition, four Quick Data Releases (Q\#) are foreseen, each of $\sim 50\,\deg^2$. }  
  \label{Fig:timeline}
 \end{centering}
\end{figure}

With this design, Fisher matrix analyses forecast that EWS $3\times2$pt datasets will obtain a DE FoM of 500 for a non-flat $w_0-w_a$~CDM model in an optimistic setting
\citep[defined by the range of scales used; see][]{euclid_coll_istf_2020}. It is interesting to compare \Euclid $3\times2$pt forecasts with the constraints  obtained by DES, which can be considered as representative of ongoing Stage III surveys. DES Year~1 constraints for a $w_0-w_a$~CDM model \citep{DES19} are intermediate results, where the uncertainties obtained for $w_0$ and $w_a$ are a factor of 10 larger than the \Euclid forecasts in the pessimistic setting (roughly translating to a factor of 100 in the FoM). 
The DES DE FoM from the DES Year~3 analysis of a more complete dataset  is not yet available, but we can use instead the results on $S_8 = \sigma_8 (\Omega_{\rm m/0.3})^{1/2}$ to compare Stage-III constraints with Stage-IV \Euclid forecasts.

Using a $3\times2$pt dataset and marginalizing over 25 nuisance parameters and 7 cosmological parameters of a flat $w$CDM model, DES Year~3  gets $3.2\%$ errors on $S_8$ \citep{DES21}.
Using the Fisher matrix for $3\times2$pt obtained in \cite{euclid_coll_istf_2020} for a $w_0-w_a$~CDM model, and adding additional 10 nuisance parameters to represent the uncertainty on the shear multiplicative bias of each redshift bin (in order to increase the number of nuisance parameters to have a fair comparison with DES), we  marginalize over a total of 25 nuisance and   cosmological parameters, getting $1.25\%$ errors on $S_8$ in a pessimistic setting and $0.68\%$ errors in an optimistic setting. 
Although the comparison is not completely fair,  because of the different intrinsic alignments and bias models, less nuisance parameters considered in the \Euclid forecast and a slightly different set of cosmological parameters, 
these numbers are roughly in agreement with what one would obtain by simply scaling the DES Year~3 results for the \Euclid increase in area and source number density.

The EWS has to be carried out within the six year mission baseline and will start 3 months after the launch, following a commissioning (1 month) and a Performance Verification period (2 months). 
Figure \ref{Fig:timeline} shows the \Euclid  timeline with the data release planning.
The first major `data release' (DR1), corresponding to  2500\,deg$^2$ of the EWS is planned to take place one year after $T_1$  ($T_1=14$ months after launch), the second data release (DR2), is expected to release 7500\,deg$^2$ three years after $T_1$, and the final one (DR3) will release the full survey (15\,000\,deg$^2$) six years after $T_1$. In between there will be other `quick data releases': Q1 of $50$\,deg$^2$ is planned at $T_1$, and Q2, Q3 and Q4, will take place two, four, and five years after Q1, respectively.

In addition to the main survey, a significant fraction of time will be spent to calibrate the instruments and to characterise the target galaxies. This results in some fields to be observed to greater depth  than the wide survey (typically 2 magnitudes deeper). These deep fields have great legacy value beyond the cosmological core science. While aspects of non-core science did not influence the design of the spacecraft and instruments, they are taken into account in the design of the EWS to maximise \Euclid scientific return.  In fact, it must be noticed that the large decrease in the background with the wavelength dramatically increases the SNR in the NIR bands when compared to Earth-based observations affected by airglow, which instead increases with wavelength. This makes even a small space telescope competitive with a large ground telescope which suffers from a background dominated by atmospheric emission in the NIR bands. The relative gain is such that, in order to cover the same areas planned for \Euclid and at the same depths, a ground-based NIR survey on existing facilities would need to observe for several centuries. Regarding other space-based facilities, we notice that the James Webb Space Telescope (JWST) will be in orbit as well and with its diameter of $6.5\,\mathrm{m}$ will go much deeper and faster than \Euclid although only on very small areas (JWST field of view is 75 time smaller than the Euclid one). Hence the two facilities are complementary and, moreover, JWST will likely benefit from targets selected from the Euclid Surveys.

The challenge is to fit all these observations into a finite time allocation set by the limitation of the mission, which is six years, whilst fulfilling a wide range of constraints, which are reviewed in detail in this paper. Part of the survey optimisation involves selecting the best areas of sky to use, which in turn relies on a good model of the properties of the observable sky, such as Galactic extinction and the zodiacal background. We also need to model the distribution of (bright) stars, as their stray light lowers the observed galaxy number density.

This paper focuses on the design of the EWS, while the deep fields are described in a companion paper (Scaramella \emph{et al.}, in prep., hereafter [Sc21]). The EWS design takes into account the main backgrounds which impact any large area survey, the sequence of operations, the many limitations to the pointing of the telescope. The EWS is at an advanced stage, fulfilling the key survey requirements over the full mission. Survey scenarios at this stage therefore show the detailed feasibility of the mission but are subject to further optimisation. Nevertheless, the results we present and their discussion are instructive and useful for any future large area survey from space or ground which aims to combine imaging and spectroscopy. 

The paper is organised as follows. The spacecraft is described in Sect.~\ref{sec:spacecraft}, followed by a summary of \Euclid's instruments in Sect.~\ref{sec:instruments}. In Sect.~\ref{sec:dither} the reference observation sequence (ROS) is introduced, including a study of dithering scenarios.  Models of zodiacal light, stray light effects, and other environmental properties, define the `region of interest' (RoI) used as input for the implementation of the Euclid Reference Survey Definition (RSD). These effects and the properties of the resulting RoI are presented in Sect.~\ref{sec:keyparams}, where we also discuss complementary ground-based observations. Section.~\ref{sec:calibs} describes the implementation of the calibration program. Observations of sample characterisation fields and the EDS are briefly mentioned in this context. The construction of the EWS is presented in Sect.~\ref{sec:wide}. We present the most recent outcome of the survey optimisation (mid 2021) in Sect.~\ref{sec:RSD}. This solution is a good proxy for the actual survey. We conclude in Sect.~\ref{sec:summary}.

In the Appendix we provide a list of acronyms used in this paper.
\section{The spacecraft and telescope}\label{sec:spacecraft}

\subsection{The spacecraft}\label{sec:sc_overview}

The spacecraft comprises a service module (SVM) and a payload module (PLM), connected by an interface structure designed to maximise thermal decoupling. The PLM includes the main instruments, the folded beam optical components of the telescope, the radiators, and the fine guidance system (FGS). The SVM provides the main Spacecraft services: Power Generation, conditioning and distribution, Sun shield and Solar Array, telecommunication with ground (Low and High Gain antenna), Attitude and Orbit Control System (including FGS) and support the Instruments Warm Electronics. 
Details are given in \cite{RedBook} and \cite{Racca16}.

 \begin{figure}[!htb]
	\begin{centering}
\resizebox{\onecolspan}{!}{ 	
\includegraphics[scale=0.5]{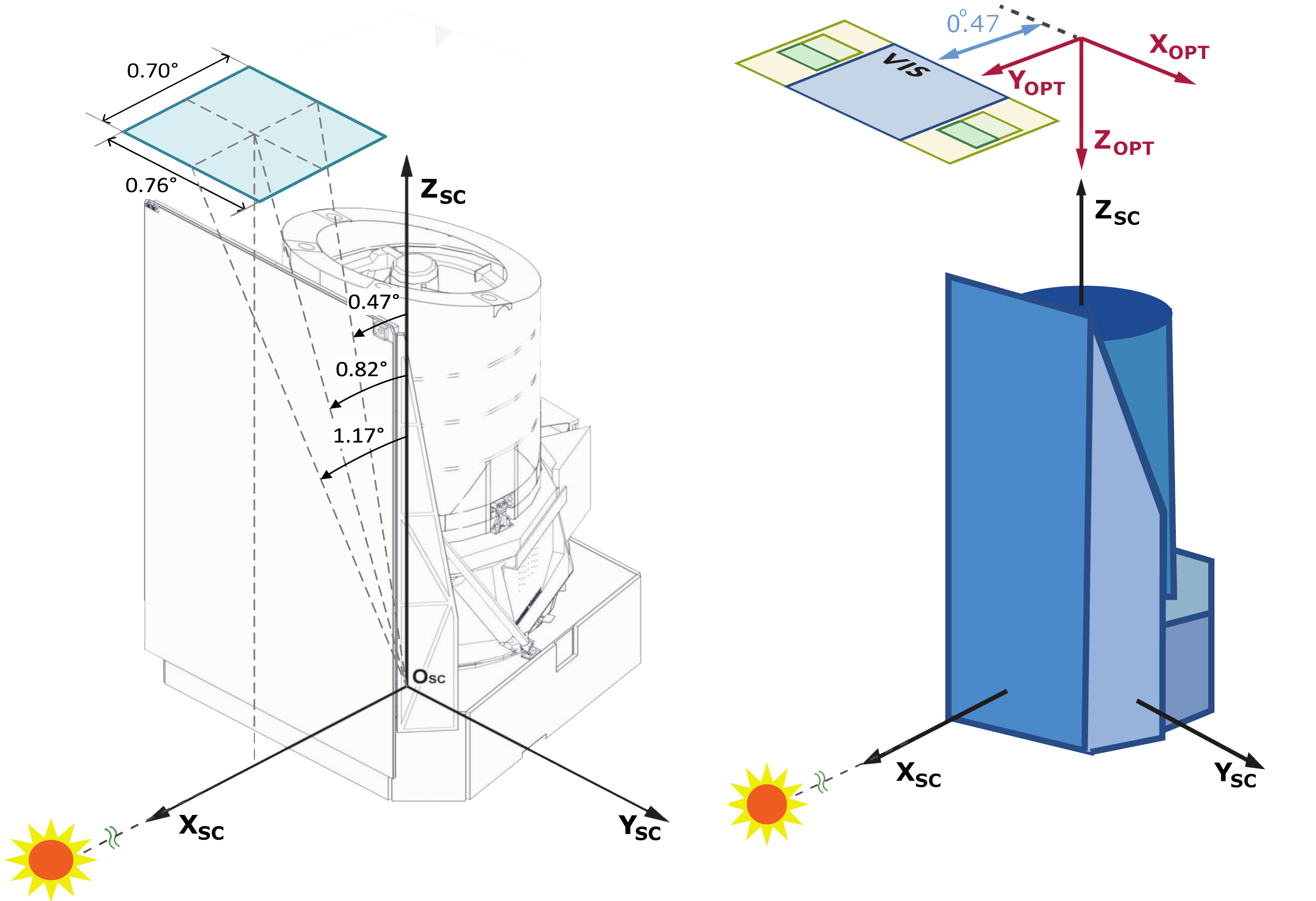} 
 }
\caption{\emph{Left panel:} \textit{Euclid} Spacecraft Reference Frame.  $X_{\rm SC}$ points toward the Sun disk center. The edge and center of the joint FoV are offset by \ang{0.47;;} and \ang{0.82;;}, respectively. The longer side of the FoV is typically aligned with ecliptic meridians during observations.
\emph{Right panel:} \textit{Euclid} Optical Reference Frame. Notice that for the latter the $X_{\rm opt}$ $Y_{\rm opt}$ plane is defined as looking onto the sky with $\hat z$ towards the spacecraft. There are four additional chips used as fine guidance sensors (FGS) placed on each side of the VIS FoV.}
\label{Fig:spacecraft_FOV_and_ref_frames}
\end{centering}
\end{figure}

\Euclid has severe constraints in pointing to ensure maximal thermal stability, which are described in this paper and limit the standard operations. Therefore it is important to describe in detail the attitude of the spacecraft. The \Euclid Spacecraft Reference Frame $(O_{\rm SC}, X_{\rm SC}, Y_{\rm SC}, Z_{\rm SC})$ is defined as follows (see Fig.~\ref{Fig:spacecraft_FOV_and_ref_frames} for a graphical representation): 
\begin{itemize}
	\item[\textbullet] $O_{\rm SC}$: origin is at the point of intersection of the longitudinal launcher axis with the launcher adapter interface plane (the plane of separation of the spacecraft from the launcher); 
	\item[\textbullet] $+Z_{\rm SC}$ is 
	in the direction perpendicular to the launcher interface plane, positive in the direction of the launch; 
	\item[\textbullet] $+X_{\rm SC}$ is in the launcher interface plane, directed to a physical mark on the interface ring nominally aligned with the solar array such that the $+X_{\rm SC}$ vector is perpendicular to the solar array and pointing towards the sun;
	\item[\textbullet] $+Y_{\rm SC}$ is in the remaining direction of the right-handed orthogonal triad.
\end{itemize}
The orientation of the telescope optical reference frame, projected onto the sky, is also specified in Fig.~\ref{Fig:spacecraft_FOV_and_ref_frames}.
The field of view (FoV) Reference Frame is centred on the centre of the FoV itself and is such that 

$X_{\rm FoV}= -X_{\rm OPT}$ and  $Y_{\rm FoV}=Y_{\rm OPT} -0\fdg 82$, 
taking into account the shift of the edge of the FoV of $0\fdg 47$, and its half size of of $0\fdg 35$ (see Fig.~\ref{Fig:spacecraft_FOV_and_ref_frames} and Table~\ref{tab:FoVsizes}). 

\subsection{The telescope}\label{sec:telescope}

\Euclid's PLM \citep{Racca16} is designed around a three-mirror anastigmat Korsch design telescope with silicon carbide (SiC) mirrors and truss \citep{Korsch72, Bougoin16}. 
The sizes of the telescope components are: primary pupil $R_1 = 0.6$~m, primary mirror (M1) stopper $R_2$ = 0.1975 m, spider arm mean length $R_3$ = 0.44 m, spider arm thickness $L$ = 0.012 m. This provides a total collecting area of
$A=\pi \left(R_1^2-R_2^2\right)-3 R_3 L = 0.99\,\mathrm{m}^2$.
\Euclid has two instruments onboard, the visible imager (VIS; Sect.~\ref{sec:VIS}) and the near-infrared spectrometer and photometer (NISP; Sect.~\ref{sec:NISP}). The wavelength separation at $\sim920$\,nm between the two instruments is performed by a dichroic plate located at the exit pupil of the telescope. The two focal planes image the same part of the sky, allowing multiple data acquisition with a single telescope pointing (see Sect.~\ref{sec:std_sequence}). The coordinates of the focal plane as projected on the sky are shown in Fig.~\ref{Fig:spacecraft_FOV_and_ref_frames}. 

\subsection{Pointing angles}\label{sec:orbit}

\begin{figure}[htb]
 \begin{centering}
  \resizebox{\onecolspan}{!}{ %
 \includegraphics{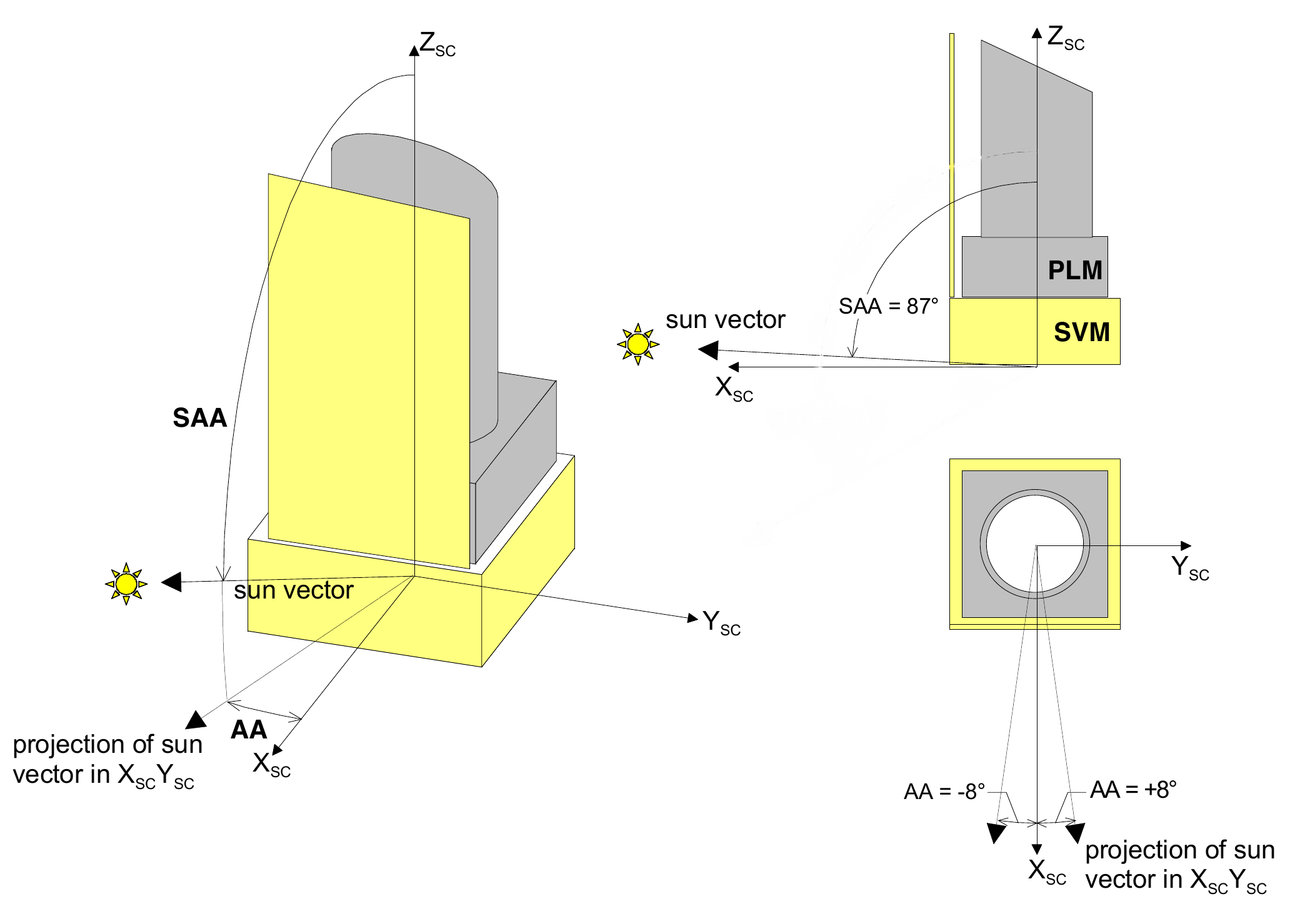} }
 \caption{Illustration of the relevant pointing angles defined in the text. In the top right image the minimal SAA (pointing towards the Sun) is indicated, while the maximum SAA (pointing away from the Sun) adopted in the survey is $110\degr$ (see also Sect.~\ref{sec:angles}). In the lower right image AA is shown; the maximum range allowed for the survey is $\pm 5\degr$, corresponding to a large margin with respect to the spacecraft capabilities.}
  \label{Fig:spacecraft_angles}
 \end{centering}
\end{figure}

The main reference frames are shown in Fig.~\ref{Fig:spacecraft_FOV_and_ref_frames}. The following angles (see  Fig.~\ref{Fig:spacecraft_angles}) have operational ranges that constrain a pointing and therefore target visibility:

\begin{itemize}
  	\item Solar aspect angle (SAA): the angle between the spacecraft's $+Z_{\rm SC}$ axis (telescope pointing direction) and the direction to the centre of the Solar disk;
	\item alpha angle (AA): the angle between the Sun vector projected onto the $X_{\rm SC}$--$Y_{\rm SC}$  plane and the $+X_{\rm SC}$ axis. It increases as the spacecraft rotates clockwise about its $+Z_{\rm SC}$ axis; 
    \item Solar panel Solar aspect angle (SPSAA): The angle between the spacecraft $+X_{\rm SC}$ axis and the direction to the centre of the Solar disk. 
\end{itemize}
%

\subsection{Orbit and operation mode}
\label{sec:slews}

\begin{figure}[thb]
 \begin{centering}
  \resizebox{\onecolspan}{!}{ %
 \includegraphics{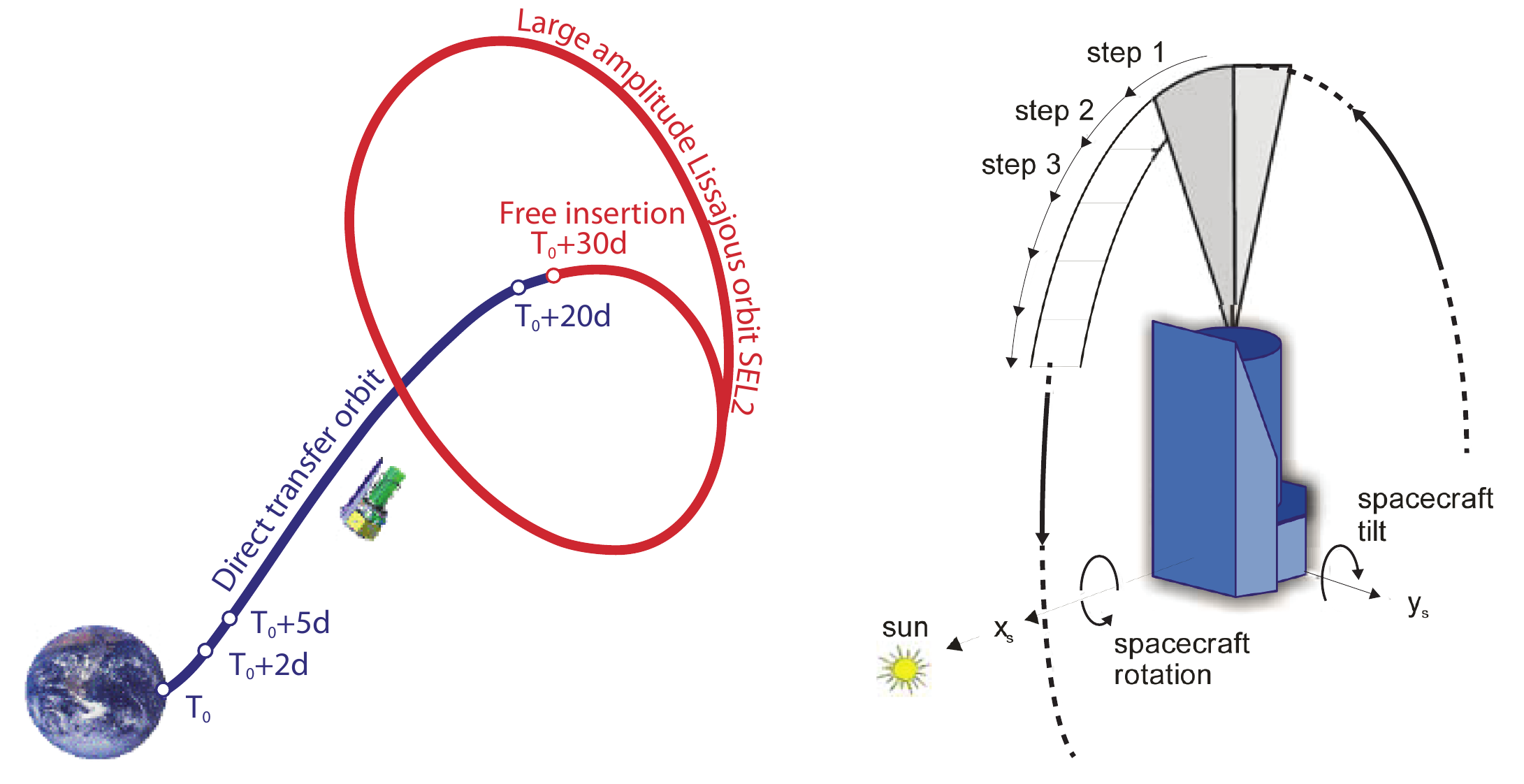} 
 }
 \caption{\emph{Left panel:} \Euclid's transfer orbit following launch at time $T_0$, and the subsequent Lissajous orbit at the Sun-Earth Lagrangian point L2 (SEL2). \emph{Right panel:} \Euclid's main step-and-stare operating mode, showing the steps as rotations around $X_{\rm SC}$. } 
  \label{Fig:orbit_and_step_and_stare}
 \end{centering}
\end{figure}

\Euclid will operate at the Sun-Earth Lagrangian point L2, following a yearly orbit with a libration within $\pm 0\fdg 41$ across the ecliptic plane (Fig.~\ref{Fig:orbit_and_step_and_stare}). The Lissajous orbit is dynamically unstable and requires regular orbital maintenance, currently planned to last one day every four weeks, i.e. $\sim3\%$ of the total mission time. This orbit offers a very stable thermal environment and maximises the visible sky at any time. 

\Euclid employs a step-and-stare mode, acquiring data on a fixed sky field and then slewing to the next pointing.
Slews come in two types, depending on the value of the eigenslew $\epsilon$ defined as the angle between one field quaternion\footnote{In 3-dimensional space, according to Euler's rotation theorem, any rotation or sequence of rotations of a rigid body or coordinate system about a fixed point is equivalent to a single rotation by a given angle $\theta$  about a fixed axis (called the Euler axis) that runs through the fixed point. Therefore the quaternion fully describes the spacecraft attitude and a single rotation quaternion relates one pointing to another.} and the next, which can be decomposed in an arc connecting two separate pointings on the sky plus a rotation around $Z_{\rm SC})$. 
Eigenslews $\epsilon\leq\ang{3.6;;}$ are considered `small slews', and $\epsilon>\ang{3.6;;}$ are `large slews'.
\Euclid has adopted a specific hybrid on-board Attitude and Orbit Control System (AOCS) architecture where slews are performed using four reaction wheels in stop and go mode, and pointing stabilisation is achieved using low-noise cold gas micro-thrusters. With this solution, slew time is reduced  thanks to the high torque provided by the reaction wheels. Slews only consume cold gas for the tranquilization transient phase, large slews requiring longer tranquilization periods.
Therefore, \Euclid's lifetime slew budget is limited to $950$ large and $2.5\times10^5$ small slews. The latter are weighted in the budget: for $\epsilon\leq\ang{1.2;;}$ they count as a single slew, and for $\ddeg 1.2<\epsilon\leq\ang{3.6;;}$ a penalty occurs proportional to $\epsilon$. The slew constraints imply that the EWS must be implemented mostly with small slews (preferably with $\epsilon\leq\ang{1.2;;}$), and that fields observed consecutively in time must be spatially adjacent.

\subsection{Transits and visibility}\label{sec:angles}

The SAA and AA ranges define how much the spacecraft can deviate from observing at `transit' meridian, which are the 
two ecliptic meridians defined by the perpendicular to the spacecraft's \Xsc axis (for a transit ${\rm SAA}=90\degr$).
The size and geometry of the Sun shield limit the SAA and AA ranges that can be used for observations. The ranges (and variations of) SAA and AA are constrained further by the fact that \Euclid needs great thermal stability to minimise temporal point spread function (PSF) variations. The SAA limits allow the telescope to `depoint' (i.e. to rotate around \Ysc) from transit to a maximum of $3\degr$ towards the Sun (${\rm SAA} = 87\degr$), and up to $20\degr$ away from the Sun (${\rm SAA} = 110\degr$), while the AA limits allow the telescope to rotate around \Zsc up to $|{\rm AA}|\leq 5\degr$.

In addition, the orbit libration mentioned in Sec.~(\ref{sec:slews}) imposes an additional $0\fdg 41$ buffer for the spacecraft orientation angles with respect to the Sun. This decreases the allowed ranges of both SAA and AA 
by  $0\fdg 41$ on each side of their range intervals.

\begin{figure}[thb]
\centering
\includegraphics[trim=1 0 0 0]{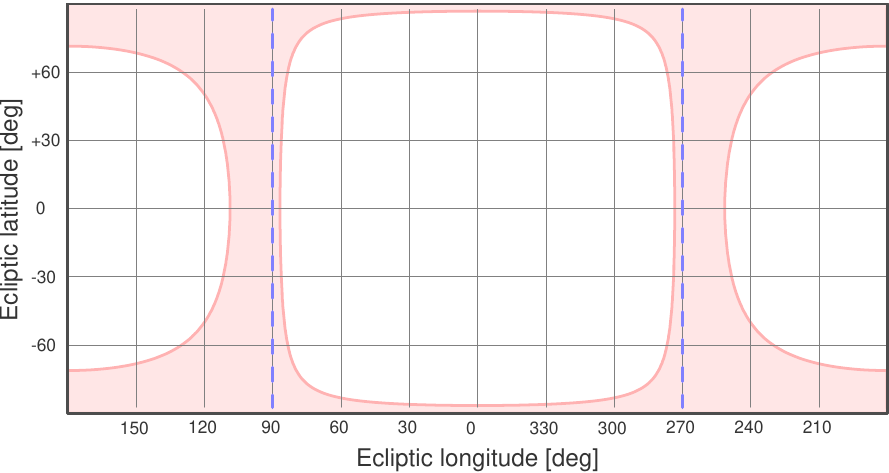} 
\caption{Strip of sky observable by \Euclid at a given epoch (in light red). The dashed blue line is the great circle orthogonal to the Sun  with the satellite observing orthogonal to the sun for a transit  at $0\deg$. The asymmetry of the red region with respect to the transit blue lines is due to the asymmetry of the usable SAA range with respect to $90\degr$ (wider when pointing away from the Sun).} 
\label{Fig:instantaneous_visibility}
\end{figure}

The allowed ranges of SAA and AA define the instantaneous sky visibility, shaped along the full circle defined by the two meridian transits (see Fig.~\ref{Fig:instantaneous_visibility}). The orbit progresses with \Euclid's revolution around the Sun, continuously changing the visible sky enabling a full sky survey. Given the symmetry of the transit meridians, the spacecraft has access to the same region of the sky every six months by pointing in the `leading' direction (towards the direction of motion around the Sun) or, flipping the telescope, six months later pointing in the `trailing' direction.

Figure~\ref{Fig:instantaneous_visibility} shows an example of the instantaneous sky visibility for a generic transit.
It is evident that two small regions located at the ecliptic poles have perennial (continuous) visibility, whereas the lowest ecliptic latitudes
can be observed only when crossed by a transit meridian. In practice, at any given time (or from a given position in the orbit) \Euclid can scan an annulus on the sky, and consequently most of the sky must be observed at or close to transit. Observations that require long and regular visibility (such as for the EDS) can only be fulfilled in a very limited area on the sky at high ecliptic latitudes.

In general, a depointing (SAA $\neq 90\degr$) induces a rotation of the focal plane with respect to the transit ecliptic meridian. To counterbalance, the spacecraft must rotate around the $+Z_{\rm SC}$  axis to keep the alignment with the transit meridian. The amplitude of this rotation must stay within $|{\rm AA}|\leq4\fdg59$ to fulfil the thermal and orbit libration constraints. This effect becomes larger with increasing ecliptic latitudes and requires ad hoc solutions for the scheduling (see Sect.~\ref{sec:polar_caps}).

\section{The instruments}\label{sec:instruments}


 \begin{figure}[htb]
   \begin{centering}
   \resizebox{\onecolspan}{!}
    {\includegraphics[]{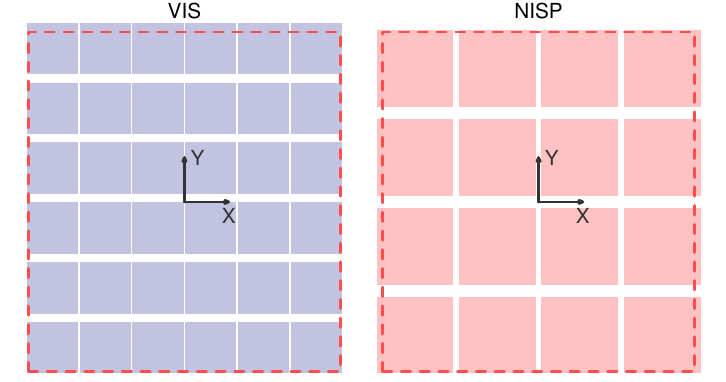}}
 \caption{\emph{Left panel}: The VIS FPA, illustrating the detector placement. The dashed line shows the joint FoV of both instruments. Two narrow strips at the extremes of the Y axis are outside the joint FoV. \emph{Right panel}: The NISP FPA, with two narrow strips at the extremes of the X axis outside the FoV (the Reference Frame is $X_{\rm FoV}-Y_{\rm FoV}$  see also Fig.~\ref{Fig:spacecraft_FOV_and_ref_frames} and the definition at the end of Sect.~\ref{sec:sc_overview}).}
 \label{Fig:VIS_and_NISP_FoV}
    \end{centering}
   \end{figure}

\subsection{VIS instrument}\label{sec:VIS}

The visible imager instrument (VIS) contains a focal plane array (FPA)
 consisting of $6\times 6$ Teledyne e2v CCDs ($4\,{\rm k}\times 4\,\rm{k}$ pixels each) with a Nyquist driven  plate scale of $0\farcs1$\,pixel$^{-1}$, yielding a field-of-view (FoV) of 0.56\,deg$^2$ including detector gaps. For details, see Table~\ref{tab:FoVsizes} and the left panel of Fig.~\ref{Fig:VIS_and_NISP_FoV}. In addition to the gaps, the central four rows in each detector serve as charge injection lines (which cannot be read out).

VIS is optimised to detect spatially resolved images of galaxies in the 550--900\,nm passband \citep[hereafter referred to as the `VIS band';] []{VIS14, VIS16}. The VIS nominal survey images (dithered; see Sect.~\ref{sec:dither}) will have at a minimum SNR of $10\,\sigma$ (average $15.8\,\sigma$) for extended sources (full width at half maximum, FWHM, such that $ {\rm {FWHM}_{ gal}} > 1.25 \, {\rm {FWHM}_{PSF}} \simeq 0\farcs 225$) at a detection limit of $m_{\rm AB}=24.5$\,mag (see Sect.~\ref{sec:roi_SNR}). These will enable accurate galaxy shape measurements for an average of 30 arcmin$^{-2}$ galaxies  over the survey area \citep{RedBook,Massey13}. Besides WL shape measurement, VIS data are also used to improve photo-$z$ estimation, by enabling optimal photometric extraction of the less resolved, complementary ground observations thanks to its diffraction-limited image quality.
To maximise the SNR for the shape measurements, the VIS band is rather broad (see Fig.~\ref{Fig:filter_shapes}), encompassing the Sloan Digital Sky Survey (SDSS) $r$ and $i$ bands, and the bluer half of the $z$-band.

The VIS central data processing unit constructs the images from the pixel data and compresses them in a lossless manner in approximately $250\,\rm{s}.$ No additional image processing will be done on board to maintain full control over systematic errors. The data will be transferred to the ground with a rate of approximately $520\,\rm{Gbit/day}$ \citep{Racca16}.

\subsection{NISP instrument}\label{sec:NISP}
The near-infrared spectrometer and photometer \citep[NISP;][]{NISP14,NISP16} contains an array of $4\times 4$ HAWAII-2RGs detectors ($2\,{\rm k} \times 2\,{\rm k}$ pixels each) with a plate scale of $0\farcs3$ pixel$^{-1}$,  undersampling its diffraction limited PSF (Fig.~\ref{Fig:VIS_and_NISP_FoV}, right panel). Table~\ref{tab:FoVsizes} shows the size of the FPA, FoV, and gaps between the detectors. Note  that in the $Y$ direction of the focal plane, the central gap ($86\farcs 1$ wide) is narrower than the two outer gaps ($101\farcs 4$ wide).

\begin{figure}[!hbt] 
	\begin{centering}
		\resizebox{\onecolspan}{!} 
		{\includegraphics{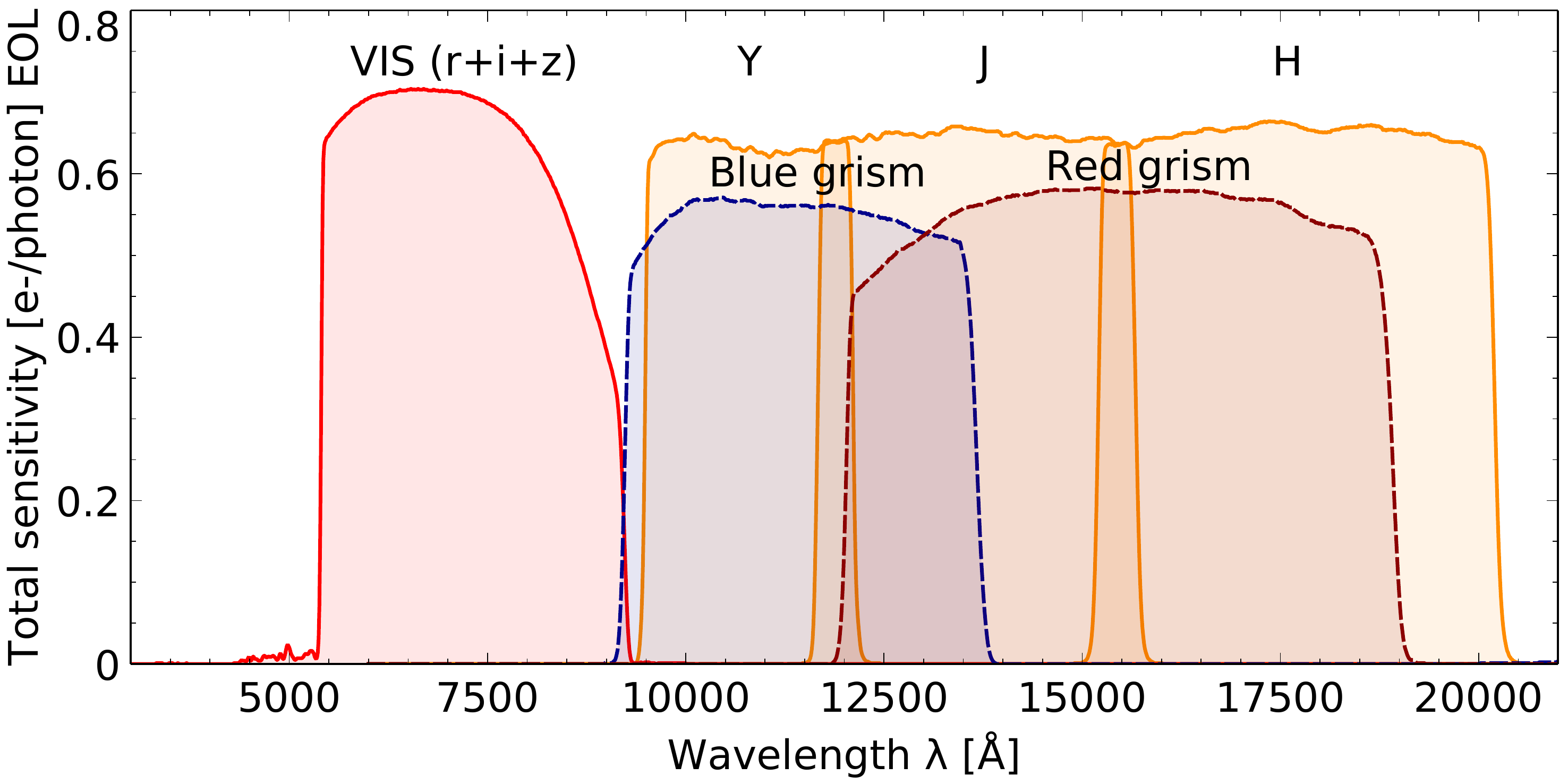}}
		\caption{Total sensitivity of \Euclid's photometric and spectroscopic bands (solid and dashed lines, respectively). The sensitivity (in electrons per photon) includes all optical surfaces as well as the detectors'  average quantum efficiency, all considered at their expected end of life (EOL) performance after six years at L2 \citep[degraded by radiation damage and contamination, see e.g.][]{Venancio20}.}
		\label{Fig:filter_shapes}
	\end{centering}
\end{figure}

NISP is designed to carry out slitless spectroscopy (NISP-S) and imaging photometry (NISP-P) at near-infrared (NIR) wavelengths (see Fig.~\ref{Fig:filter_shapes}). 
By using its grism and filter wheel assemblies (GWA and FWA, respectively), NISP can switch between slitless spectroscopy and imaging modes, which are detailed in the following. NISP will transfer data to the ground with a rate of approximately $290\,\rm{Gbit/day}$, for a total of $810\,\rm{Gbit/day}$, smaller than the spacecraft allocation of $850\,\rm{Gbit/day}$.

\begin{table}[hbt]
\caption{Sizes of the VIS and NISP focal planes and their corresponding FoVs, as well as angular sizes of detectors, detector gaps, and width of the VIS charge injection lines (see Figs.~\ref{Fig:spacecraft_FOV_and_ref_frames} and
 \ref{Fig:VIS_and_NISP_FoV}).} 
	\centering           
	\scalebox{0.85}{  \renewcommand{\arraystretch}{1.5} 
		\begin{tabular}{| c | c | c | c | c|} 
			\hline 
 &  \multicolumn{2}{c|}{VIS} & \multicolumn{2}{c|}{NISP} \\
 \cline{2-5}
 & X Size & Y Size & X Size & Y Size \\
 \hline
Focal Plane [mm] & 302.71 & 336.59 &155.85 & 164.48 \\
\hline
Plate scale [arcsec/mm]&  
\multicolumn{2}{c|}{8.33} &
\multicolumn{2}{c|}{16.70} \\
\hline
FoV [deg] & 0.700& 0.778& 0.723& 0.763\\
\hline
Detectors [$^{\prime}$] &6.82 &6.89 & 10.21 &10.21 \\
\hline
Detector gaps [$^{\prime\prime}$] &12.7  & 64.4& 50.6 &101.4 / 86.1 \\
\hline
Charge injection gaps [$^{\prime\prime}$] & N/A & 0.4 & N/A & N/A \\
\hline
		\end{tabular}
		}
	\label{tab:FoVsizes}  
\end{table}  

\begin{figure*}[!htb]
	\begin{centering}
		\resizebox{0.95\twocolspan}{!} 
		{\includegraphics{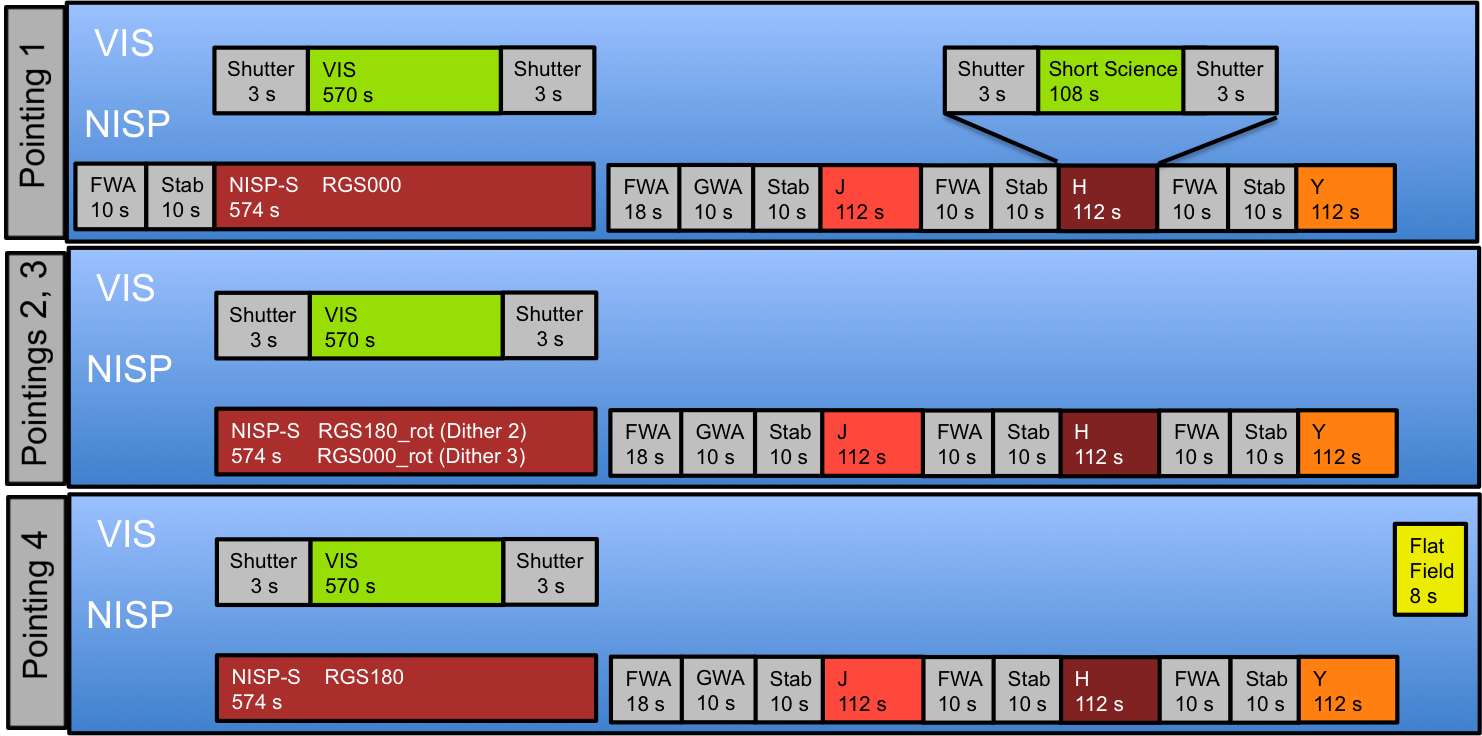}}
		\caption{The (simplified) ROS executed for each field, observing four dithered pointings. Each pointing results in a VIS, $Y$, $J$ and $H$ image, and one spectral exposure. Filter and grism wheel movements are shown, together with stabilisation times (``Stab''). Not shown are wheel movements during the dither slew, readout and processing done simultaneously with other operations and slews, overheads, and various inline calibrations.}
		\label{Fig:std_squence}
	\end{centering}
\end{figure*}

\subsubsection{Slitless spectroscopy}\label{sec:NISP-S}

\Euclid has a ``blue'' grism (BGS000) covering the 0.92--1.25\,\micron\, wavelength range, and three ``red'' grisms (RGS000, RGS180 and RGS270) covering 1.25--1.85\,\micron\, \citep{Costille16}. The blue grism is not used for the EWS observations, and only employed for part of the EDS. The numeric labels indicate the dispersion directions, offset by $90\degr$ for the red grisms. Different dispersion directions are required to disentangle the spectra of various objects in the slitless spectroscopic exposures of the EWS. Due to a non-conformity discovered in 2020 \citep{Laureijs20}, the RGS270 will not be used in the survey observations. Instead, the RGS000 and RGS180 will be rotated in the ROS by $-4\degr$ and $+4\degr$, respectively (see Sect. \ref{sec:std_sequence} for details).

The red grisms disperse the light with nearly constant spectral resolution of $1.354\,{\rm nm/px}$ which gives   ${\cal R}=\lambda / \Delta \lambda \sim 450$ for an object of diameter $0\farcs5 $. This is larger than the minimum required value of 380 to achieve an error on the measured redshift of $ \sigma (z) < 0.001 (1+z) $.
The spectroscopic observations support the GC probe and are optimised to detect the redshifted H$\alpha$ emission of galaxies at $z=0.9$--$1.8$. With a detection limit of $2\times 10^{-16}\,\ergscmsq $ $(3.5\,\sigma)$ for a typical source of size $0\farcs5$ at 1600 nm (see Sect.~\ref{sec:roi_SNR}), NISP should be able to determine spectroscopic redshifts for at least 1700~galaxies $\deg^{-2}$ on average in the corresponding wavelength range 1250--1850\,nm. This estimate, however,  strongly depends on the intrinsic luminosity function of H$\alpha$ emitters, which is still uncertain in the redshift range observed by \Euclid\citep{Pozzetti16}. Because the redshift is based on an emission line, passive galaxies will be underrepresented in the spectroscopic sample, with a bias against dense enviroments.

\subsubsection{Photometry}\label{sec:NISP-P}

Photometry will be measured for objects down to a minimum of $m_{\rm AB}=24.0$\,mag  for $5\, \sigma$ point-like source  in the $Y$, $J$ and $H$ passbands. The photometric data support the GC experiment by providing the reference images needed to extract the spectra in the (slitless) dispersed images. The NIR photometric data, however,  critically complement the ground-based observations (Sect.~\ref{sec:ground_based}) in getting accurate photometric redshift estimates, at the primary probe level mainly needed for the WL experiment and essential for many other astronomical science aspects.

\subsection{The Euclid joint field of view}\label{sec:common_fov}

The intersection of the VIS and NISP FoVs defines the \Euclid joint FoV, with the $X$ and $Y$ dimensions defined by VIS and NISP, respectively. The \Euclid FoV is $0.53$\,deg$^2$. Its borders are shown by the dashed lines in Fig.~\ref{Fig:VIS_and_NISP_FoV}, resulting from the overlap of the VIS and NISP FoVs 
aligned on an edge.
The left and right edges of the NISP FoV and the top and bottom edges of the VIS FoV are outside the joint \Euclid FoV.

\section{Observing and dithering sequences} \label{sec:dither}

\subsection{The reference observation sequence (ROS)}\label{sec:std_sequence}

\Euclid executes a highly optimised reference observation sequence (ROS; see Fig.~\ref{Fig:std_squence}) at every survey field, exploiting the instruments' inter-operability. The ROS visits four nearby pointings at every field, covering an area of $0.53\,\deg^2$ (see Sect.~\ref{sec:common_fov}) common to both instruments and fulfilling the galaxy number density and SNR requirements detailed in \cite{RedBook}. Small `dither slews' are performed between the pointings, taking 66\,s.

At each pointing, VIS takes an image and NISP a simultaneous spectral exposure with the red grism, both lasting about 570\,s (note that these times are not frozen yet). Once the VIS shutter is closed, the GWA and FWA move for the three NISP images of 112\,s each. A 2\,s margin is allocated between the end of a NISP exposure and the wheel actuation, ensuring the NISP exposure is completed before the wheel is moving to avoid compromising the last frame. Moreover, a stabilisation time of 10\,s is considered between a wheel movement and the following exposure. During the NISP imaging, VIS takes biases, flats and other calibration frames. In addition, VIS also takes a shorter science exposure of 108\,s during  the $H$-band exposure in the first pointing, in order to help with the PSF dynamic range on relatively bright stars that saturate during the standard, longer exposures. Details of these sequences are given e.g. in \cite{VIS16} and \cite{NISP16}. 

After each pointing a dither step is applied and a new grism position is selected. The ROS uses the RGS000 and RGS180 at two angles each, offset by four degrees, to allow for sufficient decontamination of the overlapping slitless spectra. 

The total duration of the ROS, including dither slews and overheads is 4214\,s. At the end of the ROS a slew towards the next field is performed. Most of these slews are small 
($\epsilon\leq\ang{3.6;;}$) and referred to as `field slews'. The slew duration is a function of the (eigen-) angular rotation. On average it is  $182\,{\rm s}$, implying a total length of $4396\,{\rm s}$ for the ROS (slews included). This is less than the upper limit of $4400\,{\rm s}$ as defined at mission system level during budget allocation. On occasion, the ERS requires larger slews that are limited to a maximum number of 950 over the full mission. In the most recent EWS solution (see Sect.~\ref{sec:RSD}), the `large slews' comprise 1\% of all non-dither slews applied.  

\subsection{The dithering strategy}\label{sec:ref_dithering}

For each field, the ROS obtains multiple exposures with dithered pointings to mitigate detector defects and cosmic rays, and to meet the required depth. The depth
will vary across the field, not only because of masked defects, but predominantly because the NISP and VIS focal planes have different detector and gap sizes (see Fig.~\ref{Fig:VIS_and_NISP_FoV}, and Table~\ref{tab:FoVsizes}). The dithering strategy between pointings, used for the ERS, must meet the following requirements:
\begin{enumerate}

	\item 95\% of the survey area shall be covered with at least three exposures in VIS. 90\% of the survey area shall be covered by at least three exposures in each of the NISP $Y$, $J$ and $H$ bands.
	
	\item 90\% of each survey field shall be covered by three or more spectroscopic exposures, and 50\% by four or more spectroscopic exposures (using different grism orientations).
	
	\item The NISP imaging of the fields covered by the NISP spectroscopic channel in the EWS shall be acquired over the whole image with depth on average fainter than $m_{\rm AB}=24.0$\,mag for $5\, \sigma$ point-like source.

\end{enumerate}

To analyse the sky coverage of a given dither pattern, we have simulated the ROS observations at the pixel level for a nearly square sky region. The latter covers several FoVs in each of the two dimensions ($N\times N$ joint FoVs, where $3 \leq N \leq 8$) to avoid boundary effects. The pixel count statistics (number of exposures per sky area) are then computed with an integration time map calculator. 

In the various dither patterns, the minimum dither step in each of the two directions is sized such as to ensure that the slew is larger than the largest of the detector gaps in that direction (which, according to Table~\ref{tab:FoVsizes} are the NISP gaps). The maximum dither step is sized to prevent gaps of one line or row of detectors to overlap with the next gaps.
In our simulations, we used slightly larger values for the detector gaps than the ones present in the as-built instruments (see Table~\ref{tab:FoVsizes}). For VIS, we used  $13\farcs 6\,$ in the $X$ direction, $67\farcs 6\,$ in the $Y$ direction and $0\farcs 4\,$ for the width of the charge injection lines, while for NISP we used 50\arcsec\ in $X$ and 100\arcsec\ in $Y$. Subsequent fields are shifted by about 0.7 degrees, setting in the simulation an overlapping of 1\% (in area) between contiguous FoVs. 
This constitutes the basic set-up of our simulations.

The dither step size is affected by errors, namely after a dither slew 
there is a pointing error of 11\arcsec at 3 $\sigma$. The overlap between contiguous FoVs is also affected by an uncertainty related to the absolute pointing error (APE), the baseline input being the requirement applicable to industry and its translation into APE.  The line-of-sight has an uncertainty off-set introduced by the rotation of the filter and grism wheels (that is compensated by the spacecraft AOCS).

We produce coverage maps in two ways. The first method is deterministic (method D). In this case, we 
consider the basic set-up and add a deterministic small shift (of 11\arcsec) to each step of the dithering pattern. The implementation of this further displacement is necessary to be safe in filling the gaps when considering 
the estimate of the
of the uncertainties in the dithering step. No
random errors are considered in this method.
In the second method (MC), we consider uncertainties on dithering steps 
and directions (a pointing error of 11\arcsec at 3 $\sigma$), on off-set repeatability, and other relevant parameters. In this method, the actual overlapping between contiguous FoVs is  affected by the APE uncertainty; we consider an  error on APE with a Gaussian random amplitude of 4.5\arcsec at $1\,\sigma$ and uniform random orientation.
Monte Carlo iterations are then run on a representative patch to extract the coverage maps and the statistics of pixel numbers for the desired area inside the simulated patch for each of the dither patterns presented in Sects.\ref{sec:minimal_pattern} and \ref{sec:S_pattern}.

 \begin{figure}[t]
 \begin{centering}
  \resizebox{0.85\onecolspan}{!}{ %
 \includegraphics{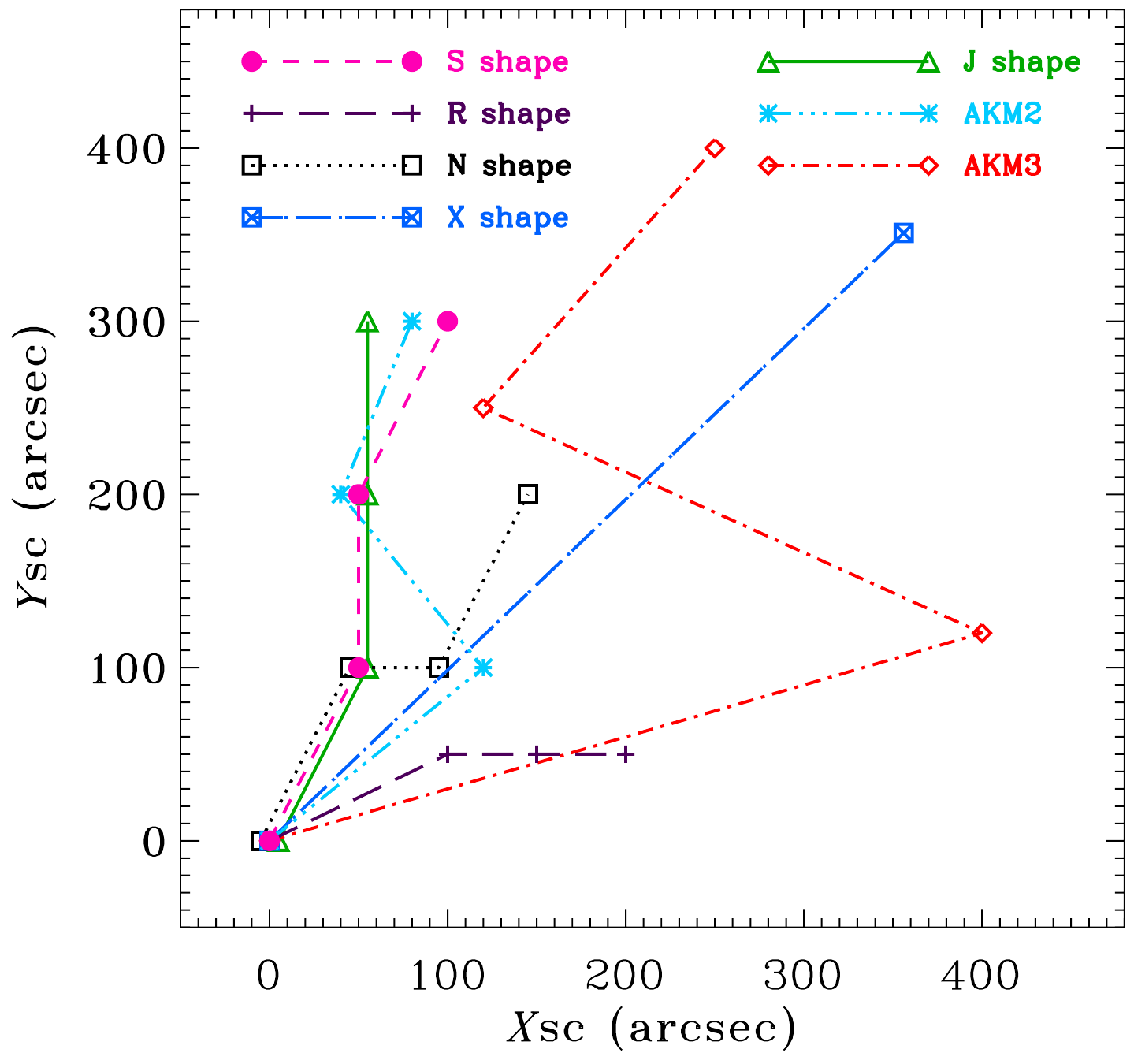} }
 \caption{The seven dither patterns that we have analysed are shown in the spacecraft reference frame (see Fig.~\ref{Fig:spacecraft_FOV_and_ref_frames}). For clarity,  a shift of $5\arcsec$ (or of $-5\arcsec$) in the $X{\rm sc}$ direction is applied to the ``J'' (or to the ``N'') pattern in the figure so to distinguish it from the ``S'' pattern. Only the ``J'' and the ``S'' patterns, which have identical first two steps, are fully compliant with the survey stringent requirements.}  
  \label{Fig:dither_patterns}
 \end{centering}
\end{figure}
\subsubsection{Minimal pattern}\label{sec:minimal_pattern}

The dither pattern defined in \cite{RedBook} was minimal in the sense that it only respected the stringent constraints on the size of a dither step. The latter is constrained to a minimum of $100\arcsec$ by the reaction wheels to prevent mechanical damage to wheel ball bearings in small rotation regime, and to a maximum of $396\arcsec$ by the size of the star catalogue available to the AOCS. The pattern (referred to as ``J'' pattern given its shape, see Fig. \ref{Fig:dither_patterns}) is defined as follows, starting from the pointing of the first exposure:

\begin{itemize}
\itemsep0em
\item Dither step 1: $\Delta X_{\rm SC} = 50\arcsec$,  $\Delta Y_{\rm SC} =100\arcsec$;
\item Dither step 2: $\Delta X_{\rm SC} =  ~~0\arcsec$,  $\Delta Y_{\rm SC} =100\arcsec$;
\item Dither step 3: $\Delta X_{\rm SC} =  ~~0\arcsec$,  $\Delta Y_{\rm SC} =100\arcsec$.
\end{itemize}
These  
are the minimum values that need to be ensured by the AOCS. As discussed earlier, in order to account for the error in the dithering step, the dither is commanded to be $11\arcsec$ larger than the minimum step. 

\begin{figure}[ht]
\centering
{\includegraphics[width=1.\onecolspan]{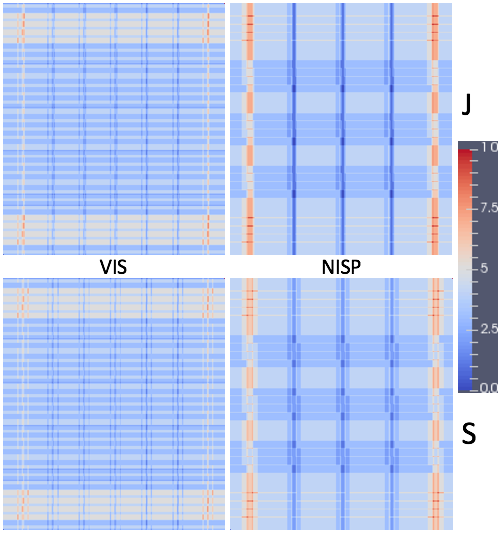} }
\caption{Illustration of the 4 stacked integration time maps for ``J'' (top) and ``S'' (bottom) dithering patterns. The left half displays the maps for VIS, and the right half those for NISP. The colour scaling gives the number of stacked exposures per map pixel. Details are given in the text.
}
\label{Fig:two_fovs_J+S}
\end{figure}

In method D, uncertainties in the telescope pointing were not considered. In total, we simulated $3\times3$ joint FoVs (i.e. nine adjacent executions of the ROS).
For a simple visualisation of the integration time map, we simulated the sky coverage with the resolution of the VIS and NISP FPAs degraded to 1\arcsec. The width of the VIS charge injection lines were increased to 1\arcsec\ to be included in the simulation. The top panels in Fig.~\ref{Fig:two_fovs_J+S} show, for VIS and NISP, the central $\simeq 50\farcm0\,\times\,56\farcm7$ of the ``J'' integration time map, i.e. a single survey field and its boundaries (containing $\simeq 10^7$ simulated pixels).

In method MC, the statistics on the number of exposures per map pixel were evaluated with the same tool, but now considering random errors on some pointing parameters.  We simulated a larger sky area ($7\times7$ joint FoVs) to allow a corresponding increase in the number of realisations and a better characterisation of the overlapping regions between the FoVs. Here, we adopted a resolution of $0\farcs2$ to accurately quantify the effect of the finer geometric characteristics of the VIS and NISP FoVs. Statistics extracted for areas of different sizes (for example of $4\fdg 0 \times 4\fdg 0$) around the central joint FoV show that the coverage requirements are fulfilled at a global level. On a field-to-field basis, the coverage can vary within the associated standard deviations.

\subsubsection{The baseline ``S'' dither pattern }\label{sec:S_pattern}

Looking for a potentially better dither pattern, we analysed six other patterns with four vertices (Fig.~\ref{Fig:dither_patterns}). Two cases were proposed by Arendt, Kashlinski and Mosley based on previous experience (private communication; labelled as AKM2 and AKM3), and four cases (``S'', ``R'', ``N'', ``X'') were taken from \cite{Markovic17}, who focused on optimising the performance of the spectroscopic survey only, without taking into account the concurrent constraints required by the imaging part. The two AKM patterns exceed the maximum step size, with only AKM3 satisfying the constraint on the joint visibility. 


\begin{table} [hbt]
\caption{The statistics (mean and standard deviation) 
on the number, $X$, of exposures per pixel for the ``S'' dither pattern: percentages of covering for individual
($X = 0, 1, 2, 3, 4$) and cumulative ($X > 4$, $X \ge 3$, $X \ge 4$) bins. The percentage of pixels with $X = 0$ or 1 is negligible or very small for both VIS and NISP. Boldface values refer to the requirements 1 and 2 of Sect. \ref{sec:ref_dithering}.
}
	\centering           
		\begin{tabular}{| c | c | c |} 
\hline
 & VIS & NISP \\
\hline
Covering & mean $\pm$ st. dev. (\%) & mean $\pm$ st. dev. (\%) \\
 \hline
$X = 0$ & $(3.57 \pm 5.71) \times 10^{-6}$ & $0.00 \pm 0.00$ \\
\hline
$X = 1$ & $0.23 \pm 0.05$ & $0.99 \pm 0.04$ \\
\hline
$X = 2$ & $4.34 \pm 0.03$ & $8.13 \pm 0.02$ \\
\hline
$X = 3$  & $47.51 \pm 0.09$ & $40.46 \pm 0.10$ \\
\hline
$X = 4$ & $36.52 \pm 0.32$ & $42.12 \pm 0.16$ \\
\hline
$X > 4$ & $11.40 \pm 0.29$ & $8.30 \pm 0.17$ \\
\hline
$X \geq 3$ & ${\bf 95.43 \pm 0.03}$ & ${\bf 90.88 \pm 0.05}$ \\
\hline
$X \geq 4$ & $47.92 \pm 0.08$ & ${\bf 50.42 \pm 0.06}$ \\
\hline
		\end{tabular}
	\label{tab:S_dither_stat}  
\end{table}  

Among the four possibilities suggested in \cite{Markovic17}, only the ``S'' pattern, which is the one closest to the reference ``J'', meets all constraints. Its statistics on the number of exposures per pixel derived from simulations performed considering the most relevant sources of uncertainty are given in Table \ref{tab:S_dither_stat}. 
The ``S'' pattern improves, particularly for NISP, upon the ``J'' pattern, decreasing the fraction of pixels with a single exposure (from $0.59$\% to $0.23$\%  in the case of VIS and from $3.47$\% to $0.99$\% in the case of NISP) while increasing the fraction of pixels with two exposures (from $3.91$\% to $4.34$\% in the case of VIS and from $4.79$\% to $8.13$\% in the case of NISP). For the ``S'' pattern, the percentages of $X \geq 3$ covering are $95.43 \pm 0.03$ and  $90.88 \pm 0.05$ for VIS and NISP, respectively, while the percentage of $X \geq 4$ covering for NISP is $50.42 \pm 0.06$ (the statistics reported in Table \ref{tab:S_dither_stat} are extracted from 500 realisations).
Thus, at least under the considered uncertainty in specifications, the requirement 1 of Sect. \ref{sec:ref_dithering} for VIS and NISP imaging is satisfied at $\sim 14.3 \,\sigma$ level and at $\sim 17.6 \,\sigma$ level, respectively, while the requirement 2 of Sect. \ref{sec:ref_dithering} for NISP spectroscopy is satisfied at $\sim 17.6 \,\sigma$ level and at $\sim 7 \,\sigma$ level for the $X \geq 3$ covering and $X \geq 4$ covering, respectively.
We note that the standard deviation of the cumulative case ($X \ge 3$ or $X \ge 4$) covering cannot be derived by a simple analytical propagation of the standard deviations of simpler cases because of the presence of mutual correlations. These, however, are taken into account in the numerical simulation.

\begin{figure}[!hbt]
\begin{centering}
  \includegraphics[scale=.54]{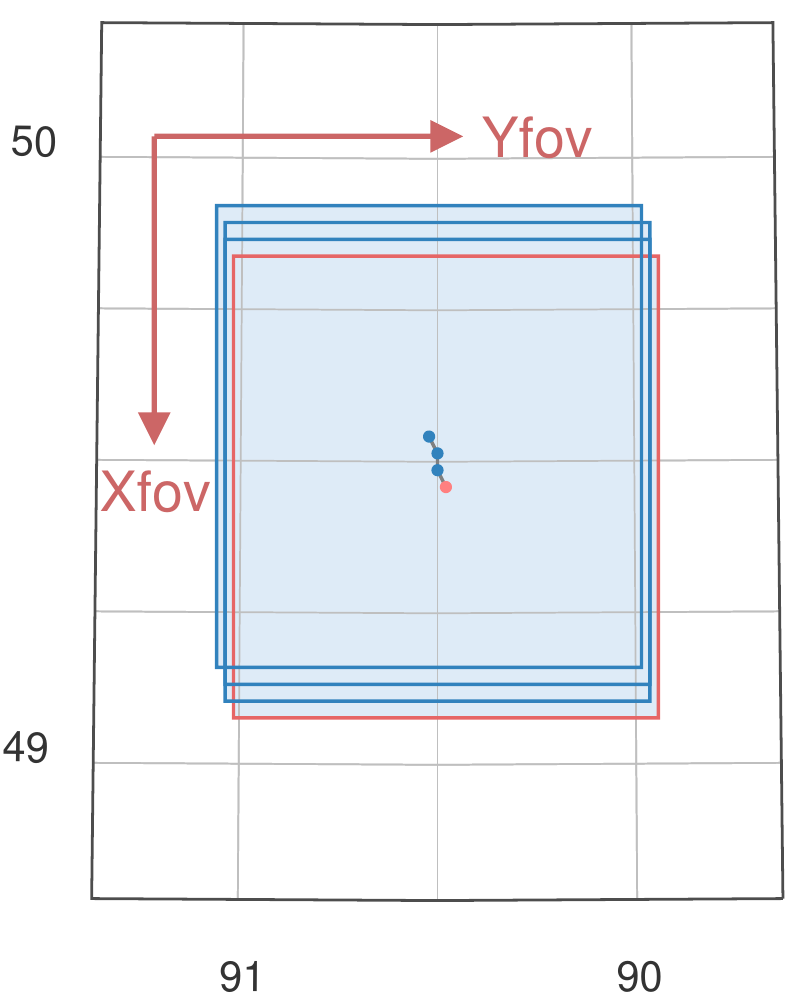}
  \hfill
  \includegraphics[scale=.54]{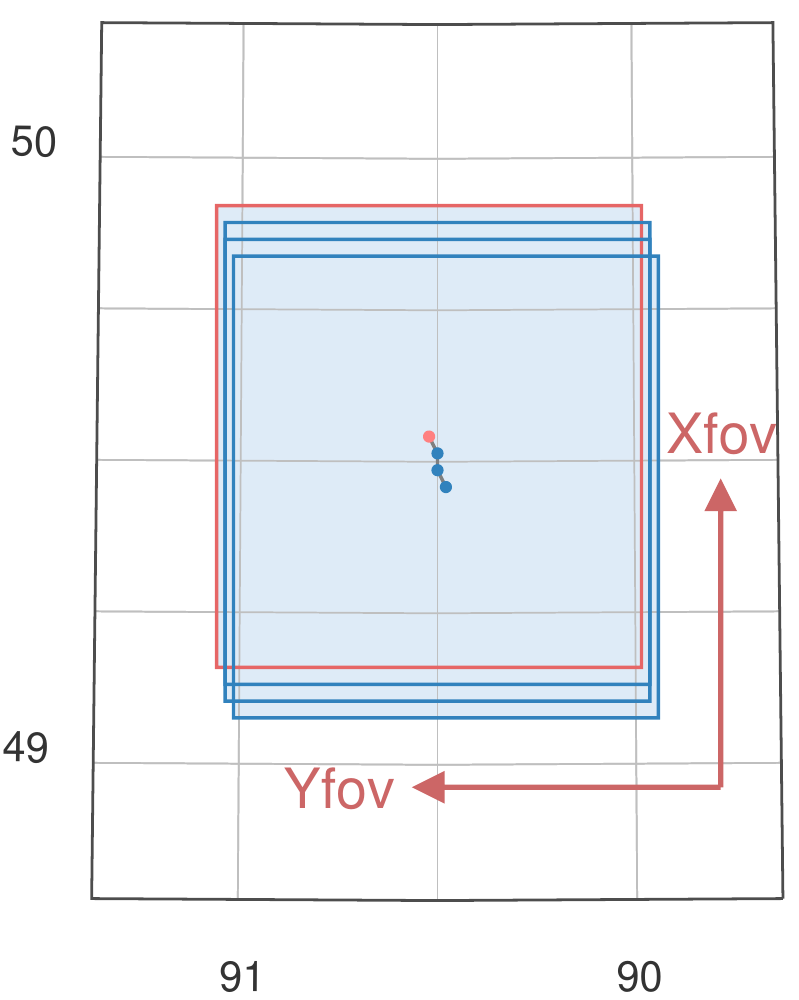}
\caption{The adopted ``S'' dither pattern projected on the sky for a leading (left) and a trailing (right) pointing. See Fig.~\ref{Fig:spacecraft_FOV_and_ref_frames} for reference frames. The red frame is the first to be observed, and the dots mark the centre of each frame. Note how the pattern on the sky is invariant if the telescope is flipped.}
\label{Fig:S_dither_on_sky}
\end{centering}
\end{figure}

The ``S'' pattern was therefore chosen as the baseline dither pattern for the ERS \citep[see also ][]{Markovic17}. 
The coverage map is shown in the bottom panels of Fig.~\ref{Fig:two_fovs_J+S}, and in Fig.~\ref{Fig:S_dither_on_sky} we display the ``S'' pattern as it appears on sky. It is defined as follows.
\begin{itemize}
\itemsep0em
\item Dither step 1: $\Delta X_{\rm sc} = 50\arcsec$,  $\Delta Y_{\rm sc} =100\arcsec$.
\item Dither step 2: $\Delta X_{\rm sc} =  ~~0\arcsec$,  $\Delta Y_{\rm sc} =100\arcsec$.
\item Dither step 3: $\Delta X_{\rm sc} =  50\arcsec$,  $\Delta Y_{\rm sc} =100\arcsec$.
\end{itemize}

We note that by increasing the number of dithers and their step size, a more uniform coverage can be achieved \citep[see e.g.][]{BHR11}. However, this needs to be balanced against the total survey area, the mission duration, readout overheads etc. Other patterns can still be studied and implemented in case of a revision of the current survey and hardware limitations.


\section{The EWS region of interest}
\label{sec:keyparams}

The constraints that drive the implementation of the Euclid Survey areas fall into three main categories (see Sect.~\ref{sec:intro}): 

\begin{itemize}
    \item{environmental;}
    \item{calibration needs;}
    \item{spacecraft constraints.}
\end{itemize}

Environmental properties are external physical constraints of the observable sky, namely zodiacal background, Galactic extinction, foreground stellar density and blinding stars. The zodiacal background and extinction affect the SNR, as does the stray light from bright stars and Galactic plane. The star density also affects the number of useful objects, because their light is dispersed and overlaps with the galaxy spectra of interest, or affects the ability of measure reliable shapes.
Moreover, an increased star density also increase the multiplicative bias in weak lensing shear estimates, if unaccounted for \citep{Hoekstra2017}.

Together, these constraints lead to the definition of the EWS RoI, a collection of four contours enclosing two larger `mainlands' and two smaller `islands'.  Each contour is defined by a series of joint segments derived from either the zodiacal light (ecliptic latitude segment), dust extinction (Galactic caps segment), and the stellar density (Galactic latitude segment).

\subsection{Environment models}\label{sec:environment}

The cosmological measurements are strongly mediated by the nature of the areas of the sky selected for the survey. In the following, we discuss the models of the environment used in the survey planning. 

\subsubsection{Zodiacal background}\label{sec:zodiacal}
A large amount of literature is available on this subject and several models and estimates have been proposed over the years. Updating previous work done by the DUNE consortium \citep{Dune}, we have used a lean and conservative model of the zodiacal background obtained by combining the spectral dependence proposed by \cite{Aldering01} for the proposal for the SNAP satellite \citep{SNAP} and the angular dependence found in \cite{Leinert88}. This `basic' model assumes a cylindrical symmetry with respect to the Sun. 
In this time-invariant and symmetrical model, the zodiacal background flux density $\zeta$ depends solely on wavelength $\lambda$ and ecliptic latitude $\beta$,
\begin{equation}
    \zeta(\lambda, \bbeta) = f_\lambda(\lambda) \, g(|\bbeta|).
\end{equation}

For the spectral dependence, \cite{Aldering01} suggests
\begin{equation}
f_\lambda(\lambda) = 
\begin{cases}
\kappa & ,\,\lambda\in[0.41\,\micron; \lambda_{\rm ref}] \\
\kappa\,10^{-0.730\,(\lambda-\lambda_{\rm ref})/\micron} & ,\,\lambda\in[\lambda_{\rm ref},2.2\,\micron]
\end{cases}
\end{equation}
with $\lambda_{\rm ref}=0.61$\,\micron. The normalisation constant $\kappa$ fixes the flux density to the North Ecliptic Pole (NEP; $\beta=\ang{90;;}$), such that
$\kappa = 1.76 \times 10^{-18}\,\mathrm{erg}\,\mathrm{cm}^{-2}\,\text{\AA}^{-1}\,\mathrm{arcsec}^{-2}$, corresponding to $m_{\mathrm{AB}}=23.05$\,mag\,arcsec$^{-2}$. For comparison, this is just 3\% lower than the value given by \cite{Leinert88} for the NEP at $0.50\,\micron$.

For the dependence on ecliptic latitude $\beta$, we have $g(|\bbeta|)$ as a dimensionless, monotonically decreasing function over the interval $\beta\in[\ang{0;;}, \ang{90;;}]$, normalised to $g(\ang{90;;})=1$. \cite{Leinert88} report values for $\zeta$ (their table 17) as a function of $\beta$ and elongation from the Sun, for a wavelength of $0.50\,\micron$. We reproduce their values for an elongation of $\ang{90;;}$ (applicable to \Euclid) in Table~\ref{tab:simple_zodi_lat_table}, showing that at $\beta=\ang{20;;}$ ($\ang{10;;}$) the zodiacal background is 2 times (3 times) higher than at the NEP. This dependence on latitude is in good agreement with values measured by the SMEI satellite \citep{Buffington16}. \Euclid uses fixed integration times and thus does not compensate for increased background. We therefore limit the EWS to $|\beta|\geq\ang{10;;}$, corresponding to a reduction of 20/30\% in  VIS/J SNR compared to the NEP when taking into account also the stray light (see Sect.~\ref{sec:brightsources} and Fig.~\ref{Fig:SNR_maps}). This still allows for a suitable number density of detected galaxies for WL and GC averaged over the survey.
\begin{table} [!hbt]    
	\caption{The ecliptic latitude dependence of the simple zodiacal background model of \cite{Leinert88}. Here, $\zeta$ is in units of $10^{-8}\,\mathrm{W}\,\mathrm{m}^{-2}\,\mathrm{s}^{-1}\,\micron^{-1}\,\mathrm{sr}^{-1}$.
		}
	\centering               
	\renewcommand{\arraystretch}{1.1}
	\resizebox{\columnwidth}{!}{  %
		\normalsize
		\begin{tabular}{| c | c | c | c | c | c | c | c | c | c | c | c |}
			\hline  
			&  &  &  &  &  &  &  &  &  &  &   \\               
			$ |\bbeta|$ & \ang{0;;} & \ang{5;;} & \ang{10;;} & \ang{15;;} & \ang{20;;} & \ang{25;;} & \ang{30;;} & \ang{45;;} & \ang{60;;} & \ang{75;;} & \ang{90;;} \\

			&  &  &  &  &  &  &  &  &  &  &   \\           
			\hline
			&  &  &  &  &  &  &  &  &  &  &   \\           
			$ \zeta(0.50\,\micron, \bbeta)$ & 259 & 251 & 225 & 193 & 166 & 147 & 132 & 104 & 86 & 79 & 77 \\
			&  &  &  &  &  &  &  &  &  &  &   \\
			$g(\bbeta)$ & 3.36 & 3.26 & 2.92 & 2.51 & 2.16 & 1.91 & 1.71 & 1.35 & 1.12 & 1.03 & 1.0 \\
			&  &  &  &  &  &  &  &  &  &  &   \\           
			\hline
		\end{tabular}
	}
	\label{tab:simple_zodi_lat_table}
\end{table}

\begin{table*}[hbt]
\caption{The normalisation of the basic zodiacal model at the average wavelength of Euclid bands (see Fig.~\ref{Fig:filter_shapes}). } 
	\centering           
	\small
		\begin{tabular}{| c | c | c | c | c | c | c |}        
			\hline  
			band & $\lambda_*$ &  $ f_{\lambda} (\lambda_*)$ & $m_{\rm AB}$ & ${\rm F}(\lambda_*) = \lambda_* f_{\lambda}(\lambda_*)$ & ${\rm F}(\nu_*) = \nu_* f_{\nu}(\nu_*)$ & ${\rm F}(\nu_*) = \nu_* f_{\nu}(\nu_*)$\\    
			& [$\micron$] & [$\ergscmsqA\, \mathrm{arcsec}^{-2}$] & $[\mathrm{mag\,arcsec}^{-2}]$ & $[{\rm nW}\,{\rm m}^{-2}\,{\rm s}^{-1}\, {\rm sr}^{-1}]$ & [{\rm MJy sr}$^{-1}$] & [${\rm \mu}$Jy\,arcsec$^{-2}$]\\               
			\hline             
			VIS  & 0.716 & 1.47$\times 10^{-18}$ & 22.90 & 448 & 0.107 & 2.515\\ 
			$Y$  & 1.080 & 8.62$\times 10^{-19}$ & 22.67 & 367 & 0.132 & 3.103\\ 
			$J$  & 1.367 &  4.92$\times 10^{-19}$ & 22.68 & 286 & 0.131 & 3.079\\ 
			$H$  & 1.770 & 2.50$\times 10^{-19}$ & 22.85 & 188 & 0.111 & 2.609\\ 
			blue grism & 1.145 &  7.15$\times 10^{-19}$ & 22.66 & 348 & 0.133 & 3.126\\ 
			red grism & 1.550 & 3.62$\times 10^{-19}$ & 22.74 & 238 & 0.123 & 2.891\\ 
			\hline          
			\hline          
			Normalisation & 0.610 & 1.76$\times 10^{-18}$ & 23.05 & 456 & 0.093 & 2.186\\
			\hline          
		\end{tabular}
	\label{tab:simple_zodi_filter_table}  
\end{table*}  

This basic model results in a background that is constant in time, depends on the ecliptic latitude as in \cite{Leinert88}, and follows an exponential decay for $\lambda\geq0.61$\,\micron. 
%
%
In  Table~\ref{tab:simple_zodi_filter_table} we report corresponding numeric values and magnitudes for the various \Euclid bands. Here we used the simple average wavelength of a band, defined as
\begin{equation}
\lambda_* = \int \lambda\,T(\lambda)\,{\rm d}\lambda\; / \int T(\lambda)\,{\rm d}\lambda\,,
\end{equation}
where $T(\lambda)$ is the end-of-life (EOL) throughput (see Fig.~\ref{Fig:filter_shapes}). 
For background-limited observations, we could instead weight the integrals in Eq.~(4) by the background flux density; this would slightly change the values of $\lambda_*$, as we will discuss in a separate paper.


Better fits to the dust emission detected by the COBE Differential Microwave Radiometer, can be obtained by considering more sophisticated models, such as the ones of \cite{Kelsall98}. These models include a slab model that is not centered on the ecliptic plane and add inhomogeneous clouds. These result in a zodiacal background that depends not only on the wavelength and direction of observation, but also on the observation's epoch. In such models the minimum background no longer coincides with the ecliptic poles. Instead, it circles the poles with a yearly period, and the value at the NEP has a corresponding sinusoidal variation with a 20\% peak-to-peak variation \citep[see also][]{Pyo12}. 

 \begin{figure}[!hbt]
 \begin{centering}
 \resizebox{0.98\onecolspan}{!}{ 
 \includegraphics{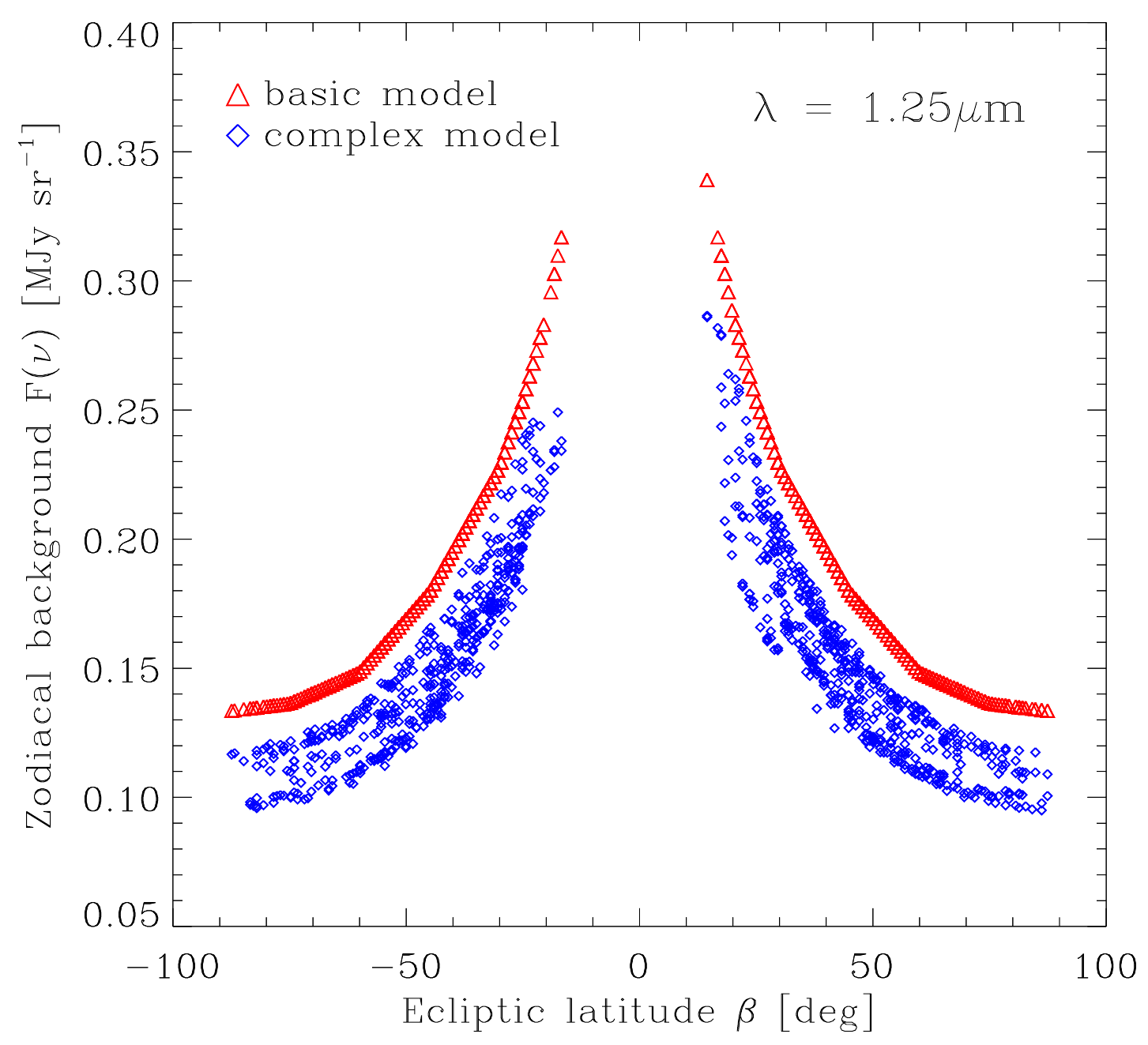} 
 }
 \caption{
 A comparison of the zodiacal emission at $1.25\,\micron$. For a random subset of the pointings in the reference survey  we show results from the conservative basic model (red triangles) and  the estimates by the more complex model (blue diamonds). The latter \citep{Maris16} depends on the observation time and angular direction, and it no longer has cylindrical symmetry. This causes the scatter in the blue points. The Euclid $YJH$-bands show a similar behaviour.}
  \label{Fig:zodi_chk}
 \end{centering}
\end{figure}


We have implemented such a model \citep{Maris06, Maris16}, which includes time dependence, differences in trailing and leading directions and possible deviations of the pointing directions from the orthogonal direction to the Sun. The model is evaluated for $10^3$ random fields of the reference survey binned in ecliptic latitude, and shows scattering due to different observation epochs. It predicts for all bands on average $15-20$\% lower background values than the basic model, and 30\% less flux at the NEP (see Fig.~\ref{Fig:zodi_chk}). We use this more complex model for more precise estimates done {\it a posteriori} once a reference pointing solution is obtained using the basic model as an input.

We have adopted the basic model as our baseline to define the ecliptic latitude exclusion zone presented in Sect.~\ref{sec:roi}, because it is more conservative, whilst providing reasonable margins. This model is also the reference model adopted for Mission Performance Evaluations, and hence \Euclid might detect a slightly larger number density of usable galaxies than our current predictions. We note that among the different models in the literature some could yield background values larger than the adopted basic model, because of a different normalisation 
\citep[see e.g.][]{Wright98}.

\subsubsection{Galactic extinction} \label{sec:extinction}

Extinction by interstellar dust is estimated from the $E(B-V)$ reddening maps\footnote{In particular
{\tt Ebv\_xgal\_ns2048\_REL5.fits}, found at\\
\url{http://hyperstars.lmpa.eu/mamd/planck_dust_model.html}.}
produced by the \cite{Planckdust14}. The map's resolution of $5^\prime$ is high compared to the linear size of \Euclid's FoV of $\sim \ang{0.7;;}$ $\sim \ang{;44;}$.  
We smoothed the map with a 2\,deg wide Gaussian kernel, such that the segments of the RoI boundary that are determined by $E(B-V)$ have a comparable smoothness as the segments determined by Galactic and ecliptic latitude. 

To apply rigidly the original $E(B-V)=0.08$\,mag limit of \cite{RedBook} 
would cause highly convoluted Region of Interest (RoI, see Sect.~\ref{sec:roi}) contours and  holes inside the contiguous survey areas. To achieve a larger RoI with compact regions, 
a slightly adjusted upper limit of $E(B-V)=0.09$\,mag was chosen, while allowing local excursions up to 0.17\,mag to simplify contours and avoid local holes. 
These settings define a first version of the Galactic exclusion zone.  The introduction of the ecliptic exclusion zone (Sect. \ref{sec:zodiacal}) in the two resulting Galactic caps, divides them into two larger mainlands and two smaller islands, as discussed in Sect.~\ref{sec:roi}.
The median value of $E(B-V)$ over the RoI is $0.037$\,mag (more statistics are given in Sect.~\ref{sec:roi_SNR}). 
This approach meets the performance requirement on mean galaxy number density, while preserving a connected survey that optimally complements ground-based data (Sect. \ref{sec:ground_based}). 

We also use the smoothed $E(B-V)$ to estimate the SNR in the RoI. To this end we must compute the total extinction in magnitude for an \Euclid band of central wavelength $\lambda$:
\begin{equation}
-2.5 \logten{(F_{\rm obs}/F_0)} = A_\lambda \; 3.1 \; E(B-V).
\end{equation}
Here, $F_{{\rm obs}}$ is the observed flux and $F_0$ is the flux in the absence of extinction. The total extinction in the $V$-band is quantified by $A_{V} = R_V \; E(B-V)$
where $R_V=3.1$ parameterises the dust extinction in our Galaxy.
We infer the extinction scaling coefficient $A_\lambda$ with respect to the $V$-band from the dust extinction curves of \cite{Gordon03} for the Euclid channels (VIS band, $Y$, $J$, $H$, red grism band) based on their central wavelength. Results are given in Table~\ref{tab:extinction_table} for the NEP with $E(B-V)=0.07$\,mag.

\begin{table}   [hbt]
\caption{Extinction scaling coefficients $A_\lambda$ for the \Euclid passband central wavelengths, and total extinction $A$ at the NEP.} 
	\centering           
	\scalebox{0.99}{ 
\setlength{\tabcolsep}{6 pt}
\renewcommand{\arraystretch}{1.3}
		\begin{tabular}{| c | c | c | c |}        
			\hline  
band & $\lambda \, [\micron]$ & $A_\lambda$ & $A$ [mag]\\
	\hline
VIS	& 0.72&		0.68 & 0.148\\
$Y$ &	1.10 &		0.34 & 0.073\\
$J$ &	1.40 &		0.23 & 0.050\\
$H$ &	1.80 &		0.16 & 0.034\\
red grism &	1.60  &	0.18 & 0.039\\
\hline
		\end{tabular}
	}      
	\label{tab:extinction_table}  
\end{table}  

 \begin{figure*}[!htb]
 \begin{centering}
  \resizebox{\twocolspan}{!}{ %
 \includegraphics{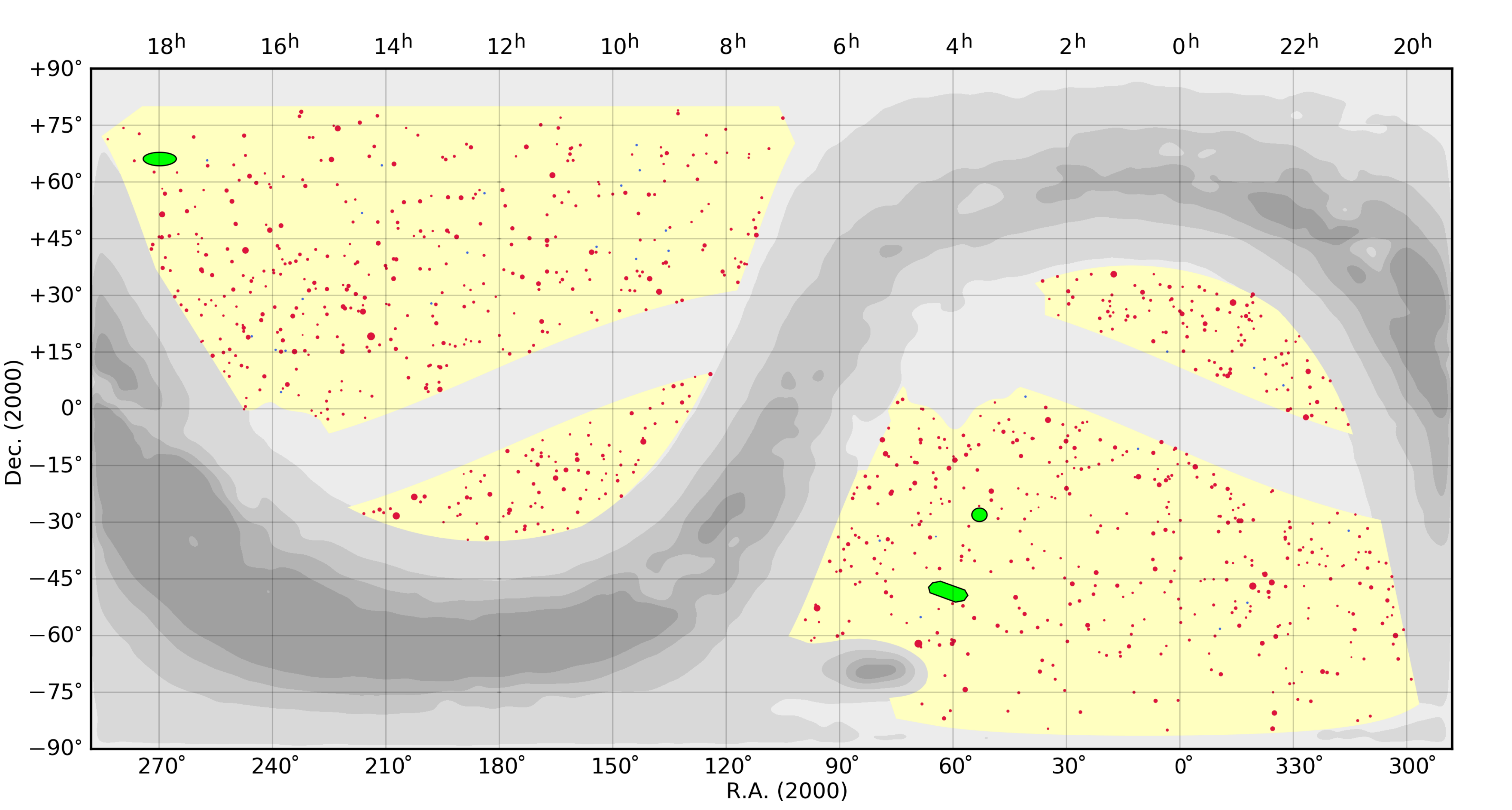} }
 \caption{The 1034 blinding stars present in \Euclid's RoI. There are 1003 stars over the RoI which have $Y$, or $J$, or $H\,<\,$4 AB mag (in red), and 31 stars with $i\,<\,$4 AB mag and all three bands $Y, J, H\,>\,$4 AB mag (in blue). The size of the dots scales down with the magnitude from $-2$ to $+4$. Those nearby bright stars are evenly distributed over the RoI, uncorrelated to the Galactic latitude. The three Euclid Deep Fields (see Sect.~\ref{sec:deep_fields}), free of blinding stars, are shown in green.}
  \label{Fig:blinding_stars}
 \end{centering}
\end{figure*}

\subsubsection{Bright sources and stray light} \label{sec:brightsources}

The EWS allows for $15\%$ of its area to be lost due to various effects such as dead pixels, cosmic rays, etc. This masking budget allocates $2\%$ for bright stars and an additional $2\%$ for their ghost images. Stars brighter than $m_{\rm AB}\sim 17.5$\,mag for VIS and $16.0$ mag for NISP will saturate the detectors for the baseline integration times, and we refer to them as `bright' stars. 

Due to image persistence constraints for NISP, stars with $m_{\rm AB} \leq 4$\,mag must never be present within the FoV for all three photometric bands. We also apply this rule to the VIS instrument due to stray light considerations as discussed below. In the following, we refer to both cases as `blinding' stars. We selected three catalogs in the literature that include all the brightest stars in the sky and sample the wavelength domain of the two instruments: $i$-band for VIS from the ATLAS All-Sky Stellar Reference Catalog \citep{Tonry18}, $Y$-band for NISP from the spectrally matched magnitudes of stars from the Tycho2/2MASS catalogs \citep{Pickles10}, and $J$ and $H$-band for NISP from the 2MASS Point Source Catalog \citep{Skrutskie06}. In total, there are 1034 blinding stars within the RoI (Fig.~\ref{Fig:blinding_stars}). 1003 of those stars were first selected through the $Y$\,<\,4 or $J$\,<\,4 or $H$\,<\,4 AB mag NISP criteria, and 31 stars with $Y, J, H$\,>\,4 mag were added through the $i$\,<\,4 mag VIS criteria (there are 275 $i$\,<\,4 mag stars within the RoI, the majority already selected through the NISP criteria). Survey fields with blinding stars will be skipped on all instruments, amounting to $3\%$ of the RoI area. Bright (non-blinding) stars will be observed but locally masked during data reduction. The observed area lost in this way is accounted for in the pre-allocated masking budget, which does not incorporate areas skipped on purpose because of the presence of blinding stars.

In the following we show how stray light from stellar sources inside and outside the FoV affects the observations and dictates the RoI Galactic latitude threshold. Stray light is generated in various ways, such as reflections on mechanical structures, scattering on contaminated optical surfaces and their intrinsic surface roughness, by diffraction on the edges of mechanical parts such as baffles, spider arms, or by multiple internal reflections between optical elements. Stray light creates an additional background contribution that can be diffuse or structured (ghosts). 

Up to 2014, the stray light was expected to be a minor fraction of the total diffuse background (this was initially specified to be less than 20\% of the zodiacal background at the NEP). However, with lessons learned from the Gaia mission, a study on stray light contamination was carried out by ESA and the industry for \Euclid. \cite{Venancio20} have studied the stray light aspects extending the analysis from pure in-field (dominated by the mirrors particulate contamination) to the far out-of-field domain (dominated by the internal structural multiple reflections). Both can contribute significantly to the stray light level, as out-of-field stray light, though largely attenuated by diffuse scattering process, integrates over the full sky and becomes dominant when getting close to the Galactic plane. They found that particulate contamination on the mirrors will be the main contributor: stray light in some sky areas can become comparable and even exceed the local zodiacal background. Consequently, \Euclid must stay clear of the Galactic plane, since sources both inside and outside the FoV (in-field and out-field stray light, respectively) contribute.


 \begin{figure}[t]
 \begin{centering}
  \resizebox{\onecolspan}{!}{ %
 \includegraphics{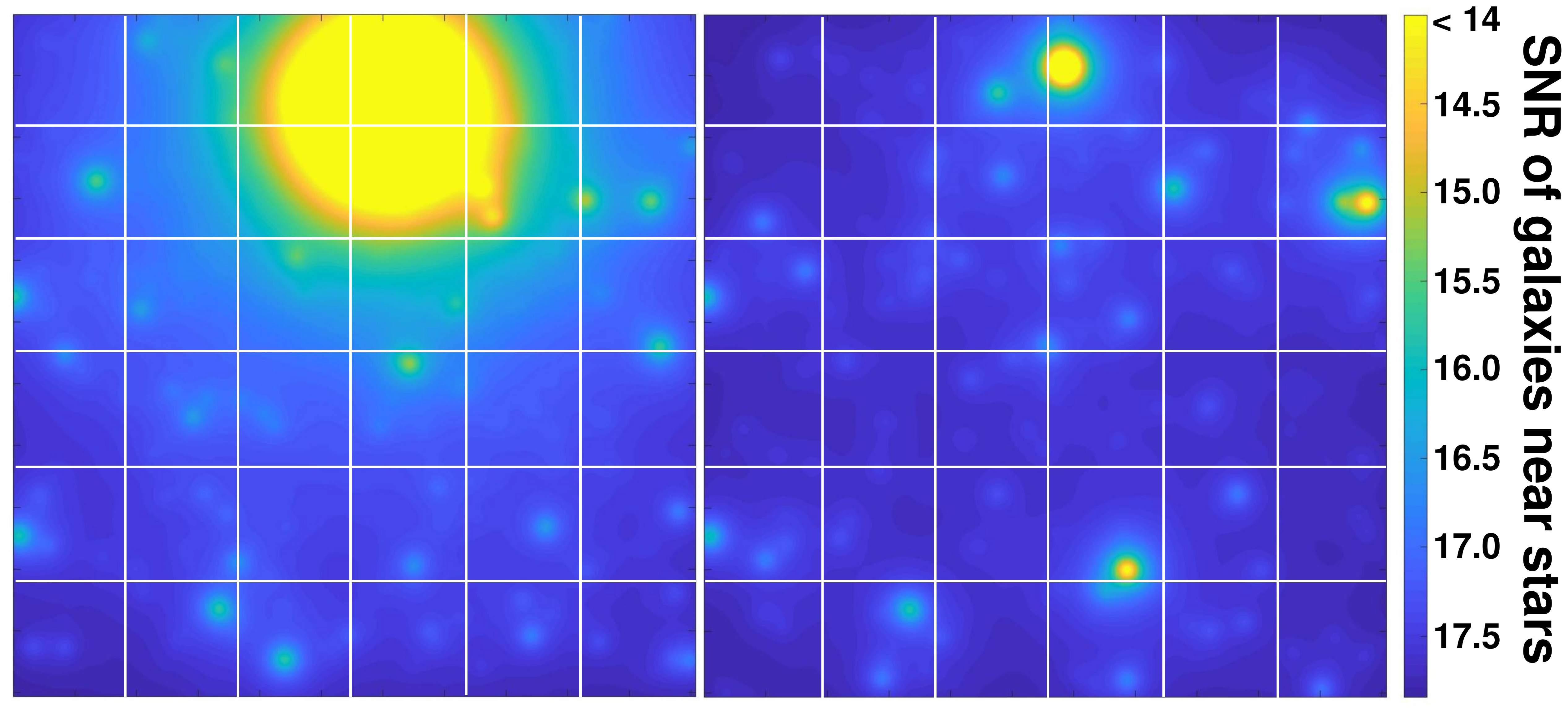} }
 \expandafter\caption{Impact from bright stars on the SNR achievable by a target galaxy $m_{\rm AB}=24.5$\,mag (extended source) in VIS when placed over a full Euclid FoV ($\sim \ang{0.7;;} \, \times \ang{0.7 ;;}$, with a scheme of the 6$\times$6 CCD mosaic layout superimposed) in a typical  part of the Euclid RoI. The colour scale is the SNR of a target source placed in the field at that position. The degradation in SNR when the target galaxy is placed closer and closer to a bright star is evident. \emph{Left panel}: a $m_{\scriptstyle\rm VIS}=4$\,AB mag star has effects on  almost 50\% of the FoV, however science is impacted on less than 10\% of the FoV, the solid yellow area where the target galaxy is at ${\rm SNR}<14$ (see text). \emph{Right panel}: common stars of magnitude  $m_{\scriptstyle\rm VIS}=8$, $9$, and $10$  have a very limited impact overall with respect to the FoV size: less than 0.1\% of the FoV lost on average across the RoI.}
  \label{Fig:SNR_VIS_brightstars}
 \end{centering}
\end{figure}


\begin{figure*}[!hbt]
	\begin{centering}
		\resizebox{\twocolspan}{!}{ %
			\includegraphics{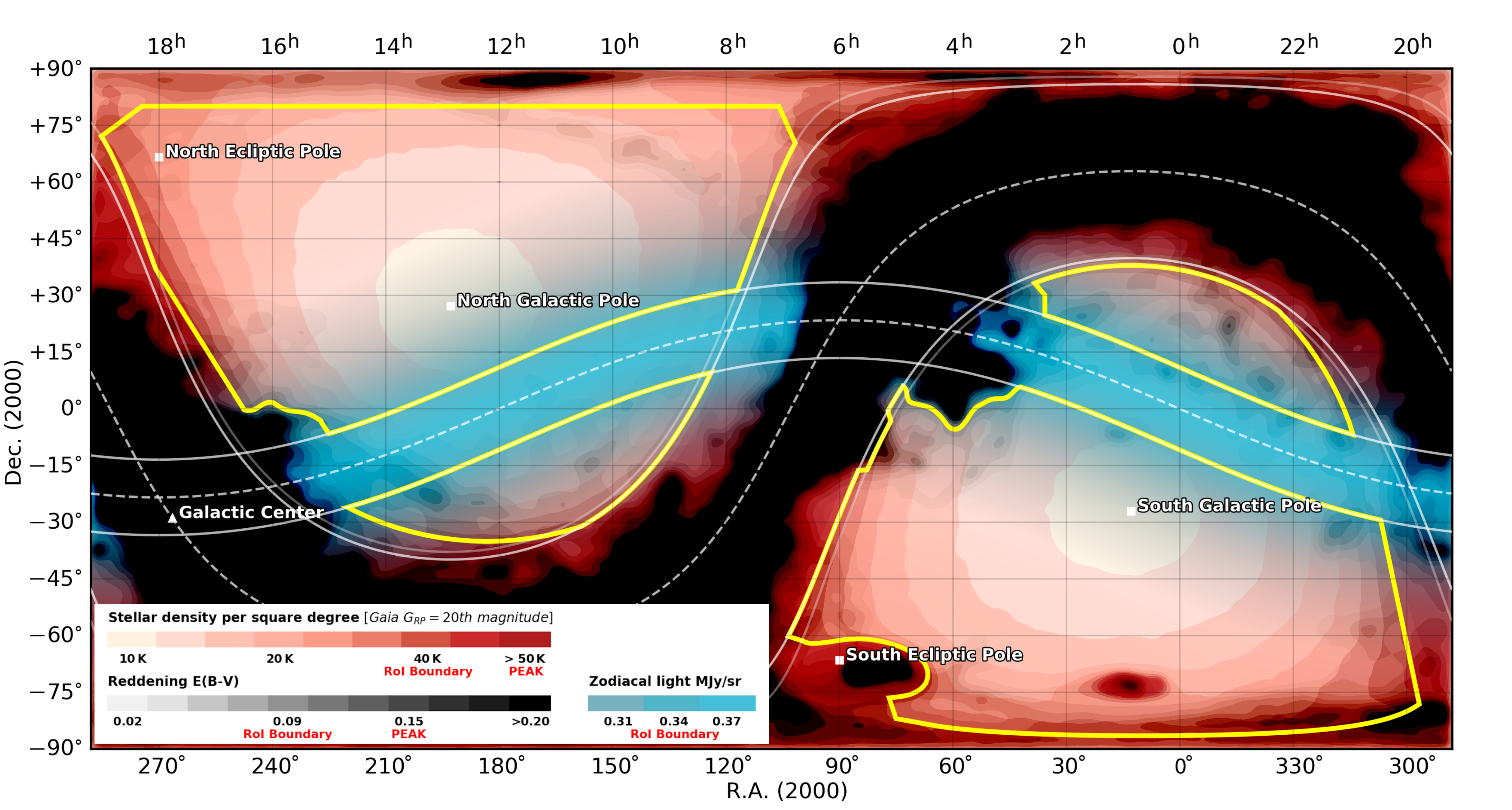} }
		\caption{The RoI outline (17\,354\,deg$^2$) with the accepted ranges of the stellar density, dust extinction, and zodiacal light.} 
		\label{Fig:roi_inputs}
	\end{centering}
\end{figure*}

The normalised diffusion irradiance profile \citep[NDI,][]{Venancio16} describes the profile of the scattered light in the telescope focal plane for a point source at a given position either within the FoV or up to 20\,degrees away from the optical axis. \Euclid's Korsch optical design \citep{Korsch72} effectively suppresses scattered light. For \Euclid and its enhanced baffling \citep{Venancio20}, in-field stray light will impact the SNR of faint galaxies from our science goal defined limit of $m_{\mathrm VIS}=24.5$, causing them to fall below the minimal value of ${\rm SNR}=10$. To ensure such SNR level is realized through the mission science pipeline involving all the steps of data processing and signal extraction, the system team in charge of scaling the mission design built margins by adopting a conservative goal of ${\rm SNR}=14$ using their own internal radiometric SNR metric. 
The SNR of a $m_{\scriptscriptstyle\rm VIS}=24.5$ galaxy degrades from the mission median value of 17.5 to 14 if the galaxy is at an angular separation of $6.8, 4.1, 2.5, 1.5$, and $1.0$ arcminutes
from a bright star of $m_{\scriptstyle\rm VIS}=4$, $5$, $6$, $7$ and $8$\,AB\,mag, respectively.
The left panel in Fig. \ref{Fig:SNR_VIS_brightstars} illustrates the case for a $m_{\mathrm VIS}=4$ star, the impacted area representing just 9\% of the entire field-of-view, a truly remarkable performance made possible by the Korsch optical design. For fainter stars, of magnitude 9 and 10, the radius of SNR degradation to a level of 14 is reached at  a $8$ and $1.5$ arcseconds radius, respectively. As shown in the right panel of Fig. \ref{Fig:SNR_VIS_brightstars} the impact of these stars is negligible beyond the core of the PSF.  The average density of the 8, 9, and 10 magnitude stars \citep[]{Zakharov13} over the RoI amounts to $0.6$, $1.6$, and $3.8$ in-field stars per FoV for the VIS: their collected impact will be limited to a tenth of a percent of area loss of the Euclid FoV on average across the RoI. Accounting for the sparse 4 to 7\,AB\,mag stars does not alter these statistics. A comparable performance is expected from the NISP as the mission design drove an NDI dominated by the telescope, not by the instruments.

In summary, for individual bright stars \Euclid will skip observing tiles in which at least one of the four exposures will contain a $m_{AB} \leq 4\, \mathrm{mag}$ star. For a handful of extremely bright stars also nearby tiles will be skipped, according to an avoidance radius for the tile center set at a level of stray light yielding a 15\% degradation in SNR (NDI model). Areas affected by in-field stars fainter than $m_{AB} = 4\, \mathrm{mag}$ will be masked during the data reduction phase, as well as  ghosts originating from the dichroic.

All the stars outside the telescope field of view will also contribute globally to the background level of stray light. Their combined effect is to add a diffuse cumulative component that depends on the pointing direction of the telescope. In consequence, this effects scales  with the Galactic latitude, the NDI defining an intensity ratio of the collected brightness of the Galaxy. The out-field stray light 2D map adopted in this paper is a Besan\c con model of the Galaxy flux \citep{Robin12,Robin14}. The model is scaled at the relevant wavelength  to match at the one percent level the out-field stray light level computed at 12 selected points across the whole sky. The levels were estimated by the system team for the 2018 Mission Critical Design Review (MCDR) effort.
\begin{figure*}[!htb]
	\begin{centering}
	\includegraphics[scale=.63]{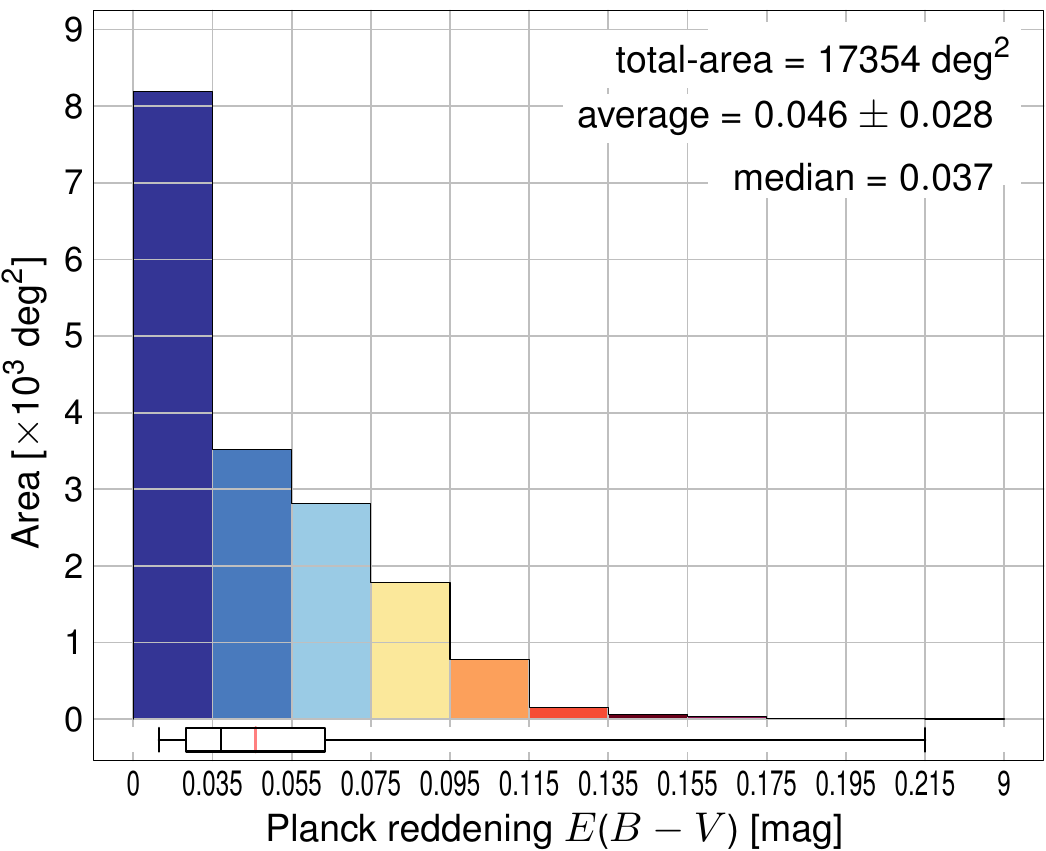}
	\hfill
	\includegraphics[scale=.63]{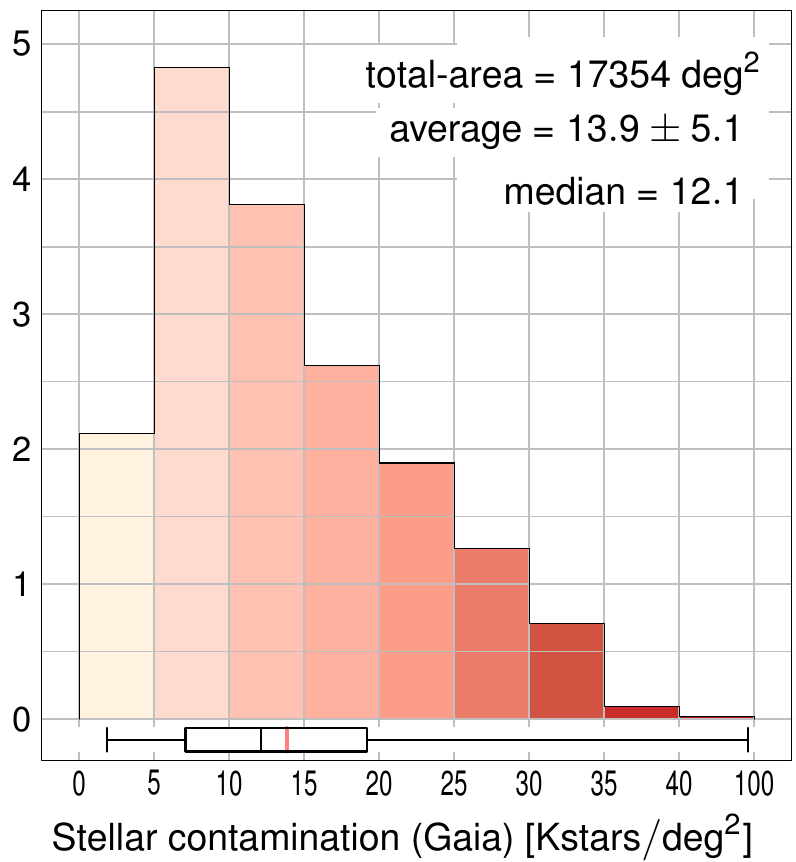}
	\hfill
	\includegraphics[scale=.63]{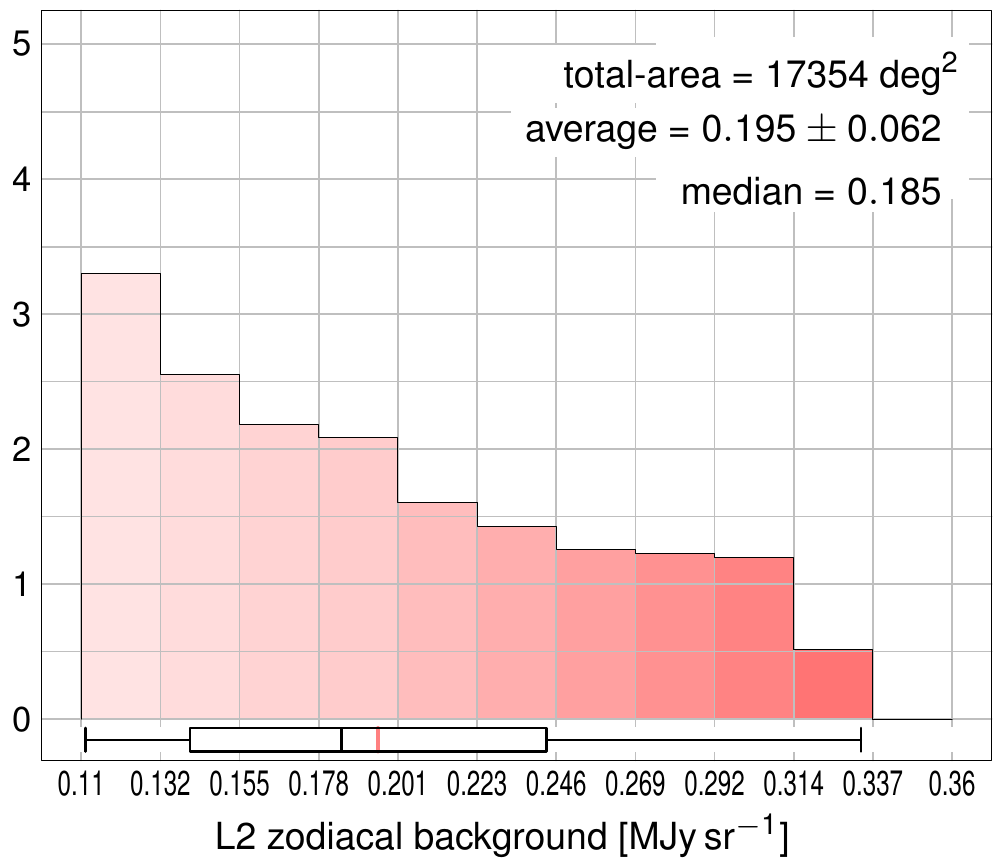}
	\caption{Distribution of reddening values (\emph{left}), stellar density (\emph{middle}) and zodiacal background (\emph{right}) across the RoI. Below each graph, the box-and-whiskers plot marks the mean (red line), median (black line), interquartile range (empty box), plus the minimum and maximum values.}
		\label{Fig:roi_histo}
	\end{centering}
\end{figure*}
A spacecraft stray light model based on the estimation of the NDI was then used by the system team to describe the stray light due to the diffusion in the telescope and gauge its effects on the PSF and local background level on the focal plane \citep{Venancio16,Venancio20}. Assuming the entrance of the telescope is illuminated by a distant point source (collimated light), then the NDI is defined as the ratio of light irradiance (power per unit area) on the image plane to the source irradiance in object space at the entrance of the telescope.
The NDI is computed for both VIS and NISP using the ASAP optical software \cite[]{ASAP14}, a ray-tracing program that uses a statistical Monte Carlo approach. The computation is done with telescope and instrument optical and mechanical models and associated contamination assumptions.
Then the NDI is applied on the sky for a mesh of pointing directions on the sky (with sampling equal to the \Euclid FoV) over the full sky in order to estimate for each possible pointing direction the cumulative out-of-field stray light maps.

 Conservative estimates of these contaminants established by the mission system team are adopted for the background and associated noise computations presented in this paper. This drives in particular the RoI definition with respect to Galactic latitude (from a minimum of $|\bbeta|\geq\ang{23;;}$ to nominal $|\bbeta|\geq\ang{25;;}$), and how close the EWS can get to the Galactic bulge, refining the Galactic exclusion zone.
 
 Finally, an additional concern is that due to stray light: \Euclid must avoid pointing within a circle centered on the position of Solar system planets,  Mars and Jupiter having the largest radius, of 13\,deg.

\begin{figure*}[!hbt]
	\begin{centering}
		\resizebox{\twocolspan}{!}{ %
			\includegraphics{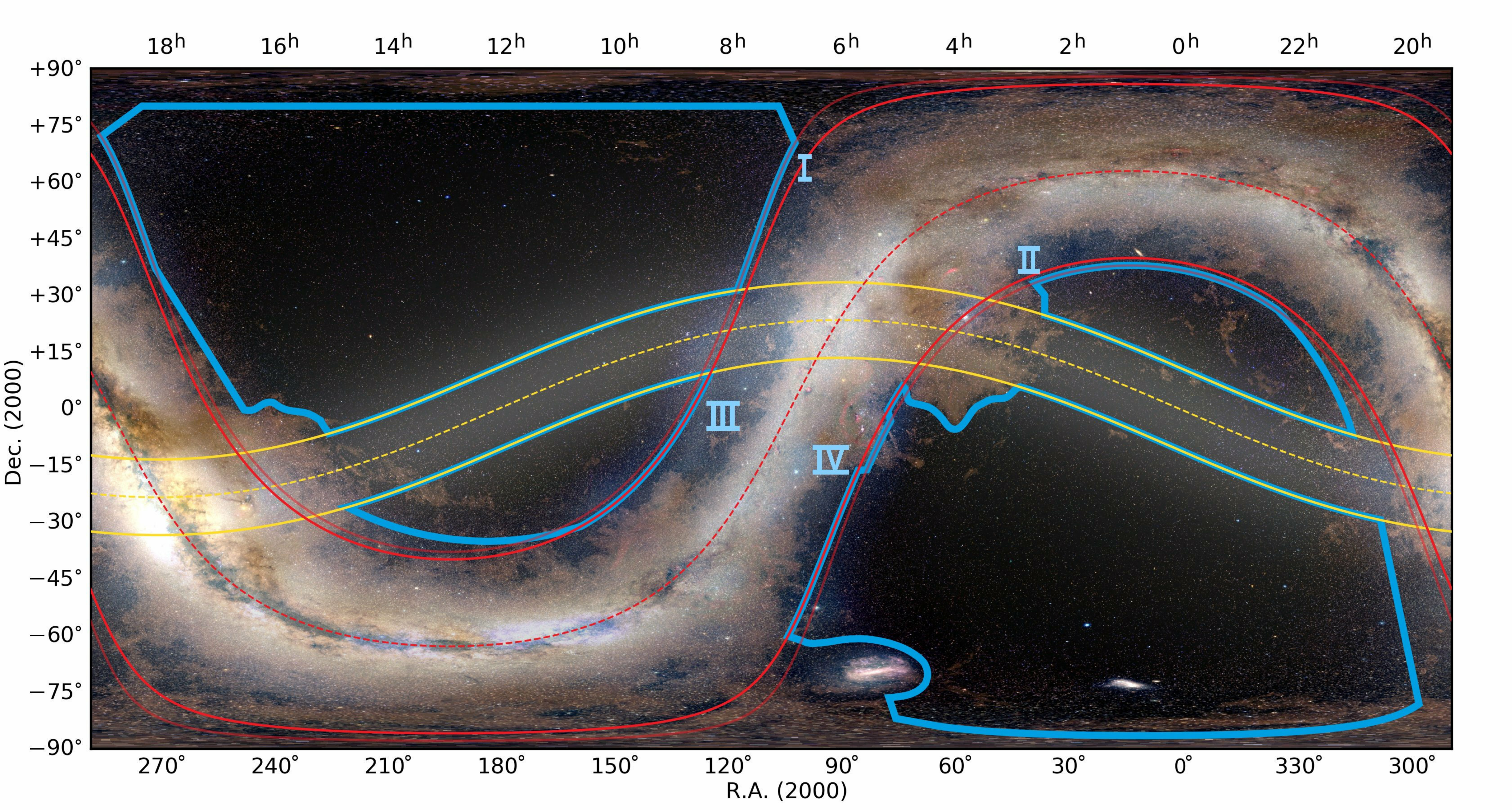} }
		\caption{The sky constraints as experienced by \Euclid from L2, and the RoI avoiding the worst regions: the four quadrants of the RoI are in blue. The solid yellow lines trace the ecliptic latitude threshold of $|\beta|=\ang{10;;}$. The red solid lines trace the galactic latitude threshold of $|b|=\ang{23;;}$. For declination $\delta\geq+\ang{30;;}$ the threshold is $|b|=\ang{25;;}$ with respect to the northern ground surveys (Sect. \ref{sec:roi_four_areas}).} 
		\label{Fig:roi_outline}
	\end{centering}
\end{figure*}

\subsection{The Euclid Region of Interest} 
\label{sec:roi}

\subsubsection{The RoI four main areas}\label{sec:roi_four_areas}

The dominant factors that determine the RoI are the zodiacal background, the Galactic extinction, and stray light due to the Galactic stellar density. Minor contributors such as emission from Galactic cirrus were ignored, being at least five magnitudes fainter than the total background over most of the RoI (Sect.~\ref{sec:lsbperf}). The EWS will also skip fields containing blinding stars (Sect.~\ref{sec:brightsources}), leaving only their faint effect imprinted on the out-field stray light.

The main outline of the RoI is defined by the extinction limits, 
an ecliptic latitude threshold of $|\beta|\leq\ang{10;;}$, and a Galactic latitude threshold of $|b|\leq\ang{23;;}$. Note that for declination $\delta\geq+\ang{30;;}$ we set $|b|\leq\ang{25;;}$, since the Euclid complementary ground surveys were designed and started in 2017 with this value from \cite{RedBook}. Section~\ref{sec:ground_based} describes how the RoI is affected by these ground-based surveys.

The RoI fragments into four quadrants delimited by the yellow contours in Fig.~\ref{Fig:roi_inputs}. 
and detailed in Table~\ref{tab:roi_quadrants}. 
The RoI is best presented on this equirectangular projection. We use elliptical projections when highlighting aspects of area conservation. Our sky projections were produced with IPAC's {\tt Montage} package\footnote{\url{http://montage.ioac.caltech}} and the University of Groningen's {\tt Kapteyn} package for Python\footnote{\url{https://www.astro.rug.nl/software/kapteyn/}}. 

The environment limits and their impact on the RoI are shown in Fig. \ref{Fig:roi_inputs}, while Fig.~\ref{Fig:roi_histo} shows the distribution of extinction, stellar counts from Gaia (limited at ${\rm mag}=20$) and zodiacal background within the RoI.  In total, the present RoI encompasses 17\,354\,deg$^2$ that are compliant for \Euclid's core cosmology science. The EWS can be constructed from any 15\,000\,deg$^2$ within.
Some parts of the EWS will inevitably be of lower quality for cosmology, yet their legacy value is high. For example, the Small Magellanic Cloud (SMC) is inside the RoI, although clearly at odds with the survey constraints. Pushing further into the Galactic plane, though, would rapidly reach extinction levels unacceptable for \Euclid's core science, as is evident from Fig. \ref{Fig:roi_outline} that shows the combination of all constraints, highlighting the best parts of the Euclid sky.

\begin{table}[hbt]
\caption{The four separate quadrants of the RoI (see Fig.~\ref{Fig:roi_outline}). The total area is 17\,354\,deg$^2$.}
    \centering
    \setlength{\tabcolsep}{6 pt}
    \renewcommand{\arraystretch}{1.2}
	\begin{tabular}{| c | l | c |}
	\hline  
    \strut Quadrant & \hfill Name & Area [deg$^2$] \\
    \hline  
  \FancyRoman{1} &	Northern ``mainland'' & 7142 \\
  \FancyRoman{2} &  Northern ``island''   &	1575 \\
  \FancyRoman{3} &	Southern ``island''   &	1700 \\
  \FancyRoman{4} &	Southern ``mainland'' & 6937 \\
  \hline
	\end{tabular}
\label{tab:roi_quadrants}  
\end{table}  

\begin{figure}[!hbt]
	\begin{centering}
		\resizebox{\onecolspan}{!}
		{\includegraphics{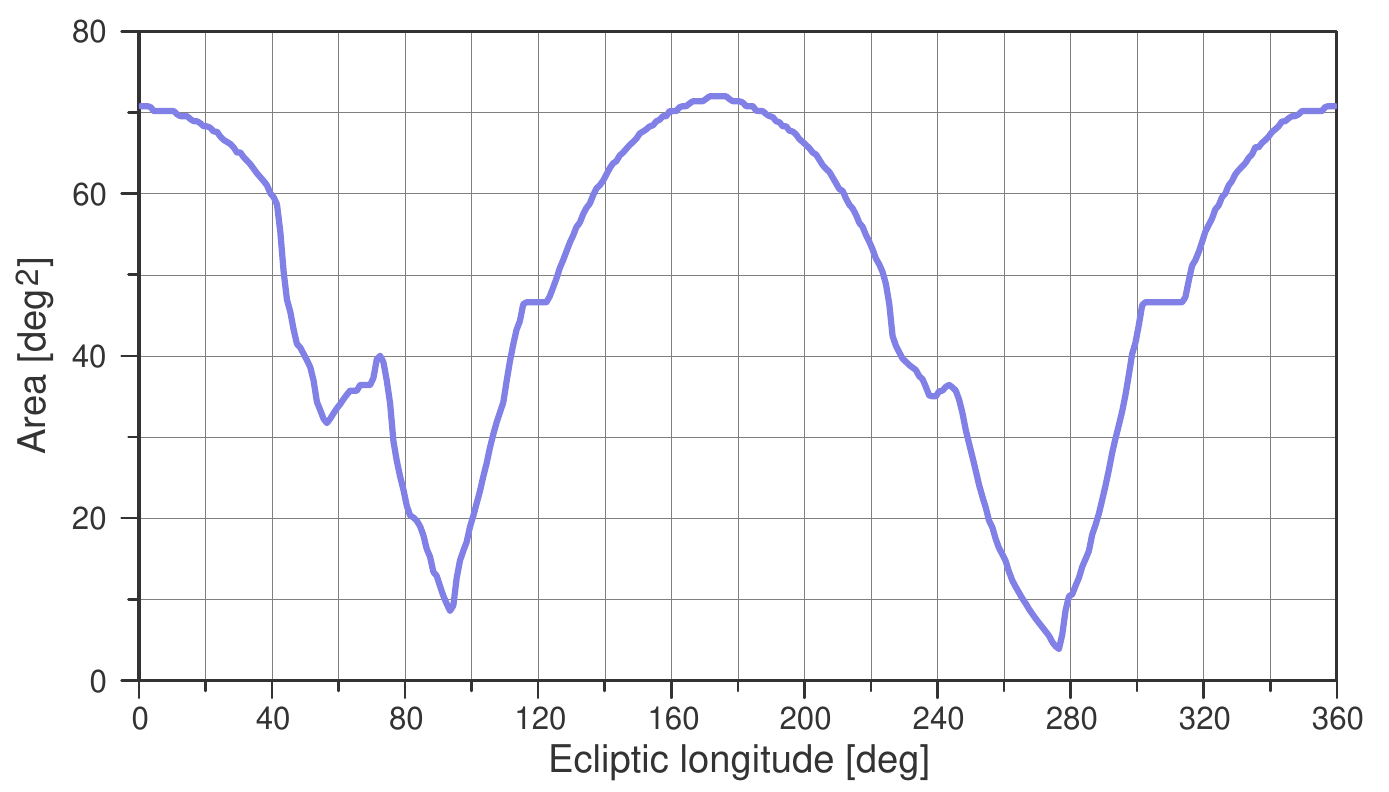}}
		\caption{
		The available RoI area as a function of the ecliptic longitude.
		The troughs are produced by the intersection of the Galactic and ecliptic planes and lead to unallocated time.}
		\label{Fig:ROI_available_area}
	\end{centering}
\end{figure}

From Fig.~\ref{Fig:roi_outline} it is clear that the area of the RoI changes with ecliptic longitude. Figure~\ref{Fig:ROI_available_area}  shows the area of the RoI as a function of ecliptic longitude (in bins of $1\degr$ of ecliptic longitude). The plot contains two global maxima, which coincide with the longitudes that cross both a mainland and an island, and  two global minima, which coincide with the intersection of the ecliptic and Galactic planes.
Given that one degree of longitude is scanned by one day of orbit and that in one day ${\sim}10\,\text{deg}^2$ of EWS are observed, in a six year mission it is only possible to observe a maximum of $60\,\text{deg}^2$ of EWS sky, per degree of longitude. In practice, this time must be shared with calibrations and EDF observations that collectively take ${\sim}20\text{\%}$ of the total time. This lowers the EWS allocation to a maximum average value of  $48\,\text{deg}^2$ per degree of longitude. From the analysis of  Figs.~\ref{Fig:roi_outline} and \ref{Fig:ROI_available_area}, it is clear that the sky in the RoI is not uniform enough to fill this quota.
Given the limited pointing range of the telescope, with observations at or close to transit, this inevitably leads to the depletion of the available (i.e., yet unobserved) sky in some ecliptic longitudes. This reveals an intrinsic limitation to the maximum efficiency attainable by the EWS, in which in some parts of the year there will be unallocated time periods that increase in duration towards the end of the mission. This is an important feature of EWS solutions, as discussed in Sect.~\ref{sec:unallocated_time}.

\subsubsection{SNR and survey depth in the RoI} 
\label{sec:roi_SNR}

Figure \ref{Fig:roi_outline}  highlights the best parts of the \Euclid sky. In the following, we compute the corresponding SNR maps, which provide the quantitative context.

For our SNR computations we take into account the following aspects at the hardware level: telescope and instruments' internal backgrounds, photometric zero points (encoding the total throughput), read noise and dark current. These are independent of sky position and were taken from the latest available ground characterisation measurements. At the environmental level, we include all-sky maps for the zodiacal background, extinction, and stray light from the Galaxy as detailed in Sect. \ref{sec:environment}. 

At the operational level, we allow for three exposures (VIS and NISP imaging) and four exposures for NISP spectroscopy, the FPA geometries, integration times, and the size of the measurement apertures. This is motivated by the fact that 90\% (50\%) of the survey area is covered with at least three (four) imaging exposures (see Table~ \ref{tab:FoVsizes}).. The SNR measurement metrics are evaluated as follows: for VIS, we consider an extended source with a total magnitude of $m_{\rm AB}=24.5$\,mag
in a $1\farcs3$ diameter aperture, capturing $94\%$ of the flux. For NISP photometry, we consider a point-like source with a total magnitude of $m_{\rm AB}=24.0$\,mag in the $Y$, $J$, $H$ bands in a $0\farcs9 \,\times \,0\farcs9$ ($3\,\times\,3$ pixel) aperture, capturing $\sim80$\% of the flux for $Y$ and $J$, 70\% for $H$. For NISP spectroscopy, we consider an emission line with a flux of $2\times10^{-16}\,\ergscmsq$ at an observed wavelength of 1600\,nm, measured in a 4$\times$4 pixel wide aperture in the dispersed images.

In this way we verify that the scientific requirements of the \Euclid project are met. Global statistics of the SNR are summarised in Table \ref{tab:SNR_RoI_table}. The median survey depths converted and scaled to a $5\,\sigma$ point like source ($5\, \sigma$ point-like source) performance metric for imaging are listed.

\begin{figure}[!hbt]
	\begin{centering}
	\resizebox{\onecolspan}{!}{ %
			\includegraphics{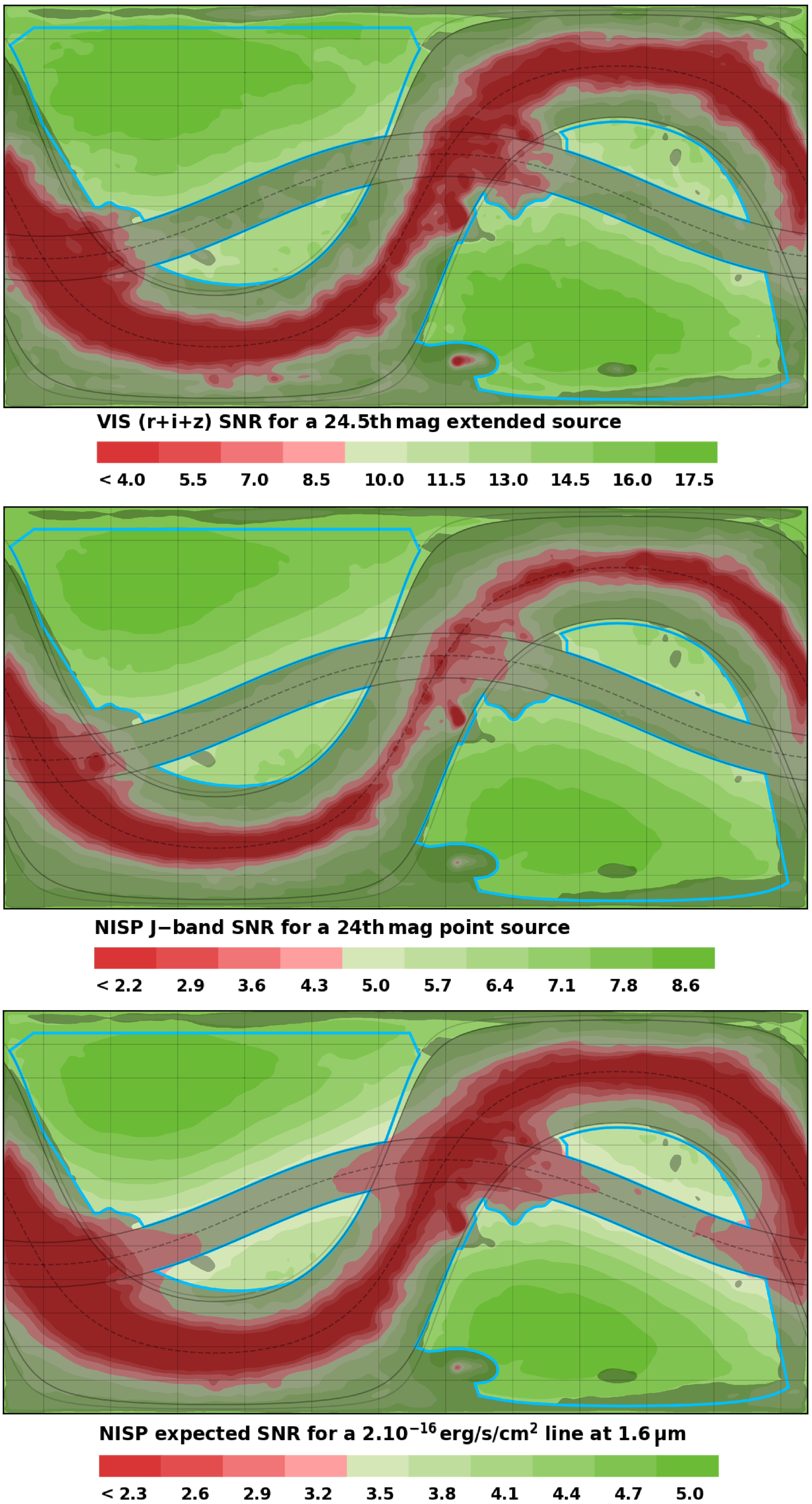}
	    }
		\caption{Top: VIS band SNR in an equirectangular celestial projection (same referential as Fig.~\ref{Fig:roi_outline}); the entire RoI is within specifications (${\rm SNR}\geq 10$, the first level of light green). Middle: NISP SNR for the $J$-band ($Y$- and $H$-bands are similar, see Table \ref{tab:SNR_RoI_table} for overall statistics). Bottom: the NISP red grism band SNR. The more convoluted isocontours of the VIS band SNR are caused by the stronger dust extinction versus the NISP. Greyed regions outside the RoI are excluded due to extreme reddening and/or high stellar density and/or high zodiacal background (c.f. RoI definition and Fig.~\ref{Fig:roi_outline}.)} 
		\label{Fig:SNR_maps}
	\end{centering}
\end{figure}

\begin{table}   [hbt]
\caption{SNR statistics for the RoI, for each channel: VIS band (boldface values refers to extended objects), NIR bands ($Y$, $J$, $H$, values refer to point like objects) and red grism band, $S$ (italic values in the last column refer to $\ang{;;0.5}$ diameter sources).  The median depth here is evaluated for $5\, \sigma$ point-like source.} 
	\centering           
	\scalebox{0.99}{ 
\setlength{\tabcolsep}{6 pt}
\renewcommand{\arraystretch}{1.4}
\begin{tabular}{| l | c | c | c | c | c |}        
\hline
  &	VIS     & $Y$     & $J$     & $H$     &	$S$ \\ 	
\hline
Minimum SNR & \bf{10.0} & 5.0 & 5.7 & 5.7 & {\it 3.2} \\
Median SNR & \bf{15.9} & 6.5 & 7.8 & 7.2 & {\it4.5} \\
Maximum	SNR & \bf{19.8} & 7.8 & 9.0 & 8.5 & {\it6.6} \\
\hline
\hline
Median depth [AB mag] & 26.2 & 24.3 & 24.5 & 24.4 & $-$ \\
\hline
\end{tabular}
}      
	\label{tab:SNR_RoI_table}  
\end{table}  

The resulting SNR maps for VIS and NISP are shown on Fig.~\ref{Fig:SNR_maps}. All four quadrants are fully green, within specifications, for all channels for their respective depth metrics (VIS, NISP-P, and NISP-S).

We note that the SNR computations do not consider the contamination of galaxy samples by stars; to this end we have introduced the thresholds to Galactic latitude. The greyed areas in Fig. \ref{Fig:SNR_maps} illustrate where a certain component (such as extinction) is out of range. These may appear inside the RoI (e.g. at the location of the SMC). Non greyed areas outside the RoI reflect an evolution of the criteria that led to the RoI definition, for example by tightening the Galactic latitude threshold from $|b|\geq\ang{23;;}$ to $|b|\geq\ang{25;;}$ after the northern ground-based surveys had been defined for $|b|\geq\ang{25;;}$.

Our more complex zodiacal model (Section~\ref{sec:zodiacal}) predicts a lower background that varies with time and position along the orbit. This modulation happens at a level far below the typical range of zodiacal background within the RoI (Fig. \ref{Fig:zodi_chk}), and hence we do not expect the median performance to change with this model.

In summary,  Fig.~\ref{Fig:SNR_maps} shows that the SNR in the VIS band exceeds the requirement of ${\rm SNR}\geq10$ over the whole RoI, with a median value of nearly 16. This gain is mostly related  to longer than required integration times, driven by the needs of the spectroscopic channel. Likewise, the $YJH$ photometric data are well above the 
${\rm SNR}\geq5$ requirement. A negligible area (less than 50\,deg$^2$) of the NISP spectroscopy is below the 
${\rm SNR}\geq3.5$ requirement (Fig. \ref{Fig:SNR_maps}. bottom). The median SNR for spectroscopy is $4.5$, a comfortable margin. Hence all specifications are exceeded, and on average \Euclid will go deeper than initially planned. 

\begin{figure}[!htb]
	\begin{centering}
	\resizebox{\onecolspan}{!}{ %
			\includegraphics{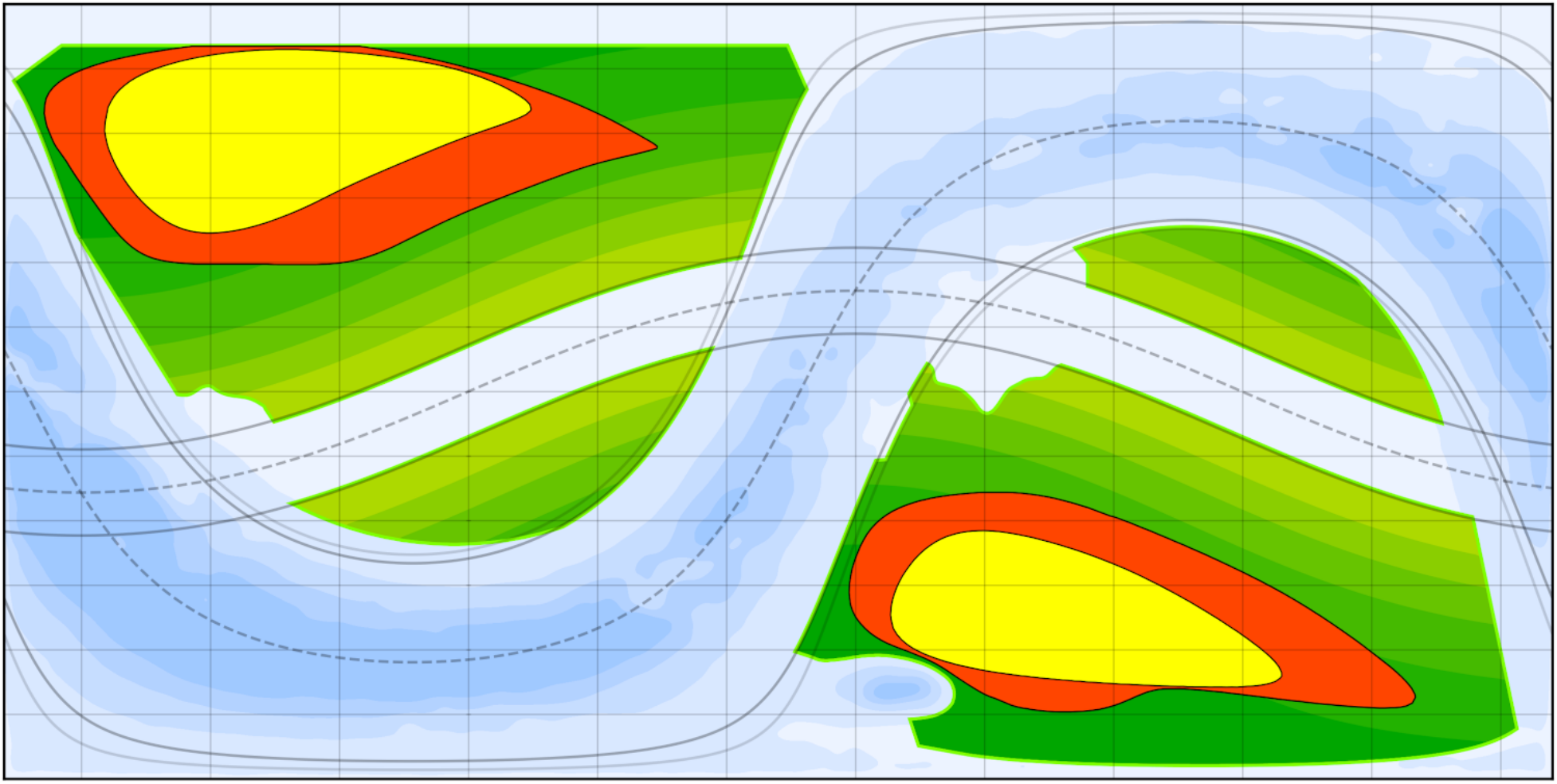} 
		} %
		\caption{Best SNR areas in each hemisphere / Galactic cap. In yellow the best 1300\,deg$^2$ in each Galactic cap, in red the best 2600\,deg$^2$ (including the yellow area). These are similar to the areas planned for the first and second data release, $2500$\,deg$^2$ and $5000$\,deg$^2$ respectively (cf. Fig.~\ref{Fig:coverage_by_year_moll}).  } 
		\label{Fig:best_SNR_areas}
	\end{centering}
\end{figure}

\subsubsection{Best SNR areas of the Euclid sky}\label{sec:best_SNR_areas}
The  areas of sky where the largest SNR can be achieved on average are offset from the ecliptic poles due to the out of field stray light from the Galaxy and the Large Magellanic Cloud (LMC). The SNR in these areas is close to the maximum values listed in Table \ref{tab:SNR_RoI_table}. Figure~\ref{Fig:best_SNR_areas} shows yellow and orange filled areas that were derived from an average of the VIS, $Y$, $J$ and $H$ SNR maps. The boundaries have been smoothed in this representation. The areas shown in yellow represent the best 1300\,deg$^2$ in each Galactic cap (or celestial hemisphere), and the orange area the best 2600\,deg$^2$ (including the yellow area). The EWS seeks to cover these best areas first. 

\begin{figure*}[!htb]
	\begin{centering}
		\resizebox{\twocolspan}{!}{ %
			\includegraphics{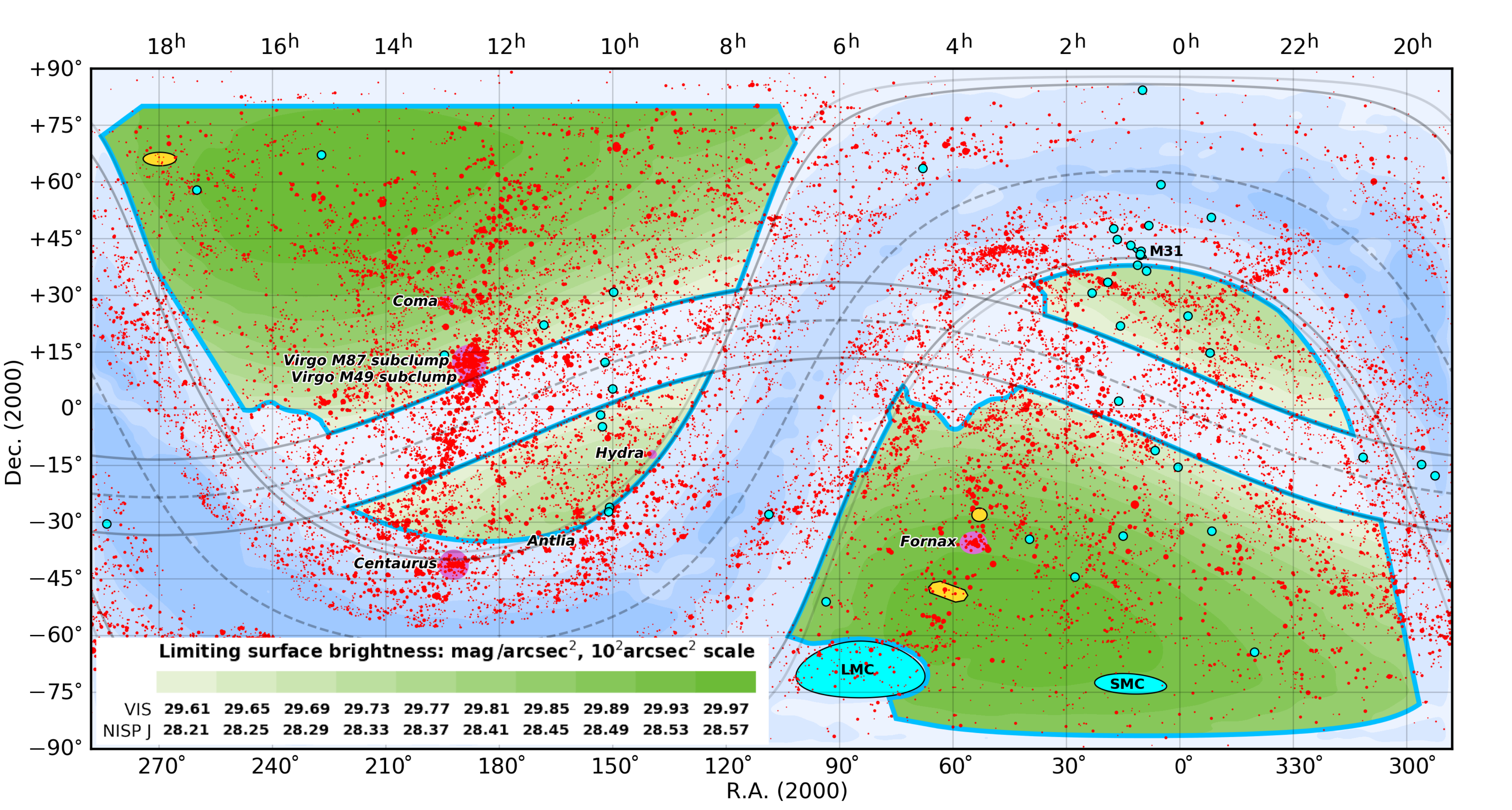} }
		\caption{Euclid VIS and $J$ band $1\,\sigma$ asinh limiting surface brightness at the $10^{\prime\prime}\times10^{\prime\prime}$ scale (LSB performance) and the nearby Universe: Local Group galaxies are represented by large cyan dots and galaxies up to $z=0.03$ by red dots, while the purple ellipses indicate the actual sky area covered by the nearest clusters of galaxies (Virgo, Coma, Fornax, Hydra are within the RoI). The RoI exhibits a 0.4 magnitude range in sensitivity for the four Euclid bands from the best area down to the ecliptic plane.}
		\label{Fig:roi_lsb}
		\resizebox{\twocolspan}{!}{ %
			\includegraphics{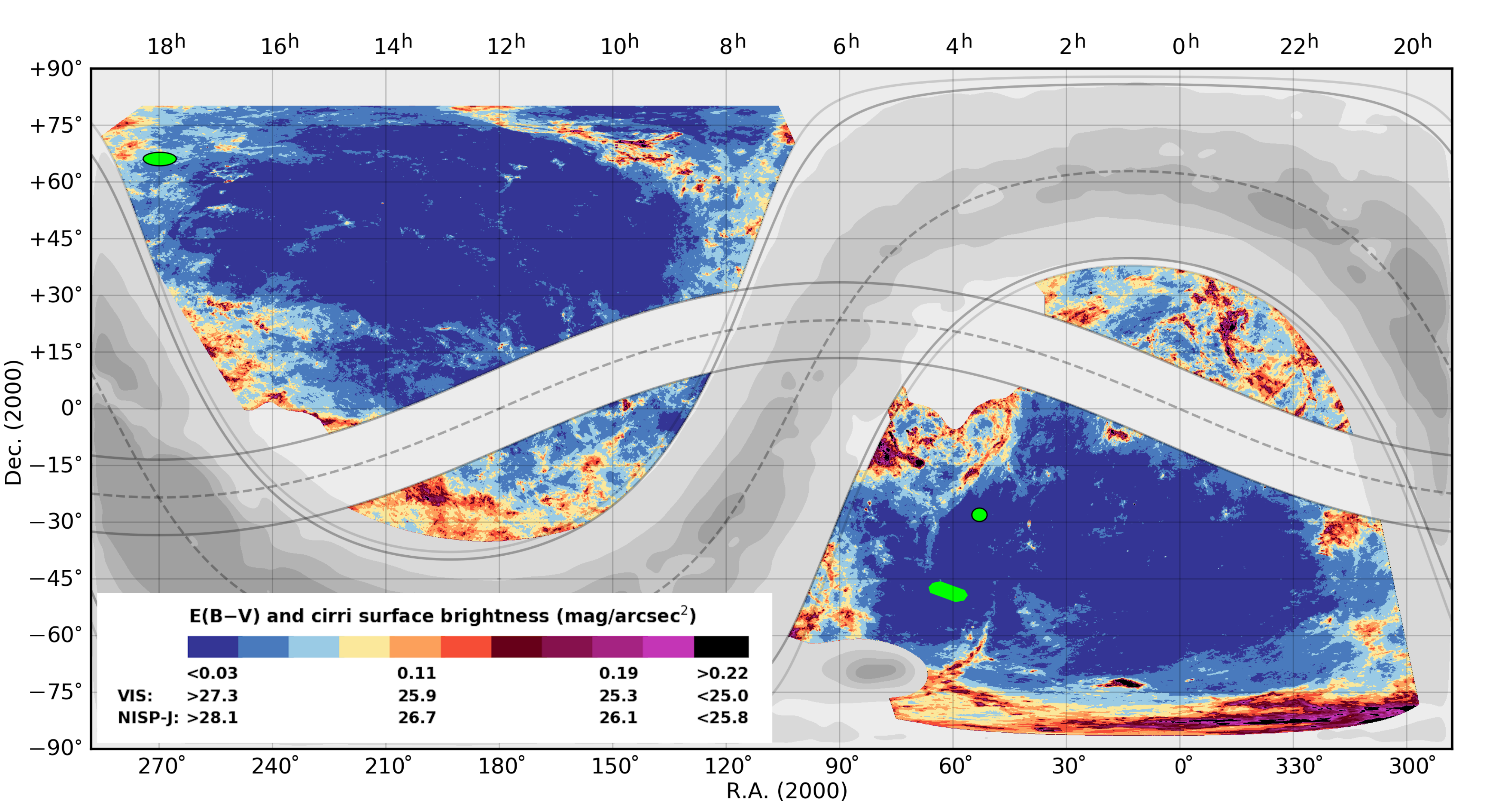} }
		\caption{$E(B-V)$ over the RoI at the native $5^\prime$ resolution of the Planck/IRAS map, and surface brightness of the Diffuse Galactic Light (cirrus) in the VIS and $J$ bands. The two islands (smaller parts of the RoI) collect the worst areas of the \Euclid sky due to the combined presence of cirrus and the proximity of the Galactic plane. The small green ellipses show the three Euclid Deep Fields.} 
		\label{Fig:roi_cirri}
	\end{centering}
\end{figure*}

\subsection{Limiting surface brightness}\label{sec:lsbperf}

As described above, we have a complete knowledge of the background reaching \Euclid's focal planes. Similarly to what we did for faint compact sources, we can derive the detection performance for diffuse emission such as non-resolved stellar populations in tidal streams of galaxies, intracluster light, and the cosmic infrared background (CIB). In the following, we will do this  considering the same noise properties as before, assuming a stack of three exposures for both VIS and NISP, as this applies to the 90\% level coverage of the imaging survey (see Sect.~\ref{sec:dither}).

We use the limiting surface brightness metric adopted by \cite{Mihos13}, which is based on the asinh magnitude introduced by \cite{Lupton99}. This conservative metric is a good description of the actual signal properties at very low $\sigma$-levels relevant for such science. It has the merit of reflecting an actual science performance: the determination of the light profile of Messier\,101 down to a surface brightness of $\mu_B=29.5$\,mag\,arcsec$^{-2}$ corresponds to the $1\,\sigma$ limit in the \cite{Mihos13} study. At $1\,\sigma$, the asinh magnitude is 0.5 magnitudes smaller than the corresponding standard magnitude.  This limiting surface brightness is computed directly from the photometric zero point of the system and the background noise property from our SNR study. Because of the very limited contamination of bright stars (see Sect.~\ref{sec:brightsources}) here we only need to consider the diffuse background. The limiting surface brightness expressed at the pixel scale is:
\begin{equation}
    \mu \; = \; {\rm ZP} - 2.5\logten(b) - a\,\asinh(f/2b),
\end{equation}
where, at the native resolution, ZP is the photometric zero point in electrons per second; $b$ is the noise per pixel in the image, assumed to be the Poisson standard deviation of the background counts per pixel, B, i.e. $b=\sqrt{B}$; $a=2.5\logten({\rm e})=1.08574$ and $f$ is the level in electrons per pixel of the extended astronomical source.

Given \Euclid's small plate scales ($0\farcs1$ and $0\farcs3$ pixel$^{-1}$ for VIS and NISP, respectively), the depth metric relation must be brought to the physical scale of common features encountered in the near-field (galaxies, streams, shells, dwarfs, etc): we adopt a generic $10^{\prime\prime}\, \times\,10^{\prime\prime}$ scale while conforming to the standard performance unit for extended emissions in magnitude per square arcsecond. We shift from the pixel scale to our scale of interest considering a square area of $n$ native pixels on the side. 
By averaging over the larger area, the estimate for
the noise
is scaled down by a factor $\sqrt{n\times\,n}$ while the zero point gets shifted by $-2.5\logten(n\times\,n)$ for flux conservation. Scaling to the magnitude per square arcsecond unit adds $-2.5\logten(p)$, with $p$ the area of the native pixel in square arcsecond (0.01 for VIS, 0.09 for NISP). For the adopted $10^{\prime\prime}\, \times\,10^{\prime\prime}$ scale ($n$=100 for VIS, 33.33 for NISP) the combined effects on the limiting surface brightness amount to $-2.5\logten(1/(0.01\times100)) = 0$ for VIS, and $-2.5\logten(1/(0.09\times33.33)) = +1.193$ for NISP (the larger the physical scale, the lower the noise and the greater the performance). 

We can now explore the \Euclid RoI at the asinh $1\,\sigma$ level $(f=b)$. The result is a map of limiting surface brightness for each band. The VIS and NISP maps show essentially the same structures (shape, amplitude, location), and in Fig.~\ref{Fig:roi_lsb} they are combined in a single map (see the color bar for the amplitudes in VIS and $J$-band).
The maximum range from the best area to the worst, at the ecliptic plane limit, is $\sim 0.4\,{\rm mag}\,{\rm arcsec}^{-2}$, corresponding to the total background level ratio of 2.25 between these best and worst areas. The median limiting surface brightness across the four bands over the RoI is: $m_{\scriptscriptstyle\rm VIS}=29.8$, $Y=28.2$, $J=28.4$, $H=28.4$ AB\,mag\,arcsec$^{-2}$, $1\,\sigma$ asinh magnitude at the $10^{\prime\prime}\,\times \, 10^{\prime\prime}$ scale ($-0.25$\,mag for the minimum performance, $+0.15$\,mag for the maximum over the RoI). We note that our various background components are conservative estimates and these levels can be considered a safe performance.
Our more complex zodiacal background model presented in Sect.~\ref{sec:zodiacal} 
indicates how intensity varies with time and position along the orbit. This will in consequence modulate the total background, hence the depth, although at a level lower than the 0.4 magnitude range depth seen here over the RoI (Fig. \ref{Fig:zodi_chk}). The median performance is not expected to change.
We also note that a depth metric  based on the integrated  Sersic radial profile over a whole galaxy, digging deep into the noise, typically adds at least two magnitudes with respect to this contrast-oriented metric. This is taking particularly into account \Euclid's pristine image quality that will enable an effective masking of the foreground and background compact sources.

For an illustration of the scientific potential, the map in Fig.~\ref{Fig:roi_lsb} features the nearby extra-galactic Universe up to a redshift of $z=0.03$: more than $10\,000$ bright ($K$-band magnitude $<12$) galaxies \citep[2MRS catalog][]{Huchra12}, including several members from the Local Group, and four nearby clusters of galaxies all falling within the Euclid RoI (some additional targets that are located outside the RoI might be observed during the unallocated time, see Sect.~\ref{sec:unallocated_time}). The three Euclid Deep Fields are shown in yellow 
on the map of Fig.~\ref{Fig:roi_lsb}. 
They are designed to be $+2$\,mag deeper than the EWS for compact sources, but the gain in depth is comparable for the diffuse emission (for specific details, see our companion paper on the Euclid Deep Fields, [Sc21]).

Such capacity at detecting faint nebulous objects will make the Diffuse Galactic Light (DGL, cirrus) an ubiquitous component of the background over the entire RoI as it averages to a level of 27.1$\,$mag\,arcsec$^{-2}$ in the VIS imaging (derived from multi-band dedicated CFHT-MegaCam observations to help characterise the RoI). Based on the scaling of the DGL's albedo \citep{Gordon04}, we derive an average of 27.9$\,$mag\,arcsec$^{-2}$ over the RoI in the NISP $J$-band. 
This extra background is shown in the map in Fig.~\ref{Fig:roi_cirri}. Note that it is included in our derivation of the limiting surface brightness, but it has no impact since it is more than five magnitudes fainter than the combination of zodiacal background and out-field stray light. 
We know that structures exist in the DGL down to the arcsecond scale even at high Galactic latitude \citep{MivilleDeschenes16}. The DGL is, however, truly diffuse over the great majority of the RoI, although as shown in Fig. \ref{Fig:roi_cirri} some parts of the \Euclid sky are inevitably worse than others. The two islands of the RoI (regions \FancyRoman{2} and \FancyRoman{3}, see Sect.~\ref{sec:roi_four_areas}) correspond to the worst parts of the \Euclid sky due to their proximity to the Galactic plane and a considerable presence of cirrus.

\begin{figure}[hbt]
	\begin{centering}
		\resizebox{1.0\onecolspan}{!}
		{\includegraphics[trim=10 0 10 0]{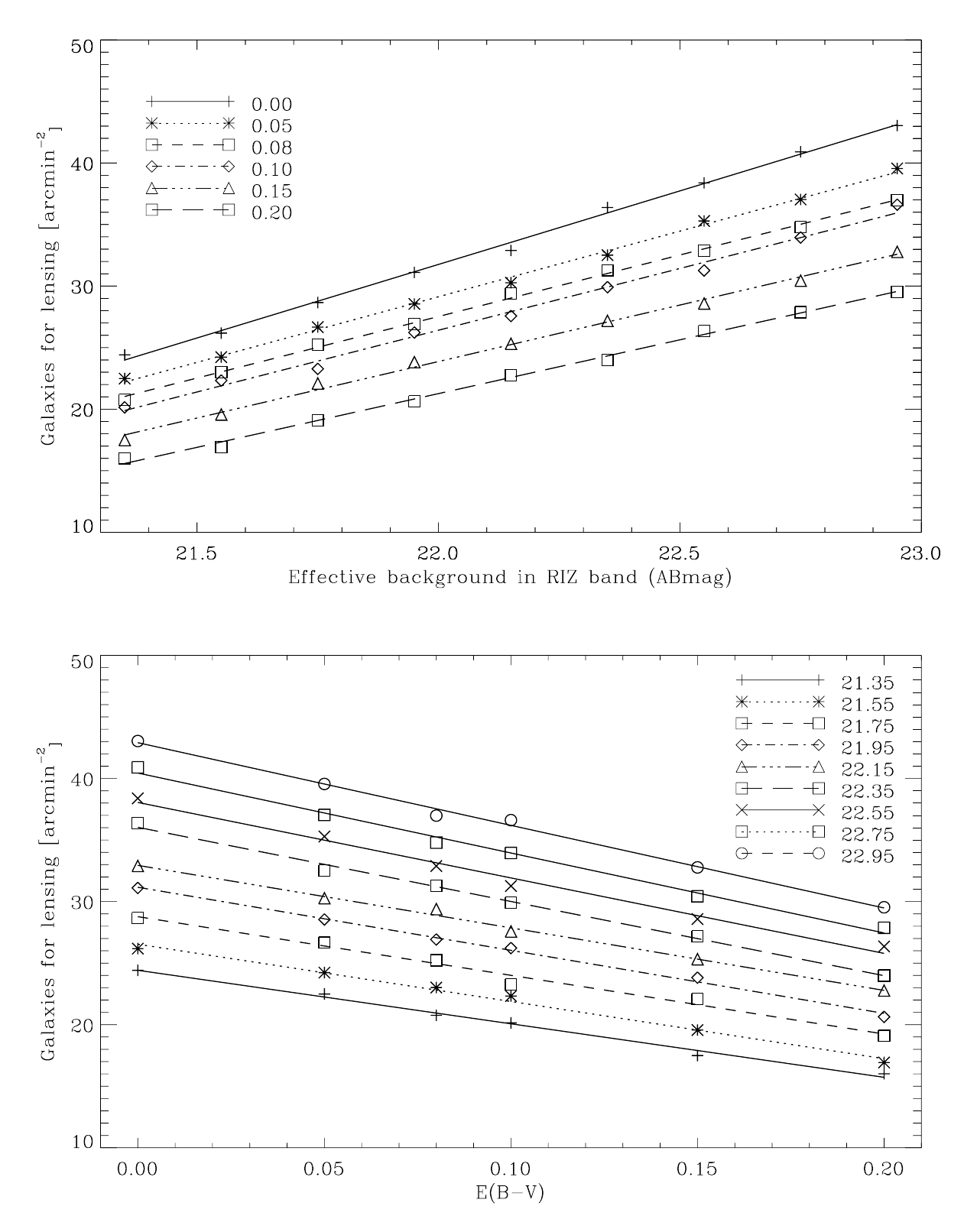}}
		\caption{Expected average number density of VIS galaxies satisfying the WL requirements (notice that the local impact of foreground galaxies or clusters is not considered here). \emph{Upper panel:} counts as a function of the local background (in mag\,arcsec$^{-2}$) for different levels of $E(B-V)$ reddening; \emph{Lower panel:} counts as a function of reddening for different levels of the background. Lines are linear fits to the simulations.}
		\label{Fig:WL_counts_laws}
	\end{centering}
\end{figure}

\begin{figure}[tb]
	\begin{centering}
		\resizebox{\onecolspan}{!}
		{ %
	\includegraphics{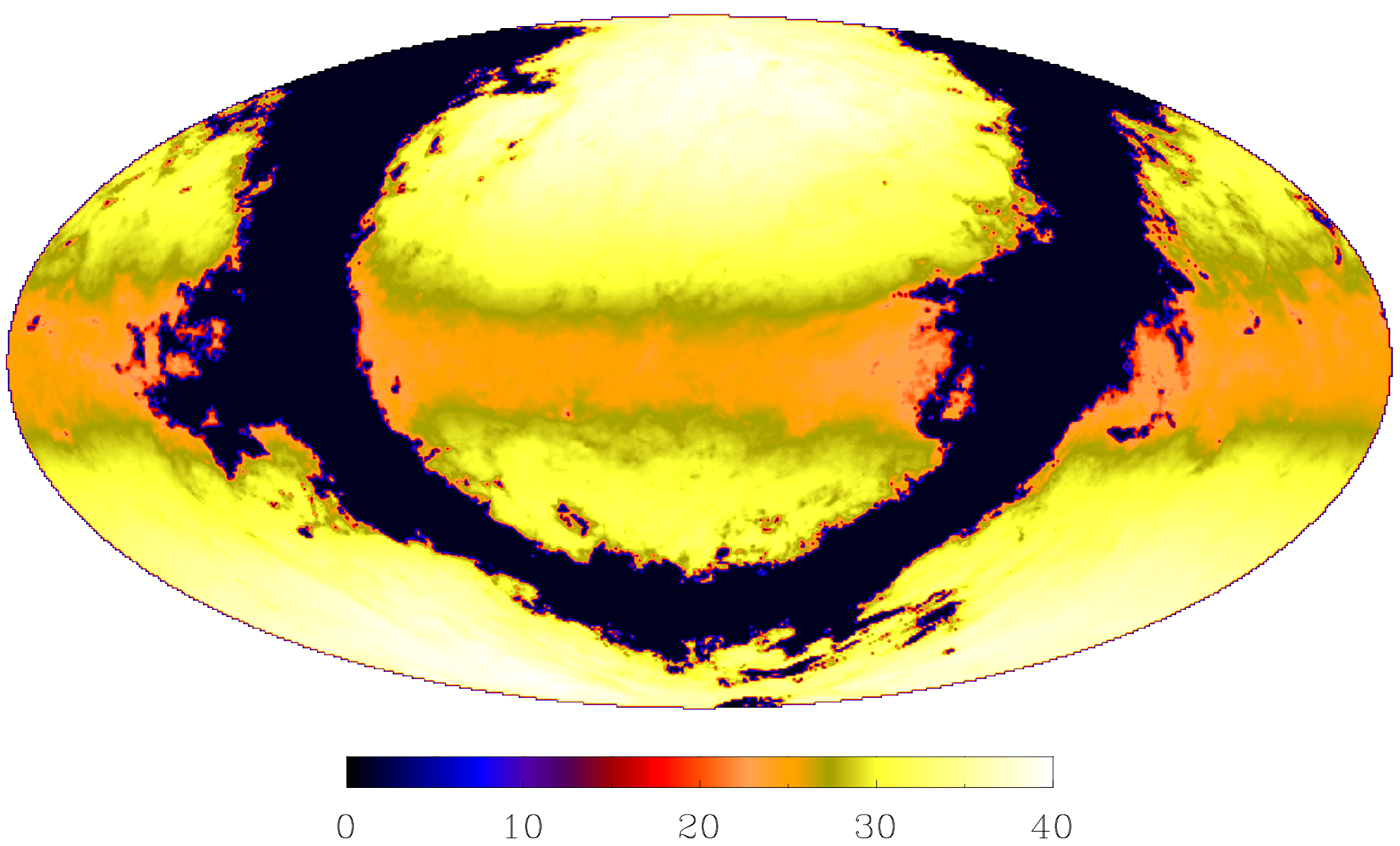}}
		\caption{
		Expected counts of WL galaxies in arcmin$^{-2}$ 
		(see text; note that the local impact of foreground galaxies is not considered here) in ecliptic coordinates with the Galactic plane removed (black). The decrease in the number counts is evident going towards the ecliptic plane, where the zodiacal background is higher. See also \cite{Scaramell15}.}
		\label{Fig:WL_counts}
	\end{centering}
\end{figure}
%
\begin{figure}[!htb]
	\begin{centering}
		\resizebox{\onecolspan}{!}{ %
		\includegraphics{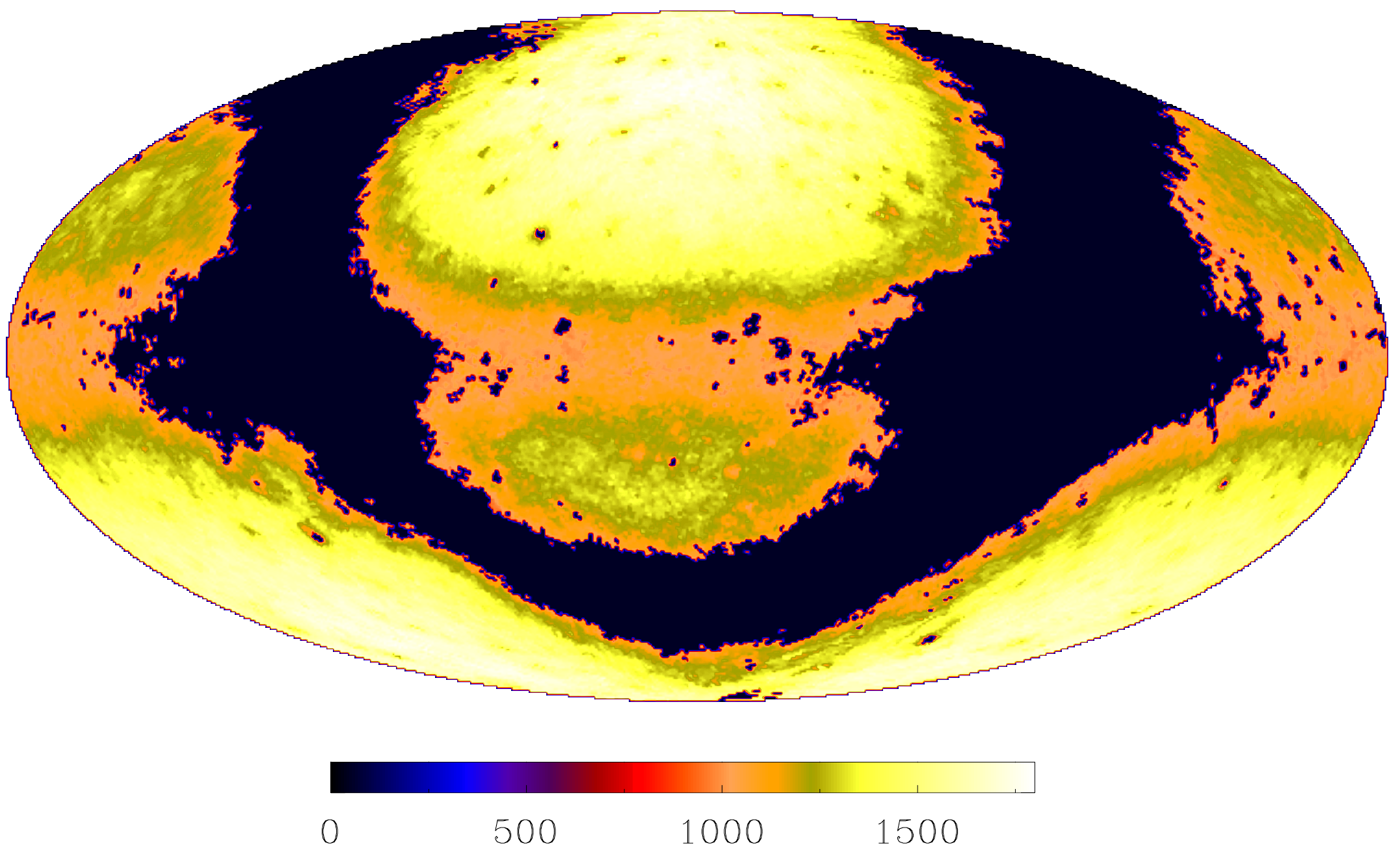} }
		\caption{
		Expected counts of GC galaxies (with reliable spectroscopic redshifts) in deg$^{-2}$ (see text) in ecliptic coordinates. Here the number counts decrease faster with latitude than for the WL counts (cf. Fig.~\ref{Fig:WL_counts}) since the impact of stars (spectra) from our Galaxy on the GC 1D spectra is more severe than the impact of stars (pointlike) on the 2D WL data.}
		\label{Fig:GC_counts_map}
	\end{centering}
\end{figure}

\subsection{Expected number of sources}\label{sec:counts}

The expected counts of galaxies that satisfy the WL requirements in terms of ${\rm SNR}>10$ \citep[as measured by \texttt{SourceExtractor},][]{SExtractor} and size (${\rm FWHM_{gal}} > 1.25\,{\rm FWHM_{PSF}}$) can be inferred using realistic image simulations. In our analysis, we used sky simulations produced with {\texttt{SkyLens}\xspace}\footnote{ http://metcalf1.difa.unibo.it/blf-portal/skylens.html}, in which the input galaxies magnitude, spectro-morphological, and redshift distributions were drawn from the HST Ultra Deep Field (HUDF). Since the HUDF is a very small field (11 square arcmin), the galaxies therein may not be representative of the mean properties of galaxies on the whole sky. Given the larger size of the COSMOS field, its counts are more robust. We have thus corrected the UDF counts such as to reproduce the magnitude distribution in the COSMOS field. We have also estimated the SNR based on the specified throughput of the VIS instrument and including Galactic extinction and the zodiacal background. Note this estimates are conservative because the latest throughput estimates for the VIS SNR are larger than the initially specified ones that we used. 

By simulating images under various conditions, and extracting the sources from them, the dependencies of the galaxy density with extinction and background were determined, as shown in Fig.~\ref{Fig:WL_counts_laws}. The results were interpolated from each pointing  to derive an estimate of the spatial map of number counts, shown in 
Fig.~\ref{Fig:WL_counts}. These estimates are currently being updated to include stray light, which is quite dependent on the local environment of bright stars and their spectral energy distribution (cf. Sect.~\ref{sec:brightsources}), and to include the latest instrument parameters and data reduction methods. 

\begin{figure}[!htb]
	\begin{centering}
\includegraphics[width=0.98\onecolspan]{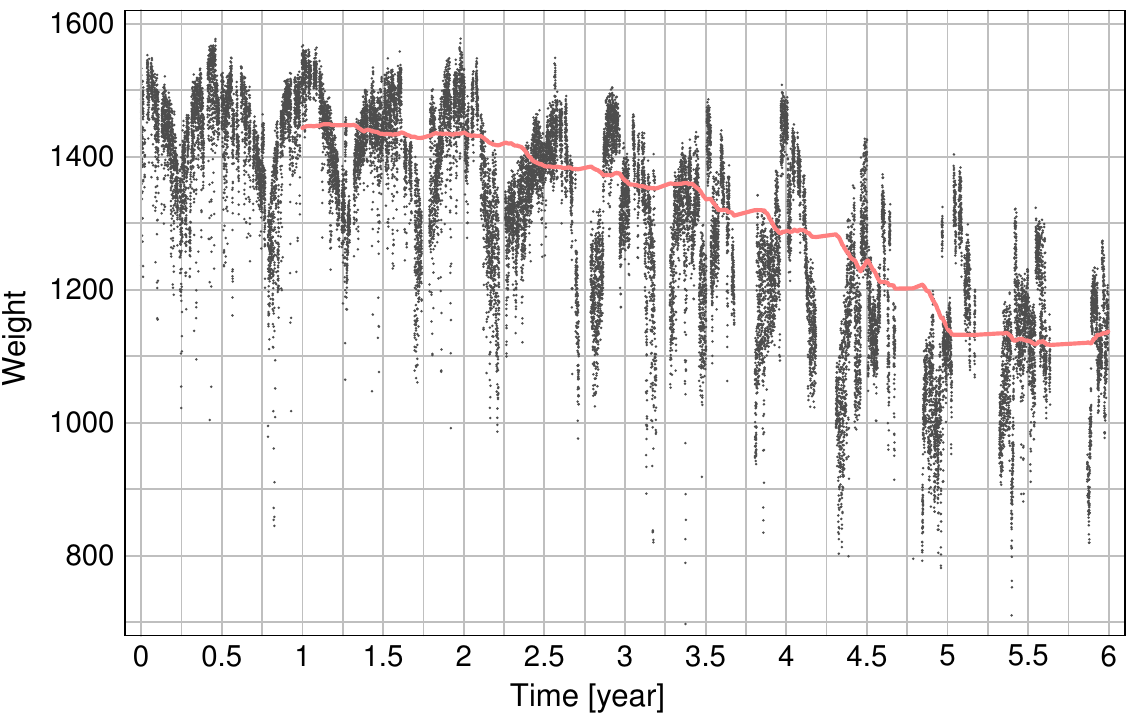}
		\caption{Weight (GC expected number of good redshifts deg$^{-2}$) of observed fields along a typical EWS progression (see Sect.~\ref{sec:wide})  with a superimposed moving median taken on the previous year (red line). The approximately regular cusps every 6 months are due to the paucity of observable fields close to the Galactic and ecliptic planes and their increased background and extinction. The decrease of the median and the upper envelope in the second half of the mission is due to a relatively lower quality of the sky areas still available for observations in those periods. In fact, the best areas are observed in the first half of the mission, see Fig.~\ref{Fig:best_SNR_areas} and Fig.~\ref{Fig:coverage_by_year_moll}}
		\label{Fig:weight_vs_time}
	\end{centering}
\end{figure}

A similar approach was carried out for GC, where spectra were simulated, then extracted and finally measured \citep{Zoubian14}, yielding a similar sky map of expected number of reliable redhifts over the sky  \cite[see][other methods are being developed]{Jamal2018}.  Figure~\ref{Fig:GC_counts_map} shows an updated version from the `Science Performance Verification' \#2 (SPV2) GC simulation (SPV is an end-to-end simulated exercise of the whole data reduction chain\footnote{A summary of the SPV2 exercise can be found in pages 4 and 5 of the EC newsletter at:   \url{https://www.euclid-ec.org/Documents/Newsletter/EC-Newsletter_issue08.pdf}\;.}).  
As for the WL case, the GC simulations are currently being updated to include the latest estimated effects of stray light, instrument characteristics and data reduction procedures. We expect that the overall fraction of recovered redshifts in simulations will increase with updated models and  throughput, coupled with  better simulations and data reduction, since the simple but up to date estimates for spectra now yield ${\rm SNR}>3.5$ almost everywhere (see Fig.~\ref{Fig:SNR_maps}).
The indicative expected number of reliable redshifts of each field (used as a relative weight) as a function of time for a typical EWS (see Sect.~\ref{sec:wide}) is shown in Fig.~\ref{Fig:weight_vs_time}. 
Only 2\% of the fields have a weight below 1000,
the consequence of an excellent survey efficiency and a good RoI selection. 
The overall normalisation of these numbers is based on model estimates of the intrinsic number of H$_\alpha$ emitters plus current estimates of contamination rate and a preliminary data reduction success rate. The latter is expected to be improved.
At present, the number of H-alpha emitters observable with \Euclid remains uncertain, but recently \cite{Bagley20} forecasted this number to be  $\sim 3300 \deg^{-2}$ in  the redshift range measurable by \Euclid. 
According to this forecast, to meet the initial \Euclid specification of $1700 \,\deg^{-2}$ would require  to reliably measure  the redshifts of half of the underlying population.

\begin{figure}[!htb]
	\begin{centering}
	\resizebox{\onecolspan}{!}{ %
		\includegraphics{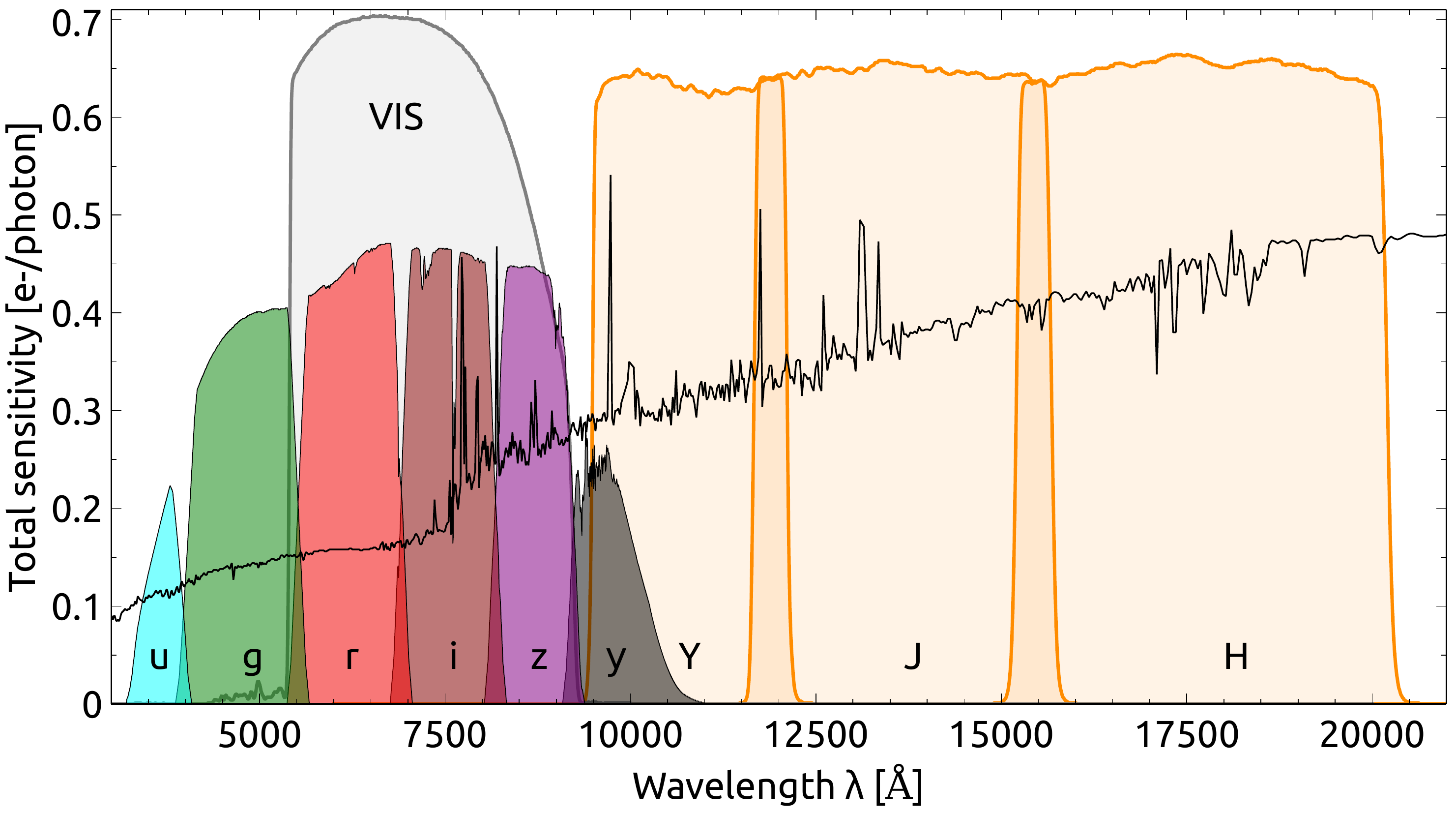} 
		} %
		\caption{A simulated spectrum of a typical (mass, star formation rate) galaxy from the \Euclid weak lensing population at $z=1$ (D. Masters, private communication), overlapping the \Euclid VIS and NISP $Y, J, H$ bands. The  Vera C. Rubin Observatory photometric bands ($u, g, r, i, z, y$) are also shown, highlighting the importance of the ground-based data to the photometric redshift derivation as the $4000\,$\r{A} break falls within the $i$-band. The total sensitivity for the ground is given at 1.2 airmass.}
		\label{Fig:photom_bands}
	\end{centering}
\end{figure}

\begin{figure*}[!hbt]
	\begin{centering}
	\resizebox{\twocolspan}{!}{ %
			\includegraphics{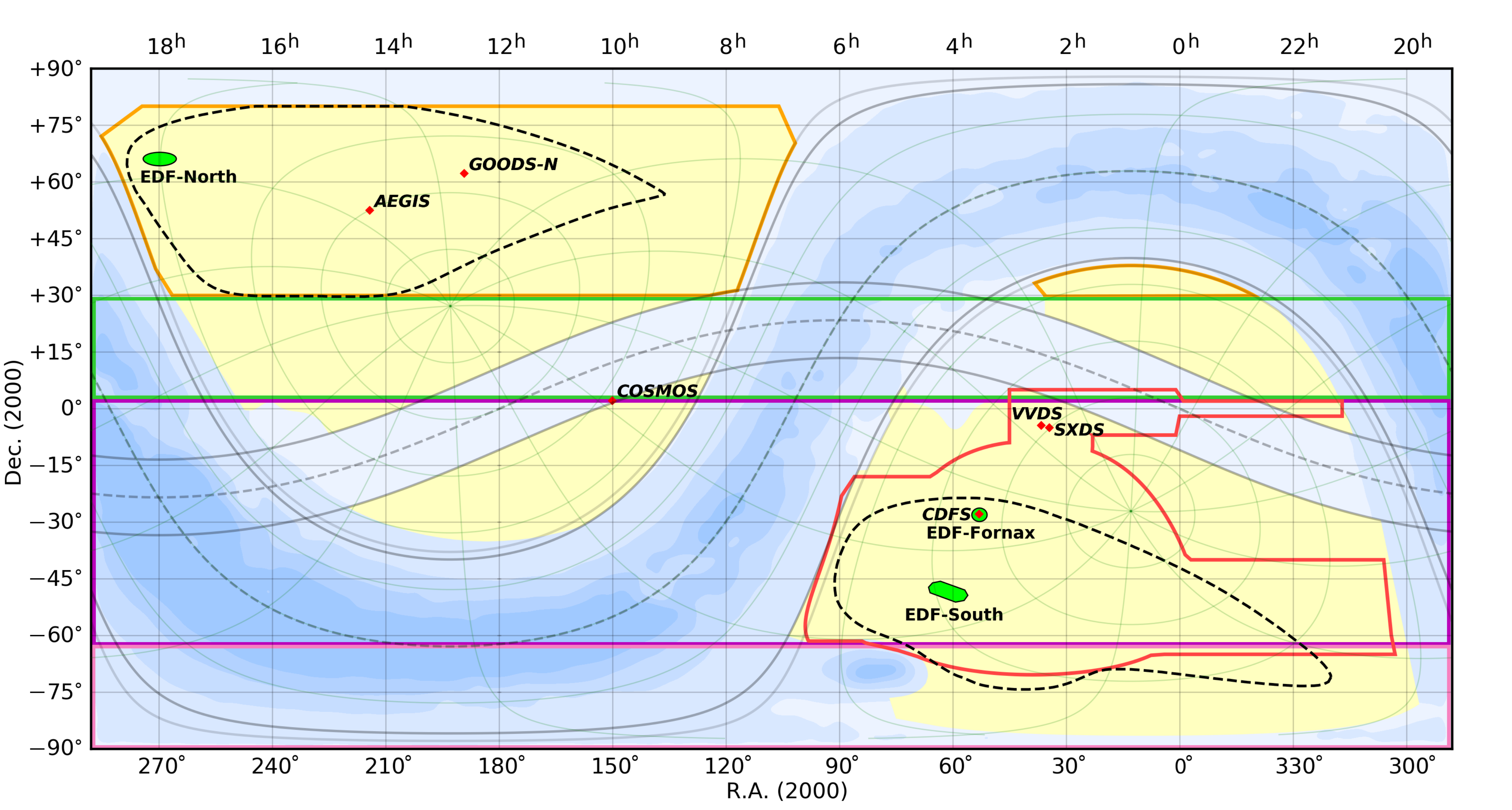} 
		} %
		\caption{Current plan (2021) of the ground-based coverage of the \Euclid sky based on the six most powerful ground-based wide-field imagers, and the \Euclid RoI (in solid yellow). Vera C. Rubin Observatory (LSST and suggested northern and southern extensions: purple, green and pink rectangles) and Blanco (DES, red) in the South. Subaru (WISHES), JST (JEDIS-$g$), Pan-STARRS, and CFHT (CFIS) in the North (orange). The best SNR 2600\,deg$^2$ areas for \Euclid in each Galactic cap (black dotted contours) are key to steer the ground efforts with respect to the mission's first  data release (DR1). The Galactic referential is shown in thin light green.}
		\label{Fig:ground_based_map}
	\end{centering}
\end{figure*}

\subsection{The EWS coverage by ground-based telescopes}\label{sec:ground_based}

Weak lensing tomography and the need to account for the contamination by galaxy intrinsic alignments require solid estimates of the redshifts of the galaxies used as sources in the weak lensing analysis. \Euclid will exploit galaxies up to a redshift of $z\sim 2$ with the majority of galaxies, at the lensing goal of $m_{\scriptstyle\rm VIS}=24.5$, lying at a redshift $z\sim 1$. A typical spectral energy distribution of such galaxy is shown on the simulated spectrum of Fig. \ref{Fig:photom_bands}. The NISP depth goals ($Y, J, H$\,=\,24.0, see Sect.~\ref{sec:NISP-P}) were scaled to capture the flux of this galaxy population at the required SNR for proper photometric redshift derivation. However, since the weak lensing imaging band through the broad VIS band ($r$+$i$+$z$) does not sample key features of a $z=1$ galaxy energy distribution, in particular the $4000\,$\r{A} break which falls within the $i$-band (Fig. \ref{Fig:photom_bands}), complementary bands are needed to reach the required redshift precision. The $g, r, i, z$ bands are critical in particular, as introduced by \cite{RedBook}. 

Large projects aiming at obtaining photometry through the Sloan bands over large parts of the sky were on the rise at the time of the \Euclid mission definition. Since photometry in those bands does not require the observing conditions of a space observatory, this critical part of the mission was left for 
external up-and-coming photometric surveys by ground-based facilities located across the two celestial hemispheres in order to reach the entire \Euclid sky. The minimal depths needed to derive photometric redshifts for the WL probe are 25.7, 25.1, 24.8, 24.6\,AB\,mag in the $g, r, i, z$ bands, respectively, for $5\, \sigma$ point-like source. These levels were first introduced in \cite{RedBook} and later on fine tuned 
to optimally match the spectral energy distribution of the $z$=1 $m_{\scriptstyle\rm VIS}$=24.5 galaxy populations anchored on the $Y, J, H$ depths of 24.0 that will be achieved by \Euclid. 

At the time of the mission selection in 2012, \cite{RedBook} commented on ground surveys that were still speculative: only the Dark Energy Survey \citep[DES,][]{DESC16} was about to start on its broad-band imaging effort, with nearly 4500\,deg$^2$ of its of 5000\,deg$^2$ goal overlapping the EWS. As of the end of 2020, the now completed DES has secured in the $g, r, i, z$ bands coverage of nearly a third of the EWS area over the south Galactic cap while the rest of the \Euclid RoI is an on-going effort. Together, the following six most powerful ground-based wide-field telescopes will eventually deliver the photometry needed by \Euclid across the $u, g, r, i, z$ bands from the northern and southern hemispheres (Fig.~\ref{Fig:ground_based_map}), 
the $u$-band being a solid bonus for photometric redshifts at any depth.

Extensive community-based actions led to the Canada-France Imaging Survey \citep[CFIS,][]{Ibata17}, which ought to cover by 2025 the northernmost 4800$\,$deg$^2$ of the \Euclid RoI in the $u$- and $r$-band, using 314 MegaCam nights on the 3.6$\,$m Canada-France-Hawaii Telescope (CFHT). Spain's 2.6$\,$m Javalambre Survey Telescope \citep{Cenarro18} should start in 2021 covering that area with the Javalambre-Euclid Deep Imaging Survey in $g$-band (JEDIS-$g$, 100 nights). Pan-STARRS \citep[USA, 2$\,\times\,$1.8$\,$m telescopes,][]{Chambers16} joined in 2018 to provide the $i$-band by 2025 as a result of their on-going Near Earth Object (NEO) search. Finally, a group of Japanese scientists joined the Euclid Consortium in 2020 through the contribution of Subaru Hyper Suprime-Cam \citep[HSC,][]{Miyazaki18} time (40 nights). WISHES (Wide Imaging with Subaru HSC of the Euclid Sky) will cover the northern area in the $z$-band, the most demanding band in terms of depth, hence requiring an 8$\,$m class telescope. The telescopes actively collecting data are now working in concert as part of the Ultraviolet Near-Infrared Optical Northern Survey (UNIONS), an independent consortium motivated by the shared effort for \Euclid, to cover the northernmost sky over the complete set of photometric bands, with a completion date around 2025. Canadian and University of Hawaii UNIONS members launched an effort in 2019 to gather $g$-band data with Subaru-HSC to complement the Spanish effort. Note that since CFHT and Subaru cannot effectively observe from Hawai'i at declinations $\delta\geq+80^\circ$, the EWS RoI has been trimmed around the equatorial pole by a few tens of degrees, a minor hit since the area was already mostly rejected due to high dust extinction.

Meanwhile the Vera C. Rubin Observatory \citep[USA,][]{Ivezic19} is approaching first light and the start of the Legacy Survey of Space and Time (LSST) should be in phase with \Euclid. When on the sky, Rubin will be the most powerful wide-field imager ever built and the \Euclid minimal depths in the $ugriz$ bands will be reached within one year of normal Rubin LSST operations over the 8000\,deg$^2$ overlapping the \Euclid RoI in the southern sky; this will supersede the DES dataset. The Rubin Observatory being such a powerful machine, the Euclid Consortium is investigating with the Rubin community a northern survey extension serving various strategic Rubin scientific niches \citep{Rhodes17}. Such an extension (3000$\,$deg$^2$ of \Euclid RoI area) would fill the $+$2 to $+$30$\,$deg declination gap between the main component of the LSST and the on-going \Euclid northernmost sky effort (Fig.~\ref{Fig:ground_based_map}).


\section{Calibration and Deep Field observations}\label{sec:calibs}
\Euclid's three major modes of observation (VIS, NISP-P and NISP-S) and their tight scientific requirements imply a thorough and extensive calibration program throughout the mission, which serves two main purposes, namely the calibration of the flight hardware (instrument calibrations, see Sect.~\ref{sec:instrument_calibrations}) and the characterisation of the target galaxies and quantification of any biases that may arise in the WL and GC experiments (sample characterisation, see Sect.~\ref{sec:deep_fields}).

\subsection{Telescope and instrument calibrations}\label{sec:instrument_calibrations}
The hardware calibration focuses on the properties of the optics, detectors, electronics, and the opto-mechanical aspects of structural components. The associated performance regarding PSF, throughput, quantum efficiency, noise and bias, is subject to change due either to  variations in operational conditions (orbit, depointing, on-board power dissipation) or to long term ageing (micro meteorite pitting, particles and UV radiation damage). 
The optical performance is furthermore affected by contamination from material outgassing. Monitoring and accurate correction of these effects is paramount to the scientific success of \Euclid and requires repeated execution of calibration observations with varying cadences. 

The hardware calibrations can be divided into on-sky (e.g. transmission) and off-sky calibrations (e.g. flat fields). The latter have the least constraints as they can be executed independently of the spacecraft's pointing; the scheduling must merely respect the required cadence within some tolerance. On-sky calibrations have the additional constraint that they must minimally disturb the thermal equilibrium of the spacecraft. To this end, a selection of targets is available, from which we choose those that optimally merge with the scientific observations.

In the following we provide a summary of the main aspects of the calibrations that impact the
building of the EWS
(excluding the additional calibration data taken during the performance verification phase, and shorter instrumental calibrations that are integrated in the ROS; see Sect.~\ref{sec:std_sequence}).

\subsubsection{Self-calibration and VIS nonlinearity}\label{sec:selfcal}
Quite important for hardware calibrations are the self-calibration observations, a block of about 18\,h observing a field near the NEP with perennial visibility (for details about this field see [Sc21]). Besides monitoring the total system transmission, these observations provide the data for a large range of additional Calibration Products. The self-calibration observations are scheduled eleven times per year, approximately on a monthly basis, and back-to-back with a VIS non-linearity sequence of about 9\,h duration. 

\subsubsection{VIS PSF calibration}\label{sec:VIS_PSF_calibs}
Another large Calibration Block is a sequence of VIS PSF  observations lasting $20.5$\,h, targeting one  of about a dozen stellar fields featuring a suitable range of magnitudes and spectral energy distributions (SEDs), while minimising Galactic extinction and polarisation effects on the PSF ellipticity. The \Euclid VIS PSF has in fact a weak dependence on polarisation of the incident light.  Most fields in the EWS and EDS have low levels of Galactic polarisation at VIS wavelengths, but we must ensure that the PSF calibration fields are also selected to have low polarisation.  Additional dedicated observations of polarised regions are separately planned to measure the PSF polarsation dependence, in orbit. 
These observations are used to routinely update and validate the VIS PSF model, and must be taken with the spacecraft in thermal equilibrium. This condition is met after about one week without large changes in the SAA and AA attitude angles, i.e. as close as possible to the values used in the preceding days of EWS observations. The VIS PSF calibration data 
hence need to be embedded in the currently executed EWS patch. In the global schedule (Fig.~\ref{Fig:full_schedule}), the longitudes of the PSF calibration fields are marked on the top row, and the related observations are shown by the red strips within the EWS observation blocks.

\subsubsection{NISP calibrations}\label{sec:NISP_calibs}
Two types of NISP calibrations are relevant in the description of the EWS. First, the NISP nonlinearity calibration requires a data-intensive special readout mode and must be done one detector at a time, requiring a total of 49 h. In the current implementation of the EWS, these calibrations are scheduled approximately every six months. In a future version of the EWS, these observations might be partially executed in parallel with one of the VIS on-sky calibrations, pending a confirmation of the instrument inter-operability.

The second set of NISP calibrations is a one-time check of the NISP-S wavelength dispersion solution ({\tt NISP-S-PN-1} in the 2nd year of Fig. \ref{Fig:full_schedule}).
This is initially obtained during the performance verification phase prior to the beginning of the survey, and repeated once, about one year after the start of the EWS. The dispersion  solution is obtained from compact planetary nebula (PN) with strong emission lines, being stepped across a larger number of positions in the NISP focal plane. The dispersion solution is transferred to the self-cal field (Sect. \ref{sec:selfcal}) to establish a set of secondary standards to monitor the stability of the dispersion. Currently, the PN can be chosen from a list of 24 suitably compact PNe, which will be down-selected to some extent pending further ground-based spectroscopy.

\subsection{Euclid Deep Fields and Euclid Auxiliary Fields}\label{sec:deep_fields}

The \Euclid schedule devotes much time to deep observations for galaxy sample characterisation.
For GC, one needs to quantify biases in  redshift measurements due to contamination and emission line misclassification (completeness-purity calibrations). For WL  one needs to quantify biases in shear estimation due to noise \citep{Viola2014}, colour gradients \citep{Semboloni2013,Er2018}, and the calibration of photo-$z$s. 
To this end, \Euclid will observe
three types of fields:
\begin{enumerate}
\item Deep observations of six well-known fields that have extensive ground- and space-based multi-wavelength photometric and spectroscopic coverage. These are hereafter known as the Euclid Auxiliary Fields (EAFs): CDFS, COSMOS-Wide, SXDS, VVDS-Deep, CANDELS/AEGIS, and CANDELS/GOODS-N.  They are covered with 1--4 Euclid FoVs, i.e. spanning 0.5--2.0\,deg$^2$. The six EAFs are observed for photometric redshift calibration and colour gradient calibration purposes. 

\item Repeated observations of two 20\,deg$^2$ fields at different times to obtain different dispersion angles to calibrate spectral confusion. These are called the Completeness-Purity-Calibration fields (CPC).

\item Deep observations of large (10--20\,deg$^2$) fields, two magnitudes deeper than the EWS, for calibration of the noise bias. These are the three Euclid Deep Fields (EDFs: EDF-North, EDF-South, and EDF-Fornax).
\end{enumerate}

Defining the exact locations and footprints of the three EDFs required substantial effort.\footnote{The properties of the three EDFs can be found at\\ \url{https://www.cosmos.esa.int/web/euclid/euclid-survey}\;.}
EDF-North, located at the NEP, is visited 40 times with the ROS to reach a depth two magnitudes deeper than the EWS, while the other two EDFs, at lower latitudes and larger local background, need more visits to reach the required depth.
To maximise synergy, EDF-North and EDF-South are chosen to have the same centers  of CPC-North and CPC-South, respectively. 

The EDFs and EAFs  will have great scientific and legacy value due to the depth of the observations. 
A detailed description of their planned observations is presented in [Sc21].

\subsection{Computing the \stageone schedule} \label{sec:first_stage}

\ECTile is the software purposely developed to generate the scheduling of the EWS. It has two major stages. In the preparatory \stageone, \ECTile computes the schedule of the various calibrations, EDFs and EAFs. We review \stageone in this section. The core of \ECTile is \stagetwo, which computes the EWS; it is described in Sect.~\ref{sec:wide}. 

The observations of calibration fields, EDFs, EAFs, and also ecliptic `polar caps' (see below), hereafter called `targets', share the same traits: most are only visible during a short interval at a given time (apart from the poles themselves), must observe a specific region (some with a specific orientation), and some must be re-observed with a fixed cadence. Given these restrictive properties, their schedule is computed in \stageone, followed by the scheduling of the EWS in \stagetwo in the remaining time available. When scheduling the targets we need to make sure to leave enough time for EWS in each year, in order to fulfil the planned public delivery of reduced data to the community (Sect.~\ref{sec:intro}) . 

We note that for the reasons explained in Sect.~\ref{sec:polar_caps}, the polar caps at high ecliptic latitudes ($|\bbeta|\geq\ang{79;;}$) are also part of the \stageone schedule. Each of the two polar caps is covered with a fixed number of patches (thirteen in the northern polar cap, six in the southern polar cap), defining targets that are scheduled with a locally fixed patch area and sequence, i.e. these two regions of the EWS are observed with the same strategy as the EDFs and EAFs.

The resulting schedule is called the `\stageone schedule', and its computation consists of three steps:

\begin{enumerate}
\item analysis of each target (strategy and visibility);
\item placement of each target in longitude and year;
\item assignment of timestamps to each placement.
\end{enumerate}

\subsubsection{Target analysis} \label{sec:target_analysis}
In general, a target has four types of constraints: 
a fixed location on the sky, a specific observation sequence, a minimum depth, and a cadence (or other time constraints). From these we compute the target's `window of visibility', i.e. the  
range of ecliptic longitudes (of the Sun) in which the target may be observed (on the leading or on the trailing side of the orbit). Typically, each target has two windows of visibility per year.

The computation of each target window of visibility reveals its flexibility in placement and, more important, possible conflicts (of placement) with other targets.
The procedure is essentially manual (aided by software tools).
Three cases are of particular interest (see [Sc21] for details).

The first case is the EDF-Fornax, which is planned to be observed gradually along the mission for a total of 52 times, taking into account the larger local background.
It can be observed twice per year. However, the window of visibility of the EDF-Fornax partially collides with that of the EDF-South, and with some orientations of the CPC-South. This means that scheduling one of these targets strongly restricts the placement of the others. To overcome this conflict, the visits to CPC-South are all scheduled in the first year, leaving those longitudes free for EDF-Fornax, in the  following years (5 times once every six months, plus a short 2 times visit on the first year, as required). Likewise, the EDF-South is observed from the second year onward (but offset from EDF-Fornax).

The second case is the scheduling of the  COSMOS and SXDS photo-$z$ calibration targets. These require an observation of a $2\times 2$ pattern that, if observed in one go, would take 15 days each. This would pose a great difficulty for the scheduling of the EWS, by adding a long interruption. In general, when scheduling the EWS, it is possible to bridge over interruptions (such as calibrations), if shorter than five days. Long interruptions are not bridgeable, forcing a interruption of the EWS build-up, something that must be compensated in the next passage (at least a half-year later). It also reduces the opportunity to place PSF calibrations (that require a week of EWS observations prior to calibration).
The solution is to observe these fields in four visits, observing half of it each time, first to an intermediate depth, and then a second time to the final depth. 

The third case are the self-calibration and VIS non-linearity calibrations, which both require approximately a monthly cadence of observations. In order to decrease the number of interruptions to the EWS observations and the number of large slews used, 
these observations are always scheduled in sequence and executed at the same sky field, defining a `recurring calibration block'

Finally,  every four weeks starting on Mondays at noon (UTC) $\pm$ 1.2\,h, there is one block of 12\,h that is reserved for spacecraft orbit maintenance operations (SOP). During SOP time survey data cannot be taken.

\subsubsection{Target placement} \label{sec:target_placement}

The second part of the computation of the \stageone schedule consists on placing each target in a table (named the `design schedule'), at a given longitude and year, striving to avoid collisions with other targets. In that table, choices have to be made such as what target goes into each year, and at what longitude to place it. There is some freedom in this procedure. For instance, many targets do not require to be observed in a specific year. However, some targets are more stringent than others; i.e., have shorter visibility windows or have a single annual visibility. Therefore, for simplicity, it is preferable to place the targets by decreasing order of perceived difficulty, filling the table year by year.
This process is performed manually. 

While being manual, the process does not need to be very precise: some overlap in longitude between targets is allowed. Any overlaps are rectified in the next step and the longitudes are converted to timestamps, transforming the design schedule into the \stageone schedule.

The design schedule is filled according to a strategy that prioritises the placement of the targets with the most constraining observing conditions. In the first year the order of placement is as follows.
\begin{enumerate}
\item	Place all ten CPC-South visits at transit longitude. These are the most demanding targets.
\item 	Place recurring calibration blocks (self-cal+VIS non-linearity) at an approximate step of $\ang{360;;}/11 = \ang{32.73;;}$. This promotes a synergy with CPC-North.
\item 	Place the ten CPC-North visits. The orientation of these targets has a step of  $\ang{32.73;;}$.
\item	Place eleven targets of the northern polar cap. The width of these patches and the angular offset was chosen to match the placement of the CPC-North targets (i.e. immediately before or after). The remaining two targets are placed in the second year.
\item 	Place all eight targets of the southern polar cap.
\item	Place the first two targets of the COSMOS-wide. This covers half the area needed on this target in the first year.
\item 	Place a $\times2$ pass  visit to EDF-Fornax, slightly offset from its ideal placement, to avoid collision with one of the CPC-South targets.
\item	Place the AEGIS target (required for the first year).
\item 	Place the two NISP nonlinearity targets. Because these are freely placeable, the choice is to place them on the slot of longitudes where there is less area of EWS within the RoI.
\item   Place a double visit to EDF-North to avoid a small gap of time between other targets.  Since it can be placed all year round, the visits to EDF-North may be used as a filler (as in this case).
\end{enumerate}
At this point, the first year is filled. It would be possible to pack more targets into it, but that would reduce the size and number of EWS windows, which in turn reduces the opportunity to place VIS PSF calibration fields. The latter need to be scheduled within EWS windows larger than seven days (see Sect.~\ref{sec:VIS_PSF_calibs}). 

With all CPC observations placed on the design schedule, the most demanding target left to be scheduled is EDF-Fornax. Unlike  the EDF-South, the EDF-Fornax cannot be reached far from transit through a large depointing because of its low ecliptic latitude. The remaining years follow a common strategy, with most of the targets (EDF-Fornax, EDF-South, recurring calibration blocks, EDF-North, NISP wavelength dispersion) being scheduled almost exactly the same way. The exception are the EAF targets that vary from year to year: In the current schedule, the SXDS is placed in the 3rd and 4th years, and the VVDS and GOODS-North in the 5th year. 

The strategy for the remaining years is as follows.
\begin{enumerate}
\item 	Place two $\times5$  passes to EDF-Fornax at their transits.
\item	Place two pairs of visits to EDF-South. One pair of one pass visit plus a $\times2$  pass visit, separated by \ang{5;;}, and a second pair of two $\times2$  pass visits, also separated by \ang{5;;} (ensuring survey-windows not smaller than $5$ days, given an average orbital progression of \ang{1;;} per day). Each group is clustered around each EDF-Fornax pointing.
\item 	Place the 11 recurring calibration blocks. The blocks are slightly offset from a nominal cadence to avoid conflicts with the EDF-Fornax and EDF-South observations.
\item   Place the two remaining northern polar targets (second year only).
\item 	Place the large block of NISP wavelength dispersion target (second year only).
\item   Place EAF targets (these vary from year to year).
\item 	Place two visits to EDF-North, one $\times4$  pass and another $\times2$  pass, near the centre of the range of EWS RoI scarcity, matching surrounding recurring calibration blocks.
\item Place the two NISP non-linearity targets.
\end{enumerate}

\begin{sidewaysfigure*}
		{\includegraphics[angle=0, width=\textwidth]{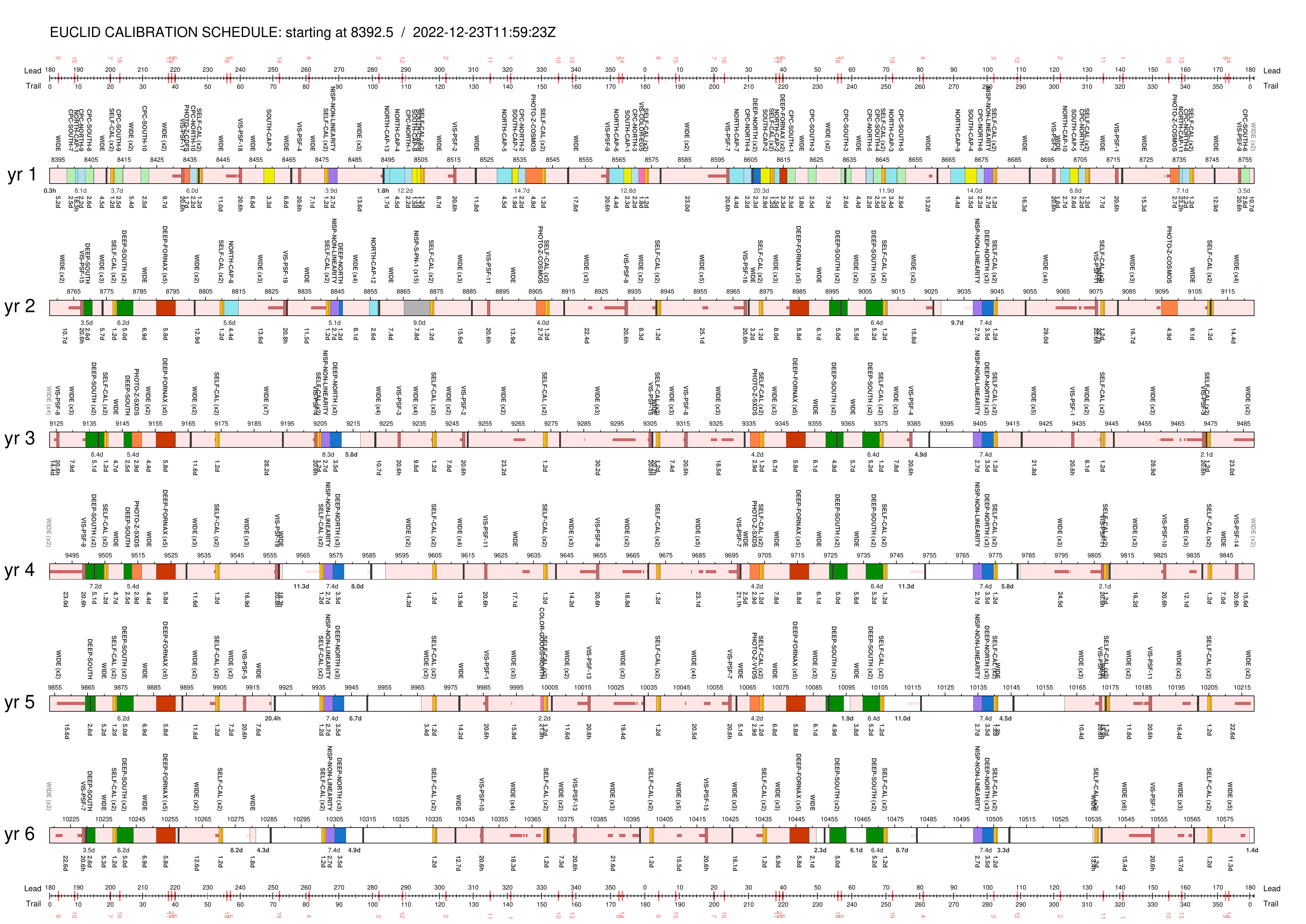}
		\caption{An example of the full schedule, including also the \stageone schedule discussed in Sect.~\ref{sec:timepstamps}. Time goes left to right for each row, which spans one year of the survey. Subsequent years go from top to bottom. See text for a description}
		\label{Fig:full_schedule}}
\end{sidewaysfigure*}

\subsubsection{Timestamp assignment} \label{sec:timepstamps}

The last stage of the computation of the intermediate schedule is the automated conversion of the year-by-year longitudes into timestamps.  
Given a starting date to the survey routine phase (currently but not frozen yet, this is expected to start on  8392.5  Modified Julian Date, or 2022-12-23T11:59:23Z), it is easy to compute the corresponding Sun longitude (\ang{271.30;;}). Then, it is a simple matter of ``reading'' the design schedule, year-wise. 

The first target following the starting longitude is a recurring calibration block at \ang{290;;}, approximately two days after the beginning. This timestamp is assigned to the first target found. Practically, this involves only computing the next timestamp when the Sun is at a given longitude. The process continues, assigning timestamps to targets of the first year, by order of longitude (wrapping around at longitude \ang{360;;}). Once the traversing reaches the initial longitude, the process continues in the second year, starting at the same longitude where the first year ended.
Once the second year is completed, the process continues on the third year and so on, until the intermediate schedule is completed. There is a possibility that the initial longitude for traversing any year coincides with the middle of a target. In those cases, the traversing continues after that target.

Besides converting longitudes to timestamps, an algorithm disentangles overlaps between targets and prevents, when possible, the occurrence of too small EWS windows.
The overlap of two or more targets (allowed in the previous stage), is fixed by offsetting those targets from their initial placement, minimising the overall offset, within the range allowed by their window of visibility. In most cases, this process is sufficient to resolve overlaps.  In case of failure, the solution is to go back to the previous stage and fix the overlap manually.
The same process is applied to eliminate the occurrence of small windows. Except that, now, the offsetting is in the opposite direction, pushing targets closer to each other. 

\subsubsection{Results}\label{sec:time_calibs}

Figure~\ref{Fig:full_schedule} shows an example of the full \Euclid schedule.
The result of the \stageone procedure, in this diagram, is the sequence of coloured boxes.  
The time allocated for each observation of a target is represented by a labelled box of a unique colour. The pink and white boxes represent the periods available to observe the EWS (at the end of the \stageone scheduling they are all still unallocated). During the \stagetwo scheduling (Sect.~\ref{sec:wide}) the EWS is scheduled in part of the available time (shown by the pink boxes), while some time periods remain unallocated (shown by the white boxes that are increasingly longer towards the final years of the survey, cf. Sect.~\ref{sec:unallocated_time}). 

The alignment of boxes with the same colour across years indicates that the respective targets are scheduled at the same time every year. 
The labels (and widths) of the boxes indicate when consecutive visits are made to the same field. The Deep Fields are often scheduled with more than one visit in a row, in particular EDF-Fornax is usually visited 5 times to efficiently use its short time visibility that occurs twice per year. 

Differently from the other calibrations, PSF calibrations are scheduled within the pink boxes, i.e., the EWS observation is interrupted to point to a PSF field (for $\sim 20.5$\,h) and then return to the same position.  This allows for a much better stability of the SAA and AA values used (that will match the ones used on that observation of the EWS), than if they were scheduled in the recurring calibration block, always together with the other targets of approximately monthly cadence. The visibility windows of twenty preliminary PSF fields are shown as  horizontal red bars within the pink EWS boxes, while the vertical red bars show the actual scheduled time of the PSF observations (the corresponding Sun longitude for each of the fields is indicated by the numbers 1 to 20 on the axes at the edges of the Fig.~\ref{Fig:full_schedule}).


\begin{table}   [hbt]
\caption{Time budget for calibration, EDFs and EAFs observations.} 
	\centering           
	\scalebox{0.9}{  \renewcommand{\arraystretch}{1.5} 
		\begin{tabular}{| c | c | c | c |}        
			\hline 
{\bf Type} & {\bf Name} & {\bf Time}& {\bf Total}\\
& & [days] & [days] \\
\hline
 & VIS PSF calibration & 38  & \\
Instrument & VIS nonlinearity & 28 & \\ 
 &	NISP wavelength dispersion & 8 & \\ 
 &	NISP nonlinearity &	32 &	106 \\ 
 \hline
EAFs &	The six fields &	34 &	 \\ 
& Self-calibration & 54 & 88 \\ 
\hline
 &  EDF-N & 35 & \\
EDFs & EDF-S & 89 & \\
& EDF-F & 61 & \\
& CPC &  48 &	233 \\
\hline
\multicolumn{3}{c}{} & \multicolumn{1}{|c|}{427} \\
\cline{4-4}
		\end{tabular}
	} 
	\label{tab:time_calibs}  
\end{table}

Table~\ref{tab:time_calibs} summarises the time allocated to make all calibration, EDF and EAF observations, which is 427 days. Note that the time allocated for the self-cal field is here included in the EAFs budget and not in the instrument calibrations, since its repeated observations will make it the deepest of the \Euclid fields.
Also note that due to the synergy between CPC and EDFs, 36 of the 48 days needed for CPC calibrations also contribute for the completion of the Euclid Deep Survey. Details are given in [Sc21].

The breakdown of the time-allocation by observing type is:
\begin{itemize}
\item[\textbullet] instrument calibrations, 25$\%$;
\item[\textbullet] auxiliary fields, $21\%$;
\item[\textbullet] deep fields, $54\%$.
\end{itemize}

\section{Computing the Euclid Wide Survey}\label{sec:wide}

The computation of the EWS is a complex optimisation problem, for which we have developed the scheduling tool \ECTile. After the preliminary \stageone described in Sect.~\ref{sec:first_stage} that determines the schedule of the various calibrations, EDFs, EAFs, and polar caps targets, \ECTile proceeds with the computation of the EWS in its \stagetwo. Before turning to a detailed description of \stagetwo of \ECTile from Sect.~\ref{sec:ectile_intro} onwards, we give a brief description of pre-\ECTile explorations.

\subsection{Early explorations}\label{sec:early_wide}

The derivation of the optimal survey is a complex process, and  \ECTile is one of many possible solutions. It is, however, worth stressing that it is the outcome of a lengthy process, in which  alternatives have been explored, but ultimately rejected. For instance, the first solutions of the EWS were delivered by industry to demonstrate the feasibility of the survey, but ignored some important additional considerations, such as observing areas with low zodiacal background first (see Sect.~\ref{sec:zodiacal}). Other solutions \citep {Amiaux12} were produced using ESA's Euclid Sky Survey Planning Tool (ESSPT) \citep{Pedro18}, which allows the user to manually place patches on a sky map and fill them with \Euclid FoVs.
In this section we provide a brief overview of the prior investigations that led to \ECTile as it is today.

\begin{figure}[hbt]
	\begin{centering}
		\resizebox{\onecolspan}{!}
		{\includegraphics{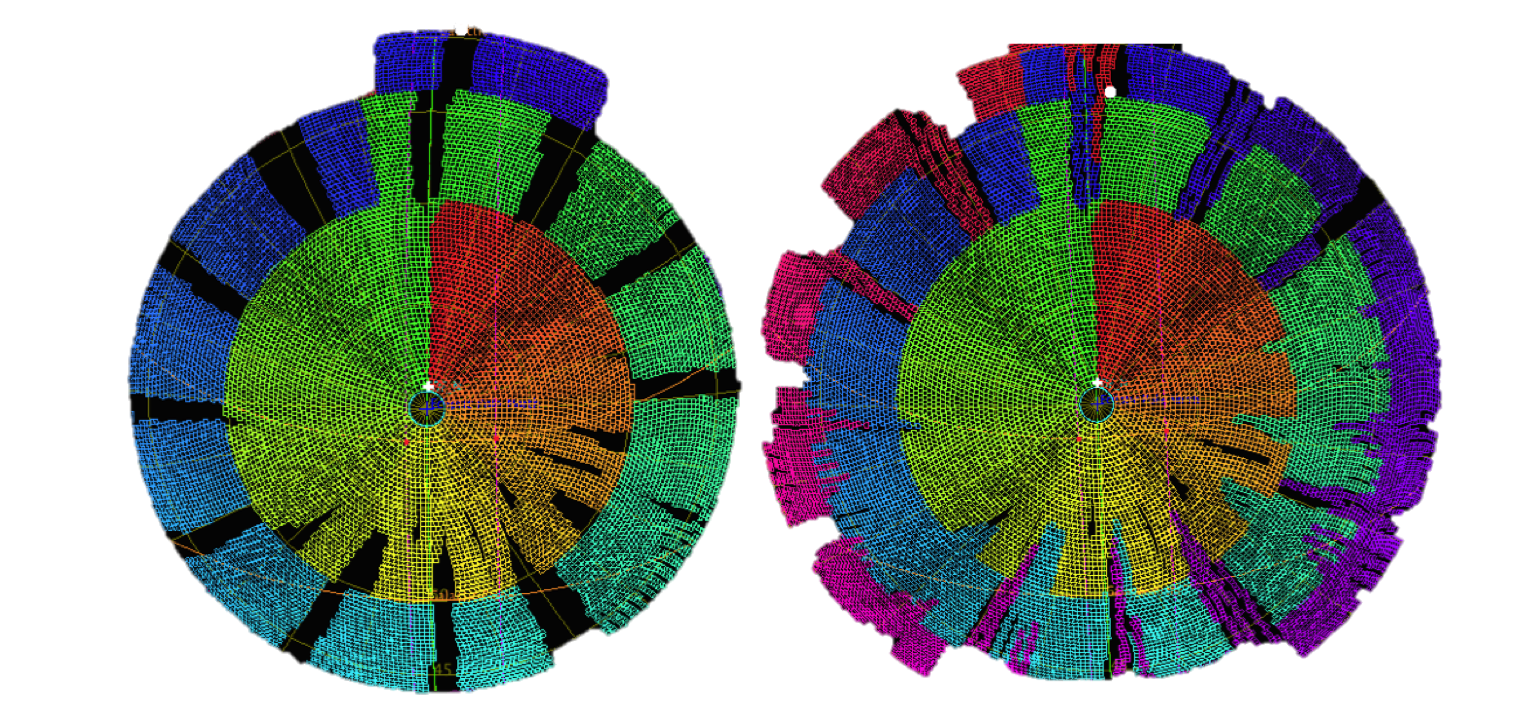}}
		\caption{An example of scheduling the EWS 
		 with a rigid tiling and requiring |AA| $< 1\degr$, covering parts of the northern ecliptic hemisphere. Different colours indicate different observing epochs. \emph{Left panel}:  The empty dark regions are caused by interruptions of the EWS.  \emph{Right panel}: The empty regions are only partially recovered at a later stage, and new gaps are formed at lower latitudes.}
		\label{Fig:maketess}
	\end{centering}
\end{figure}

To cover a maximum area with minimum overlap between single observations, a pre-determined tiling is almost unavoidable.
The early approaches \citep[see][]{Tereno15} therefore defined rigid FoVs placed parallel to the ecliptic meridians. The fields were scheduled starting at high latitudes and moving up and down along ecliptic meridians, observing in transit, that is with ${\rm SAA}=\ang{90;;}$ (and ${\rm AA}=\ang{0;;}$) when the local meridian coincides with the \Ysc direction defined in Sect.~\ref{sec:sc_overview} (see also Sect.~\ref{sec:angles}). The extent to which we move across latitudes before moving to the next longitude, defines a latitude band, to be observed in one year. Due to the convergence towards the poles, the number of fields per band decreases with latitude, and conversely the height of the bands increase with latitude.

   Whenever there is an interruption in the EWS schedule due to an observation of calibration or EDF targets, a corresponding gap is left in the band. After one revolution in the orbit, a latitude band is finished and the scheduling of the next band, on a lower latitude, starts. The gaps can be recovered in the following year when in transit again. For this, the height of the next latitude band needs to be smaller, in order to create a time buffer that allows one to cover the gaps while not creating new gaps in the lower band. This way, the missing area can be observed slightly off-transit, tilting the telescope arouns \Xsc. The tilt must be accompanied  by a  rotation around the \Zsc axis (hence changing AA) to compensate and keep the field aligned with the tiling. However, given the very stringent constraints on AA ($|{\rm AA}|< \ang{1;;}$ at the time of the early explorations) the time buffers are necessarily small, and it is only possible to partially cover the gaps.
This process is illustrated in Fig.~\ref{Fig:maketess}. It is clear that gaps in the survey are unavoidable, and attempts to fill them only leads to more gaps elsewhere. It was thus concluded that the use of a fixed tesselation and fixed latitude bands was not viable.

The AA limit was eventually relaxed to $3\degr$ at the Preliminary Design Review, (PDR) and later to $6\degr$ at the Critical Design Review (CDR), being currently fixed at $5\degr$. This makes the use of a fixed tessellation a viable approach, enabling an efficient coverage of the sky. However, the use of fixed latitude bands remains not viable and more complex strategies needed to be developed, as described in the next sections.

\subsection{Introducing \ECTile's \stagetwo} \label{sec:ectile_intro}

\begin{figure}[!hbt]
\centering
\includegraphics[scale=.75]{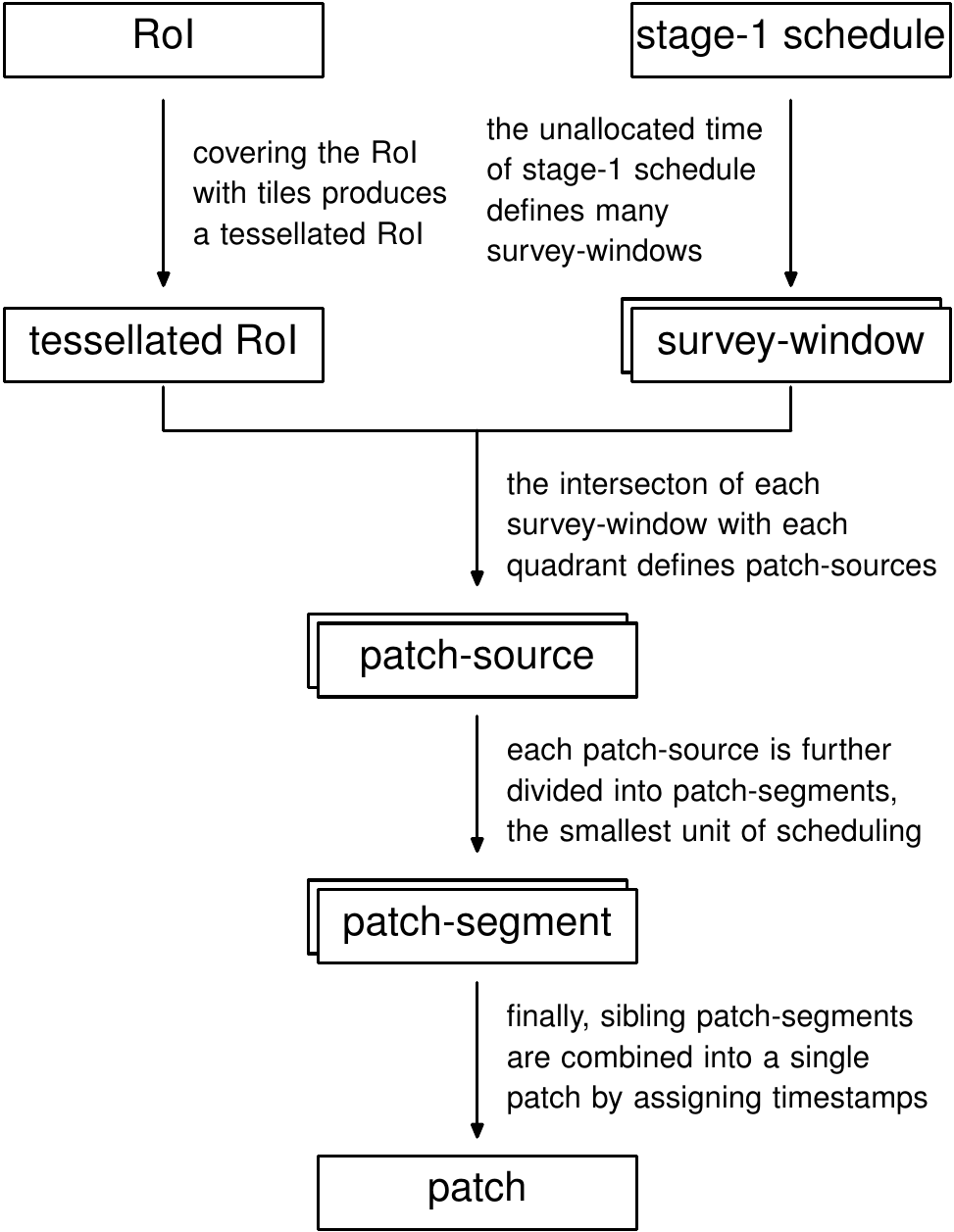}
	\caption{Steps and concepts required for the computation of the EWS, linking the RoI and stage-1 schedule to patches (defined later).}
\label{Fig:tess_to_patch}
\end{figure}

The computation of the EWS revolves around the concept of a patch, which may be loosely defined as a compact set of fields with a valid observation sequence. This is the basic building block for this computation. This section describes the steps that precede the computation of a patch, summarised in Fig.~\ref{Fig:tess_to_patch}. The first step is the computation of the tessellated RoI, a set of tiles covering the designed RoI. In parallel, survey-windows are computed from the stage-1 schedule, defining the intervals of time left to EWS observations. These two elements are then combined into patch-sources, compact sub-sets of each quadrant of the tessellated RoI within reach of a given survey-window. A patch-source is, in general, further divided into one or more patch-segments, which are sets of tiles guaranteed to be not only compact but also having their contour shaped like a lat-long rectangle (an essential property for the computation that follows it). Patch-segments from the same patch-source are then transformed into a patch by establishing upon them an ordered sequence of observations. 

\subsection{Major constraints for the EWS}\label{sec:ECTile}

In this section we present the main inputs and associated constraints for the EWS optimisation algorithm, and briefly review their impact on the current solution.

\subsubsection{Constraints due to overlap}\label{sec:1-overlap}

\begin{figure}[bt]
\centering
\includegraphics{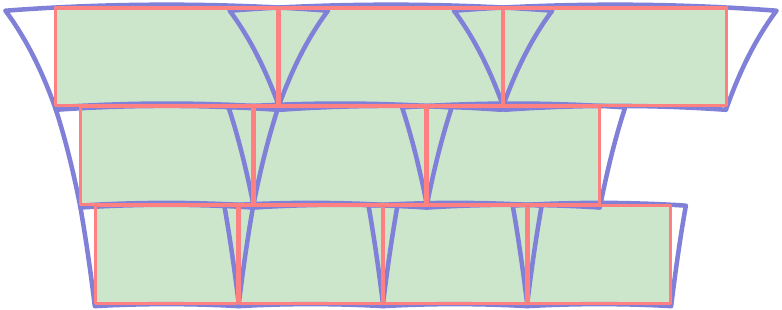}
\caption{Example of a few rows of a tile driven tessellation drawn in a cylindrical projection. Individual tiles are indicated by the red rectangles, whereas the (distorted) footprints of the FoV for non-zero latitudes are shown in blue (sizes are exaggerated for clarity).
The number of tiles per row varies with latitude.}
\label{Fig:example_tessellation}
\end{figure}

We define a tile as the largest rectangle in latitude--longitude that is completely contained in a single FoV (see Fig.~\ref{Fig:example_tessellation}). The survey area must then be observed through geometrically contiguous tiles, with an overlap of boundaries between adjacent tiles of a $0.5$\% wide strip (1\% overlap, overall), to cope with the non-null pointing error.  The goal of this requirement is to enable efficient coverage of the sky, whilst ensuring a minimum overlap between adjacent fields. This can be achieved with a tessellation of tiles laid out along parallels of latitude, with adjacent tiles on the same row and tiles between adjacent rows overlapping by $1\%$. Due to the convergence towards the poles, the number of tiles per row decreases with latitude, as shown in Fig.~\ref{Fig:example_tessellation}. Given the geometric shape of the FoV, a rectangle on a sphere, the overlap between FoVs also increases with latitude.

\subsubsection{Constraints due to SAA and AA}\label{sec:2-angles}

\begin{figure}[tb]
\centering
\includegraphics[trim=1 0 0 0]{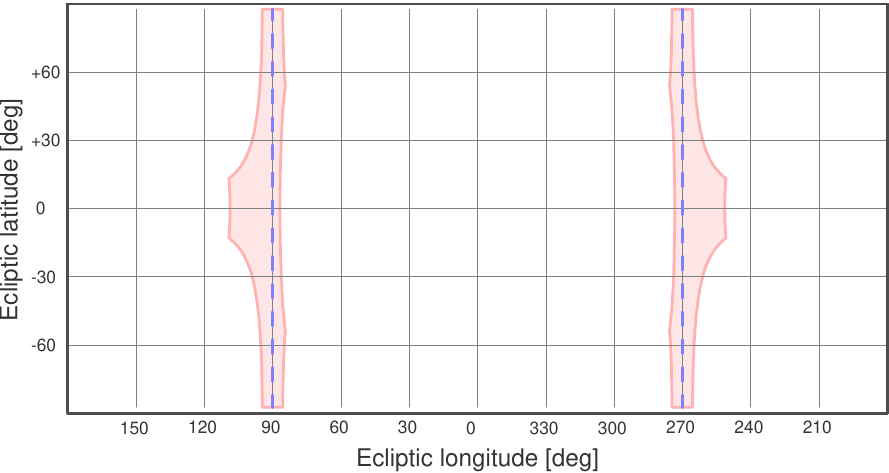} 
\caption{The subset of the tessellation observable at  given epoch, with the Sun pointing at $0\,\text{deg}$ and the transit meridians at $90\,\text{deg}$ (trailing-side) and $270\,\text{deg}$ (leading-side).  This shows that the need to observe fields aligned with meridians strongly restricts visibility (see  Fig.~\ref{Fig:instantaneous_visibility})}
\label{Fig:tess_visibility}
\end{figure}

As described in Sect.~\ref{sec:angles}, the limited range of the pointing angles implies that most of the sky must be observed at, or at least close to transit. As we motivate later, most of the EWS is observed with fields aligned with the ecliptic meridians. In general, these fields are almost never observed at transit, thus requiring a rotation around \Zsc to realign the FoV with the local meridian. 
However, the constraints on AA and SAA severely limit the extent to which a field may be observed away from transit. This is highlighted in Fig.~\ref{Fig:tess_visibility} which shows the region of the sky that is observable at a given transit. It is mostly constrained by the AA range, except at lower latitudes, where the constraint on SAA dominates.
\begin{figure}[hbt]
\centering
\includegraphics[trim=4 0 8 0]{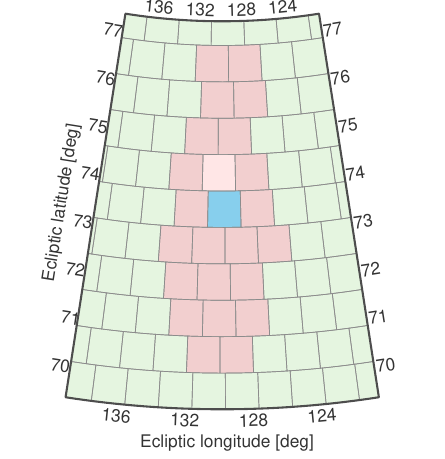}
\hfill
\includegraphics{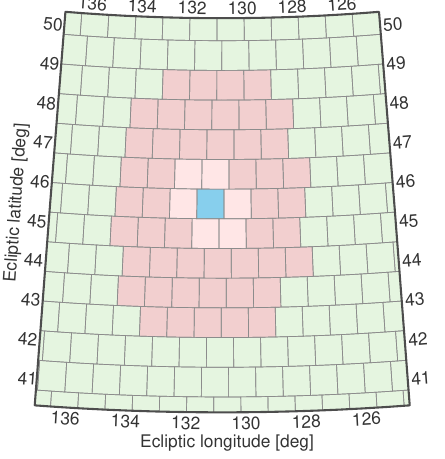}
\caption{Example of adjacent FoV slew ability at high latitude (\emph{left panel})
and low latitude (\emph{right panel}). The plots show the slew from a central tile (blue) to its 
neighbours; up to $\ddeg 1.2$ in light-red, and from $\ddeg 1.2$ to $\ddeg 3.6$ in dark-red. At high latitudes, slew movements between FoVs of a tessellation are severely restricted.}
\label{Fig:slew_ability}
\end{figure}
\subsubsection{Constraints due to slews}\label{sec:3-slews}

The limitations associated with the cost of a slew (see Sect.~\ref{sec:slews}), imply that EWS fields observed consecutively in time must also be, as much as possible, spatially adjacent to each other. In this way, large slews are mostly reserved for moving between EWS fields and calibration or Deep Fields, or between patches of the EWS.

Figure~\ref{Fig:slew_ability} shows the reach from a given field, when slewing with a small-slew within a limit of $\ddeg 1.2$ and $\ddeg 3.6$, for a field placed at two different latitudes. The example demonstrates that, at low latitude, it is possible to slew to all adjacent tiles; if considering slews up to $\ddeg 3.6$, it is possible to slew to tiles two rows away (recall that the slew can be depicted as the arc separating two different directions on the sky plus a rotation around the latter). However, at high latitude, the slew between adjacent tiles is much more limited by the size of the not small change in AA needed to keep the alignment with the local meridian of the tessellation. There, adjacent tiles on the same row are further apart, strongly limiting the field-to-field slewing which, in this case, is performed by a rotation around the \Zsc-axis. For example, given the FoV width of $\ddeg 0.701$, tiles placed at latitude of $\ddeg 79.08$ are separated exactly by $\ddeg 3.6$ of longitude. Thus, above $79\degr$, it is not possible to slew sideways between two adjacent fields (aligned with the tessellation).

\subsection{From EWS tessellation to patches}\label{sec:tessellation}

\subsubsection{The tessellated RoI}

\begin{figure}[!hbt]
\centering
\includegraphics[trim=1 0 0 0]{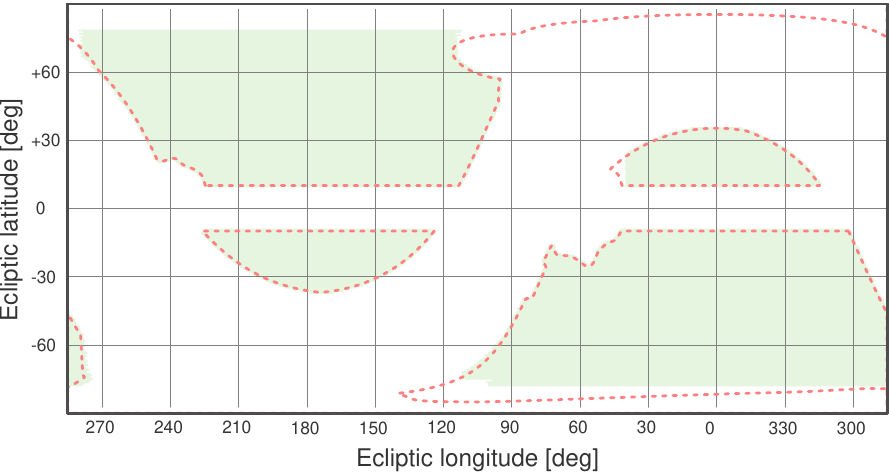}
\caption{Tessellated RoI (in green), covering the four quadrants. The dashed red line represents the RoI.}
\label{Fig:tessellated_roi}
\end{figure}

\begin{figure}[tb]
\centering
\includegraphics[trim=0 0 2 0]{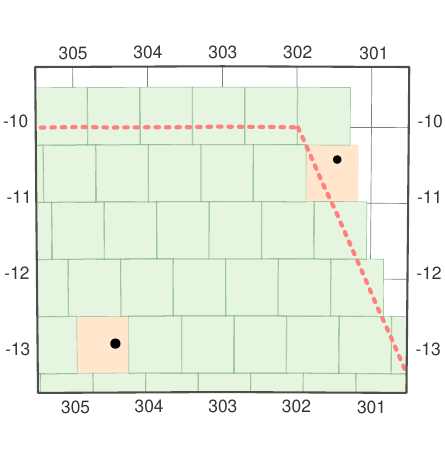} 
\hfill
\includegraphics[trim=2 0 0 0]{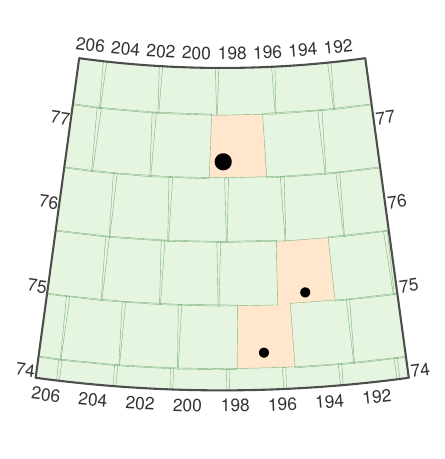}
\caption{\emph{Left panel}: Zoomed view of the tessellation for the EWS at low latitude. \emph{Right panel}: Same, but now for high latitude. The increased overlap of neighbouring tiles at high latitude is evident. Orange tiles contain a blinding star (black dots) and are skipped from observation.}
\label{Fig:ews_tessellation_example}
\end{figure}

The generation of the tessellated RoI begins with the computation of a global tessellation on the sphere, covering the sphere without polar caps from latitudes $\pm 79\degr$ towards the equator, with non-overlapping tiles aligned with the meridians. This tessellation is then filtered by selecting tiles that have at least one of their corners inside the RoI.  This represents the tessellated RoI and specifies the FoVs eligible to be observed (see Fig.~\ref{Fig:tessellated_roi}). Figure~\ref{Fig:ews_tessellation_example} shows two zoomed views, highlighting the dependence of overlap with latitude. The EWS solution schedules a large subset of the tessellated RoI, which then becomes the \Euclid `footprint'.

\subsubsection{Covering the polar caps}\label{sec:polar_caps}

As discussed in Sect.~\ref{sec:3-slews},  above $79\degr$ it is not possible to slew sideways, when traversing fields aligned with the tessellation. Indeed, at high latitudes, the centres of adjacent fields on the same row of the tessellation are separated by an eigenslew larger than $3\fdg7$. As the  amplitude of an eigenslew includes the rotations needed to align the fields and not only the angular separation between the fields centres, the part of the EWS above $+79\degr$ and below $-79\degr$, the polar caps, are not part of the global tessellation.

\begin{figure}[!hbt]
	\subfloat[\label{Fig:FIG_polar_cap_N} ] {\includegraphics[width=0.485\onecolspan]{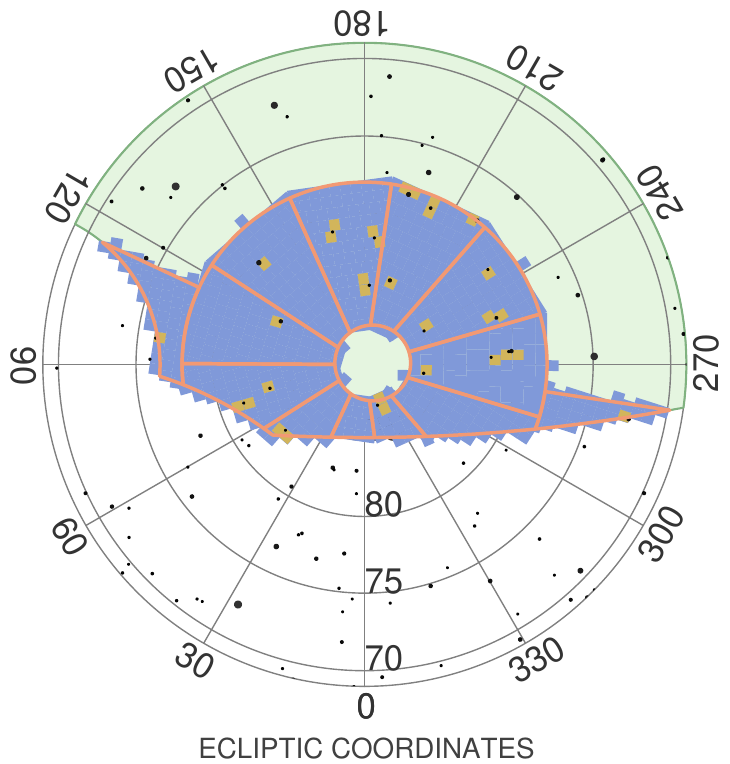} }
\hfill
\subfloat[\label{Fig:FIG_polar_cap_S} ]%
	{\includegraphics[width=0.485\onecolspan]{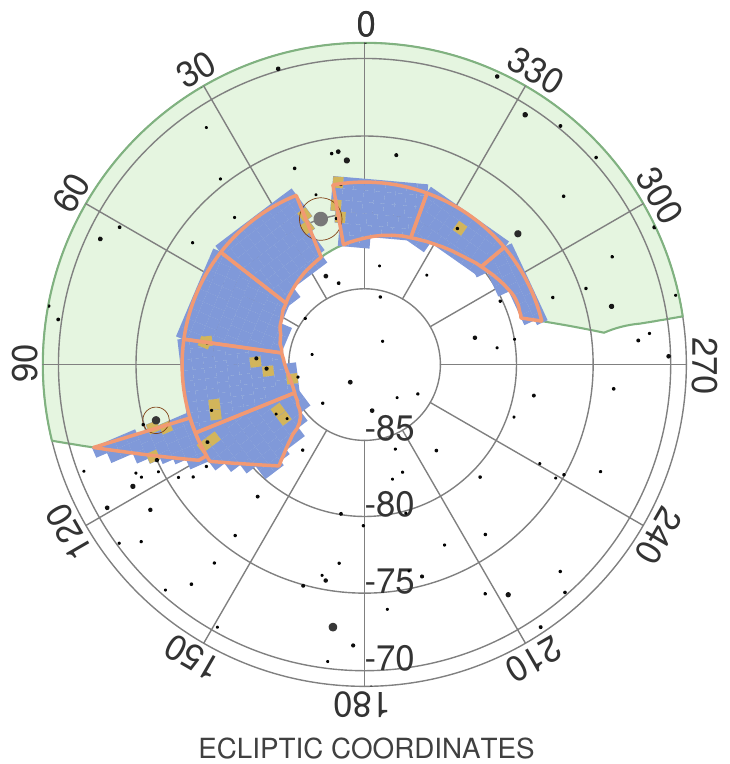} }
	\caption{Coverage of the northern and southern ecliptic polar-caps (left and right panels, respectively), with the design of polar-caps drawn in red and patches drawn in blue.	Extra patches below $79\degr$ are also added to polar-caps (two in the North and one in the South) to observe corners of the RoI that would be difficult to observe with the regular wide strategy. The presence of the Galactic plane affects both caps. In addition, the southern boundary avoids the LMC. Blinding stars are also shows; in particular the small gap in the southern cap is due to the presence of the extremely bright star R-Doradus. }
\label{Fig:polar_caps}
\end{figure}

Instead, each polar cap is covered by a fixed number of patches of fixed area, as shown in Fig.~\ref{Fig:polar_caps}. The northern cap is covered with 13 patches, while there are six patches covering the southern cap. The southern polar cap is smaller due to the presence of the LMC, which lies outside the EWS RoI.  Each patch of the polar caps is a target field scheduled during \stageone with the strategy used for observing the EDFs and calibration fields.

At high latitudes, overlap between neighbouring fields cannot be avoided, because there is longer room for a rotation of the FoV. The average FoV overlap on the polar caps is $\sim$18\%, while the average overlap on the regular wide is under 3\%. This leads to a small loss in survey efficiency.

\subsubsection{Survey-windows and patch-sources}\label{sec:patch_sources}

The next step in the computation of the EWS is to match the global tessellation with the survey-windows defined by the \stageone schedule. A `survey-window' is the span of time between consecutive calibration blocks. It defines uninterrupted time intervals available for observing the RoI. Currently, there are approximately $100$ survey-windows. As explained later, these are processed in chronological order, one at a time. But we first show how a single survey-window intersects with the RoI and how it is populated with EWS observations.

A survey-window begins at the end of the last pointing of a calibration block, and lasts until the first pointing of the following calibration block. Let $a$ and $b$ be the two pointings delimiting a survey-window, and $\Delta t$ a nominal observation time, defined as ROS time, including  the typical field-to-field slew time. The approximate number of FoVs possible to observe within a survey-window (i.e., its capacity) is then given by $n=(b-a)/\Delta t$.

\begin{figure}
\centering
\includegraphics[trim=0 0 1 0]{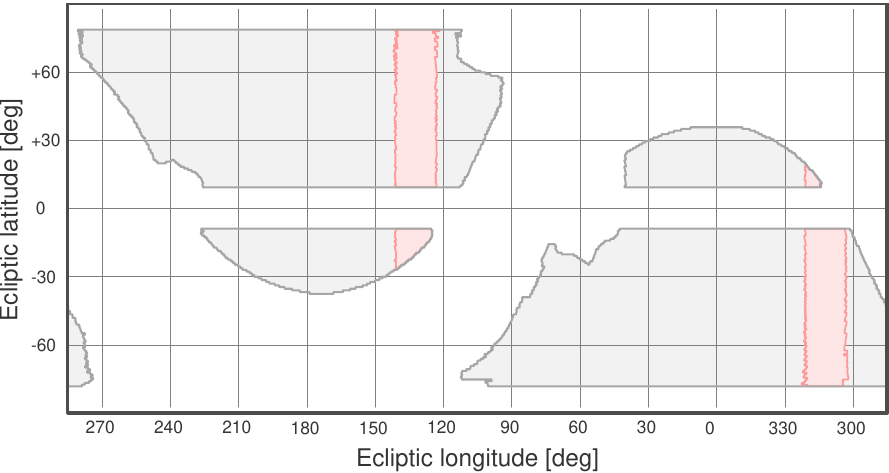}
\caption{Example of four patch-sources (in red) generated by a survey-window.  Notice how the patch-source in the northern-island intersects that quadrant only partially.}
\label{Fig:patch-source}
\end{figure}

During the slot of time defined by a survey-window, spanning from $a$ to $b$, two transit meridians scan two opposite sectors of the sphere, representing the areas within reach of the survey-window. In turn, both of these sectors intersect with two or more of the four quadrants of the RoI, identifying the eligible FoVs within reach of a given survey-window.
The intersection of each of these sectors with a single quadrant of the RoI defines a `patch-source', i.e., a contiguous subset of the tessellated RoI within reach of a survey-window (as exemplified in Fig.~\ref{Fig:patch-source}). The number of patch-sources per survey-window varies from two, intersecting only the mainlands, to four, intersecting all quadrants. In rare configurations, a survey-window intersects the same mainland twice, defining two separate patch-sources.

A patch-source provides a simplification of the wide survey time window (`survey-window'). Given the need to observe fields in sequence, ones near each other (within the slew constraints), the observations within a wide survey-window must necessarily form a contiguous compact set. As described below, these compact sets of tiles defines a patch. 

A `patch' is the building block of the survey. It constitutes a unit of observation of the wide, using some of the time budget for the wide (which may all or part of a survey-window) and covering some of the RoI. By construction, different patches do not intersect, neither in time or space. Moreover, the RoI is covered as much as possible in an orderly fashion, with patches stacking one on top of each other. In that respect, the EWS may be seen as a long sequence of patches (scheduled around the calibration blocks).

Patch-sources of the same survey-window compete for the same time, with each one (potentially) generating a patch. Hence, in general, several candidate patches are available, with the actual choice of one among the possible ones determined by several aspects. In the simplest scenario, one or more patches fill all the time available and, in that case, it suffices to choose one of them (and discard the rest). In more complex scenarios, a survey-window only intersects a quadrant partially, producing a patch-source that does not consume all the available time. In these cases, the solution is to fill the survey-window with patches from several patch-sources. One patch is selected, reducing the extent of the survey-window.  Then, the process is restarted, recomputing the patch-sources and generating a new set of patches. The process is iterated until all available time is exhausted.

\begin{figure}
\centering
\includegraphics{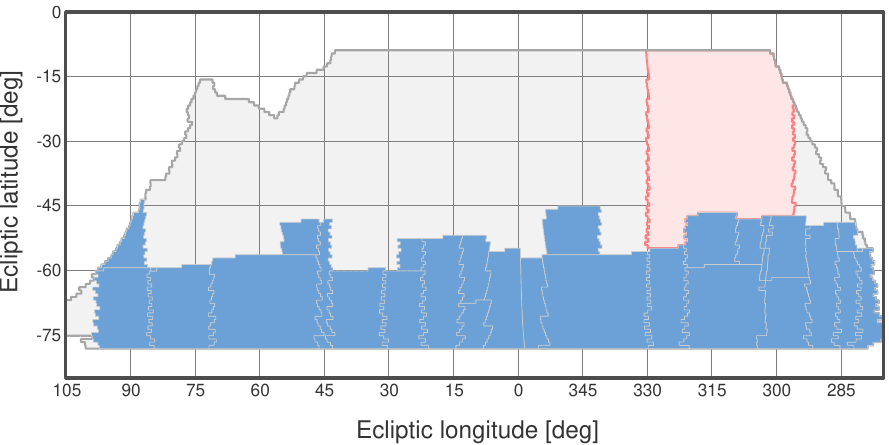}
\caption{Example of a patch-source reduced by previously scheduled patches (plot of region \protect\FancyRoman{4} only, no star skipping for clarity). Previously scheduled patches are shown in blue, with the red region depicting the patch-source of some window.  Notice how the bottom side of this quadrant acquired a lat-long shape after the stacking of some patches.}
\label{Fig:dented-source}
\end{figure}

Inevitably, due to the cyclic nature of the scanning of the sky, the selection of a patch reduces the RoI available for the generation of later patches.  This does not cause any problem and it is easily coped with by flagging observed tiles as they are scheduled, thus avoiding selecting them again in subsequent compilations of patch-sources.
However, cyclic placement of patches around the sphere creates a ``dented'' boundary of observed tiles.  Over time, this leads some of the observed regions in the RoI to acquire a boundary shaped like a polyline in latitude and longitude (see Fig.~\ref{Fig:dented-source}).

\begin{figure}[ht]
\centering
\includegraphics[scale=.58]{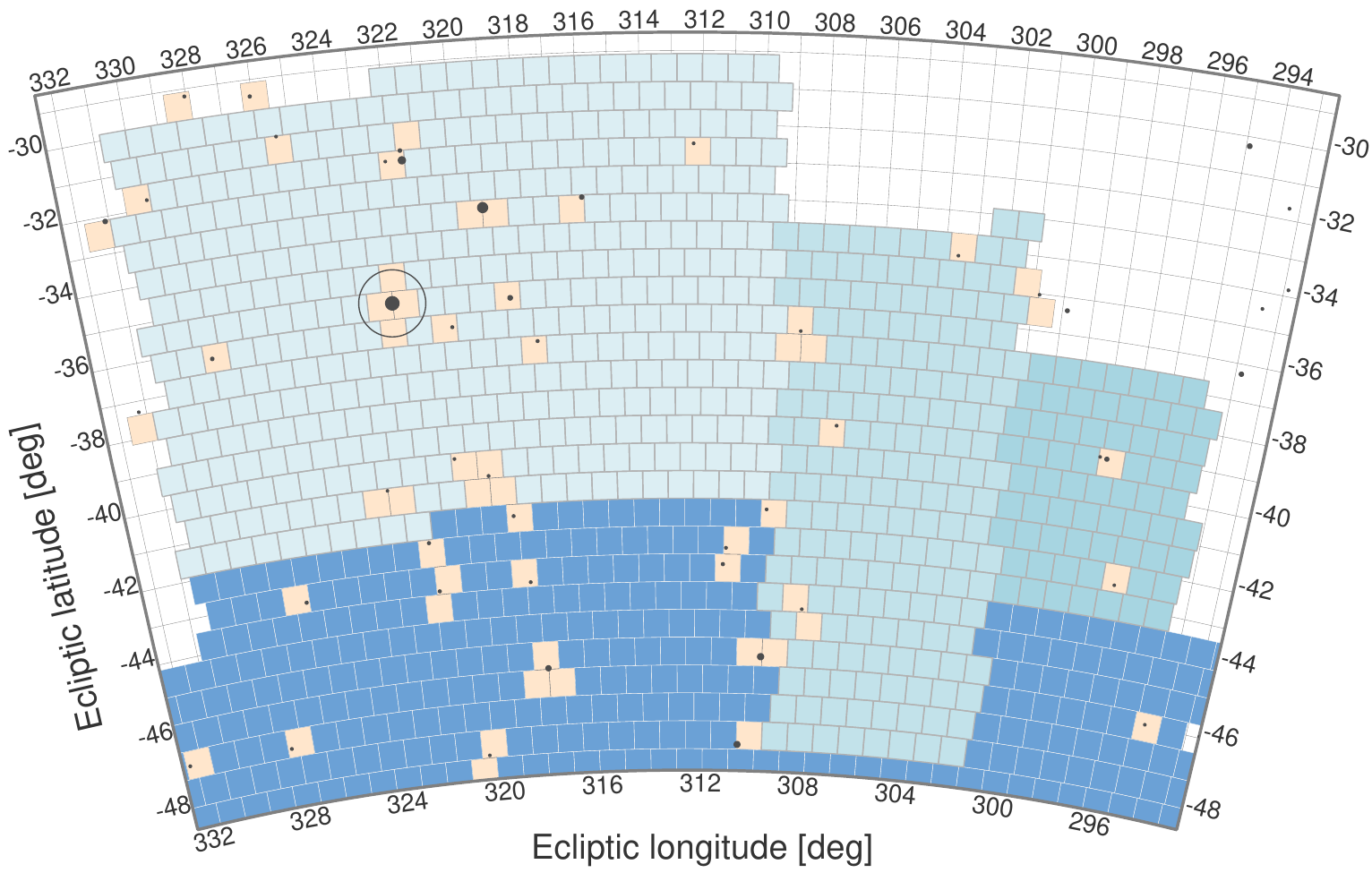}
\caption{Example of sibling patch-segments from a fragmented patch-source (of Fig.~\ref{Fig:dented-source}).  Patch-segments are displayed in shades of light blue, with previously observed patches in dark blue.  Tiles containing a blinding star are skipped for observation, here displayed in orange.  Tiles in the vicinity of an extremely bright star are also skipped (even if not containing it), as is the case of the star marked with a circle at $\sim (323, -36)$.}
\label{Fig:source_segments}
\end{figure}

\subsubsection{Patch-source partitioning}

In the previous section, the process of extracting a patch from a patch-source was simplified, for the sake of clarity.  Actually, the process is slightly more complex, requiring the definition of the concept of a patch-segment.  First, tiles are selected on the condition that tiles observed in the same time slot should also be close to each other. However, because the contour line of the unscheduled part of the RoI may become irregular, the condition on proximity might not be feasible to meet. The solution is to partition the tiles of a patch-source into patch-segments, where a patch-segment is simply a group of tiles amenable to be visited with a sequence of small-slews.

When creating a patch-segment, tiles must be selected evenly across longitude, matching the rate of fields observable per unit of time (approximately, 20 fields per day $\equiv$ per degree of longitude).  The first time a patch is extracted from a patch-source, the base of the corresponding RoI is bounded by a straight line (the side of the mainlands close to the poles), resulting in a single, possibly large, patch-segment of tiles. But, after a few iterations, the base becomes a polyline (see Fig.~\ref{Fig:dented-source}). Then, the process of selecting tiles evenly across longitude may, potentially, produce a fragmented selection, made of two or more separate patch-segments (Fig.~\ref{Fig:source_segments} shows an example of this). This fragmentation is inevitable, but it is not an obstacle. In general, a patch-source produces several patch-segments in order to fill its span of time.  In the case of multiple patch-segments, the solution is to schedule them separately, followed by a merge into a continuous single schedule.

\subsection{Scheduling patches}\label{sec:patch_filling}

This section describes the core scheduling functionality for the computation of the EWS.
We  describe two algorithms that we used. Each algorithm takes a patch-segment as input and produces an ordering of the tiles, i.e., a sequence, geometrically appropriate for observation (within the constraints of the spacecraft). We then describe how several patch-segments, properly ordered, are combined in a single schedule.

We describe first the ``look-ahead'' algorithm and later on the  ``diffusion'' algorithm that replaced the former.

\subsubsection{The look-ahead algorithm}\label{sec:look_ahead}

The look-ahead algorithm was the first successful attempt at scheduling a patch-segment of tiles, within mission constraints, allowing the generation of a compliant survey.  While superseded by the diffusion algorithm (Sect. \ref{sec:diffusion}), it gave much insight into the key factors at play, and paved the way for the design of the latest algorithm.

The look-ahead algorithm was designed around the idea that a scheduling sequence must traverse a patch-segment of tiles following a ``natural'' zig-zag scheme, monotonously across ecliptic longitudes. At its core, the algorithm traverses the patch-segment from right to left (longitude), going up and down (latitude), with minimal reversing of direction. The natural ordering is computed iteratively, moving from a given position and given direction of traversal (going-up or going-down) to the next.  If the current direction is going-up, the next tile in the sequence is the first unvisited tile of the row above found by scanning the patch-segment from right to left. Should that tile not exist, then the tile on the same row immediately to the left is selected. In the latter case, the direction is reversed from going-up to going-down, setting a flag that the top border of the patch-segment was reached. If the current direction is going-down, the choices are reversed; the next tile in the sequence is the first unvisited tile of the row below found by scanning the patch-segment from right to left or, if this does not exists, it is the tile on the same row immediately to the left. Likewise, in the later case, the direction is reversed from going-down to going-up, setting a flag that the bottom border was reached.

With this algorithm one can then define the full process. At the beginning, a starting tile is chosen (i.e. a tile on the rightmost side of the patch-segment), as well as an initial direction, and a starting timestamp. Typically, there are many tiles close to or at the same longitude as the rightmost tile; all are suitable as starting tiles. Afterwards, the algorithm computes a path to traverse the patch-segment, propagating along timestamps for the observations of each tile. 

In general, this algorithm does not cover the patch-segment completely. It may get stuck in one of three ways:
\begin{itemize}
    \item a dead-end is reached, with no unvisited tiles to jump to;
    \item an unvisited adjacent tile is identified, but it cannot be observed within the slew constraints;
    \item a non-valid observation is encountered because an observation went outside the valid range of AA or SAA when assigning timestamps.
\end{itemize}

This can be understood as follows. Due to convergence towards the ecliptic poles, patch-segments at high latitude are very asymmetric, having most of their tiles either at the top or the bottom. In this case, a sequence of observations should spend more time at latitudes with many tiles, making only occasional excursions to less populated latitudes. However, the algorithm is designed for full vertical excursions (whenever possible). Also, some patch-segments have unique geometric features; or corners odd enough to trap the single path of the traversing strategy.

The solution is to extend the algorithm in two ways: making it explore more paths and allowing changes of direction in mid-excursion.  The first should promote sequences that adapt to odd shapes, while the second should be able to cope with the asymmetry of patch-segments at high latitude. To these ends a `probing step' is implemented. Instead of blindly progressing up and down, inverting direction only at the boundaries, the algorithm first runs two probes to determine if, in the next step, the sequence should advance by going-up or going-down (i.e. whether it should continue or invert direction).

\begin{figure}
\centering
\includegraphics[scale=.65]{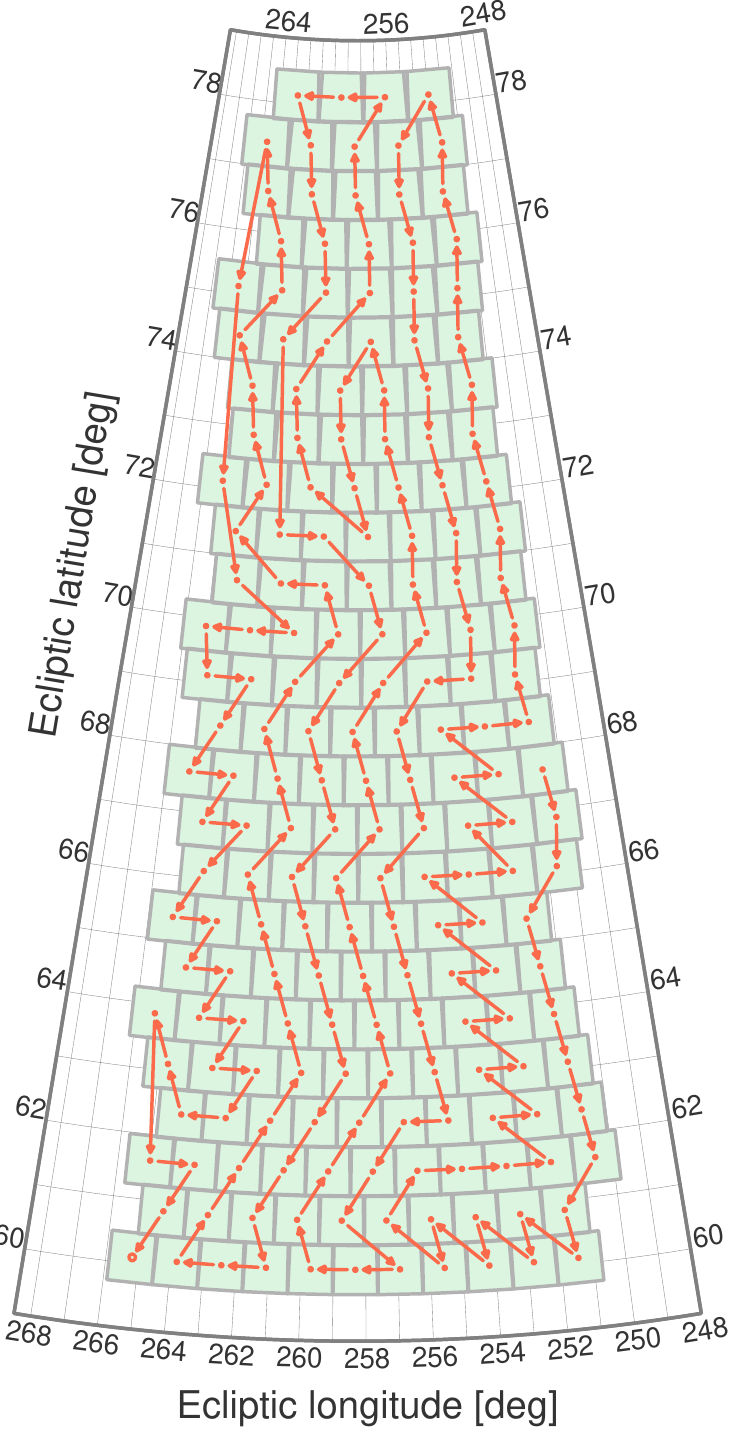}
\caption{Example output of the look-ahead algorithm.  The sequence starts at mid-height, going downwards first.  It traverses all fields running up and down, from right to left.  Notice how, sometimes, it shows an irregular behaviour, zig-zagging to the sides instead of smoothly following the expected up and down directions.}
\label{Fig:look-ahead_algorithm}
\end{figure}

The probing step is simple.  Before advancing, the algorithm first computes the natural sequence for the remaining unvisited tiles for two scenarios: first, for a natural sequence that continues going-up, and then for a natural sequence that continues going-down. The direction to take for the next step is then given by the length of the two natural sequences just computed.  If the lengths are different, it takes the direction of the longest sequence (possibly inverting direction), otherwise, it just keeps going in the same direction as before. 
The resulting sequence is obtained by applying this probing process iteratively, until all tiles are visited (returning success) or until the algorithm gets stuck (returning failure). As an optimisation, if one of the probing sequences traverses all of the remaining tiles, that sequence is taken and completes the schedule. If the above run fails, the process is repeated by trying other starting tiles (from the subset of tiles with longitude close to the rightmost 
tile).  Varying the starting point explores different configurations, greatly improving the chances of success.
Figure \ref{Fig:look-ahead_algorithm} illustrates this process.

This algorithm was used to generate the EWS from the period of time between the mission PDR to the CDR. However, following the CDR it was realised that the resulting surveys exhibited uncontrolled excursions over the full AA range (visible in Fig.~\ref{Fig:look-ahead_algorithm} and the left panel of Fig~\ref{Fig:delta_alpha}), degrading the thermal stability of the telescope and thus affecting the PSF estimation.  This was particularly acute for patch-segments at high latitude. Part of this failure is due to the restrictiveness of small slews, limited (at the time this was developed) to a maximum of $1\fdg2$. The slew range was then relaxed, allowing now for a small number of slews up to \ang{3.6;;}. However, the look-ahead algorithm is intrinsically limited by its simplicity and lack of flexibility (low number of parameters), making it difficult to accommodate new constraints such as skipping tiles with bright stars. These disadvantages prompted the development of a more capable algorithm, as explained in the following section.

\begin{figure*}[!tb]
\centering                                                                                           
\subfloat[\label{Fig:diffusion_algorithm_a}]
    {\includegraphics[scale=1.2]{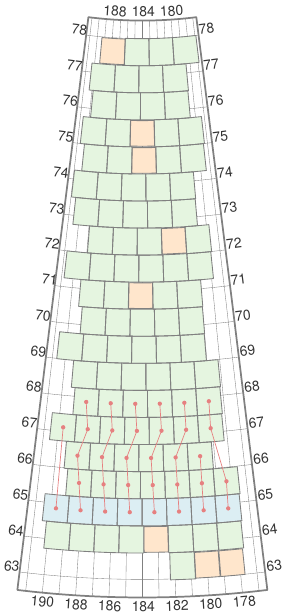}}
\hfill
\subfloat[\label{Fig:diffusion_algorithm_b}]
    {\includegraphics[scale=1.2]{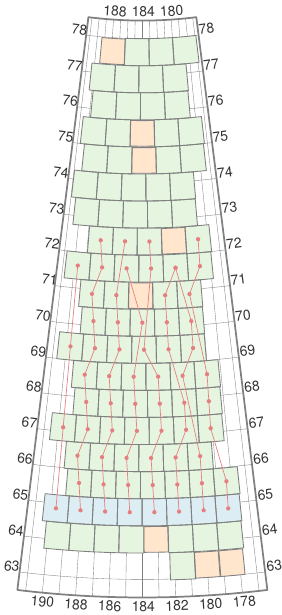}}
\hfill
\subfloat[\label{Fig:diffusion_algorithm_c}]
    {\includegraphics[scale=1.2]{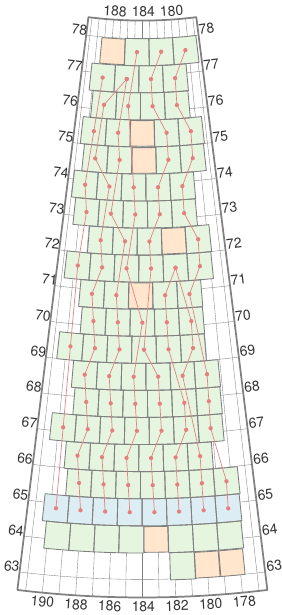}}
\hfill
\subfloat[\label{Fig:diffusion_algorithm_d}]
    {\includegraphics[scale=1.2]{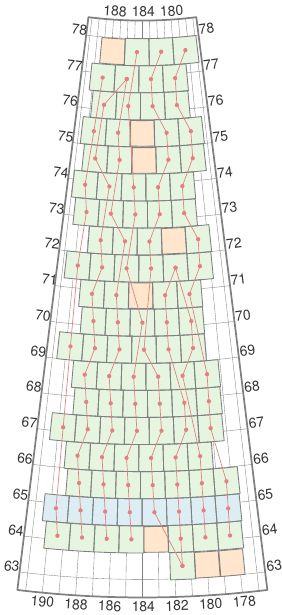}}
\hfill
\subfloat[\label{Fig:diffusion_algorithm_e}]
    {\includegraphics[scale=1.2]{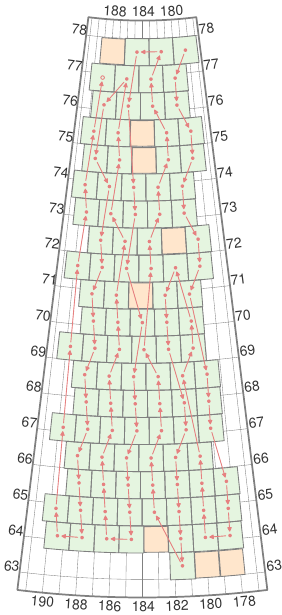}}
\caption{Example run of the diffusion algorithm.  The green and blue squares represent the fields in a patch.  The blue ones mark the base row.  The red line-segments depict the progress of the algorithm as it traverses all the fields.  Panel (a) shows diffusion after fours iterations, extending the threads upwards.  Remaining panels illustrate several points of the diffusion process: at panel (b) two threads skip across the hole left by a bright star, while two other threads merge, at panel (c) two threads merge while the remaining reach the top, at panel (d) the threads are extended two rows downwards, panel (e) shows the final sequence, after the linking of adjacent threads.  The plots are in elliptic coordinates.}
\label{Fig:diffusion_algorithm}
\end{figure*}

\subsubsection{The diffusion algorithm}\label{sec:diffusion}

Following   analysis of the full satellite structural thermal optical performance (STOP) made by \Euclid's industrial prime contractor, we analysed the impact of the spacecraft attitude on the PSF stability. It was found that that, in addition to the applicable limitations on SAA and AA, a further minimisation of the field to field variation of these angles was desirable.

The new diffusion algorithm, which is presented here, achieves this goal, whilst also facilitating the avoidance of bright stars. To minimise the angle variations, the patch-segment must be traversed in columns of alternating directions. This strategy avoids lateral slews on the same row that would cause spikes in \dalpha and \dSAA, detrimental to PSF stability. Moreover, since the slew reach is longer in the vertical direction (see Sect.~\ref{sec:angles}), moving mostly in the vertical direction increases the chances of successfully skipping across rows. This feature allows the algorithm to skip adjacent tiles (if already observed), and jump over holes created by bright stars.

To illustrate the diffusion algorithm, we consider an input patch-segment in the northern hemisphere close to the polar cap. The proximity to the pole highlights the patch-segment convergence, which is relevant in this context. The same strategy is easily adapted to other latitudes (with less convergence) and the southern hemisphere (by swapping up and down). The algorithm is divided in two steps. The first step computes parts of the final sequence, called `threads', which connects tiles along columns. The second step obtains the final sequence by tying adjacent threads together.

The computation of the threads begins by selecting the widest row of the patch-segment (not necessarily the one at the bottom), assigning a thread to each of its tiles. Then, the threads are computed in parallel, sequentially joining tiles of the current row with tiles of the row immediately above, until the top row is reached. The algorithm then returns to the starting point (the widest row), extending the threads downwards and thus completing the threads.

The double step approach is needed because, in general, the widest row is not at the bottom or at the top. Typically, a patch is bounded by a lat-long rectangle, like the examples of Fig.~\ref{Fig:look-ahead_algorithm} and Fig.~\ref{Fig:diffusion_algorithm}.  However, patches alongside the RoI boundary may get asymmetrical shapes, acquiring some of the shape of the adjoining boundary.  In those cases, the widest row is usually some row in the middle.

The computation of threads is based on two parameters of the rows: row length and row capacity. The length of a row $i$, $\rlen(i)$, is the number of tiles of that row.  The capacity of a row $i$, $\rcap(i)$, is the maximum length of a row or of any row above it, defining the minimum number of threads that must cross a given row.
If $\rcap(i) = \rlen(i)$, then all threads visit all tiles in row $i$.
If $\rcap(i) > \rlen(i)$, then $\rcap(i) - \rlen(i)$ threads do not touch row $i$, and must skip it, but participate in some row above it.  Due to the convergence towards the poles, the patch-segments in the northern hemisphere funnel on the upper part. Similarly, the rows capacity reduces with increase in latitude.

Let $k$ be the row (or one of the rows) of largest length, being also the starting row for computing the threads. Each thread is initialised with exactly one tile, of row $k$. Let $\rthreads$ be the initial number of threads. The process then begins by extending the threads, now at row $k$, to the tiles of row $k+1$, guided by the following criteria that depend on $\rlen(k+1)$ and $\rcap(k+1)$:
\begin{enumerate}
\item If $\rlen(k+1) = \rthreads$, then we have the simplest case with a one to one correspondence.
It suffices to extend each thread to row $k+1$ by making an ordered assignment, from right to left (ensuring the threads do not cross);

\item If $\rcap(k+1) < \rthreads \;\land\; \rcap(k+1) > \rthreads - 2$, then it is not possible to extend all threads; the number of tiles of the row above is smaller than the number of threads. In this case, we extend the threads closely aligned to some tile of the row above it. This is performed by trying all combinations of ordered (non-crossing) assignments between a subset of $\rlen(k+1)$ threads and all of the $\rlen(k+1)$ tiles of row $k+1$. For each tried combination that was accepted, we compute the cumulative variation in longitude (``verticality'') of the thread. The combination with the lowest variation in longitude is selected, extending $\rlen(k+1)$ threads (and leaving the rest unchanged),

\item If $\rcap(k+2) < \rthreads - 2$,  it is also not possible to extend all the threads but, unlike the previous case, it is possible to eliminate a pair of threads (leaving $\rthreads-2$ threads). Actually, it is necessary to eliminate threads to avoid the risk of having threads stalled (i.e., not reaching the top of the patch). A pair of threads may be eliminated if they are adjacent in the list of threads.  To eliminate them, it suffices to merge the two threads together, short-circuiting their paths, by connecting the two adjacent threads to the same tile (of row $k+1$, creating an inverted ``v''). This case is slightly more complex than the previous one. Now, it needs to try all combinations of ordered (non-crossing) assignments between a subset of $\rlen(k+1) + 1$ threads and all of the $\rlen(k+1)$ tiles of row $k+1$, considering that pairs of consecutive threads extend to the same tile (merging those threads). Like before, all combinations are scored against verticality, selecting the one with lowest cumulative variation in longitude. It extends $\rlen(k+1) - 2$ threads, leaving $\rthreads - 2$ active threads, and merges two threads (ending their progress).

\end{enumerate}

\begin{figure*}[hbt]
\begin{centering}
\includegraphics[scale=0.62,trim=8 0 8 0]{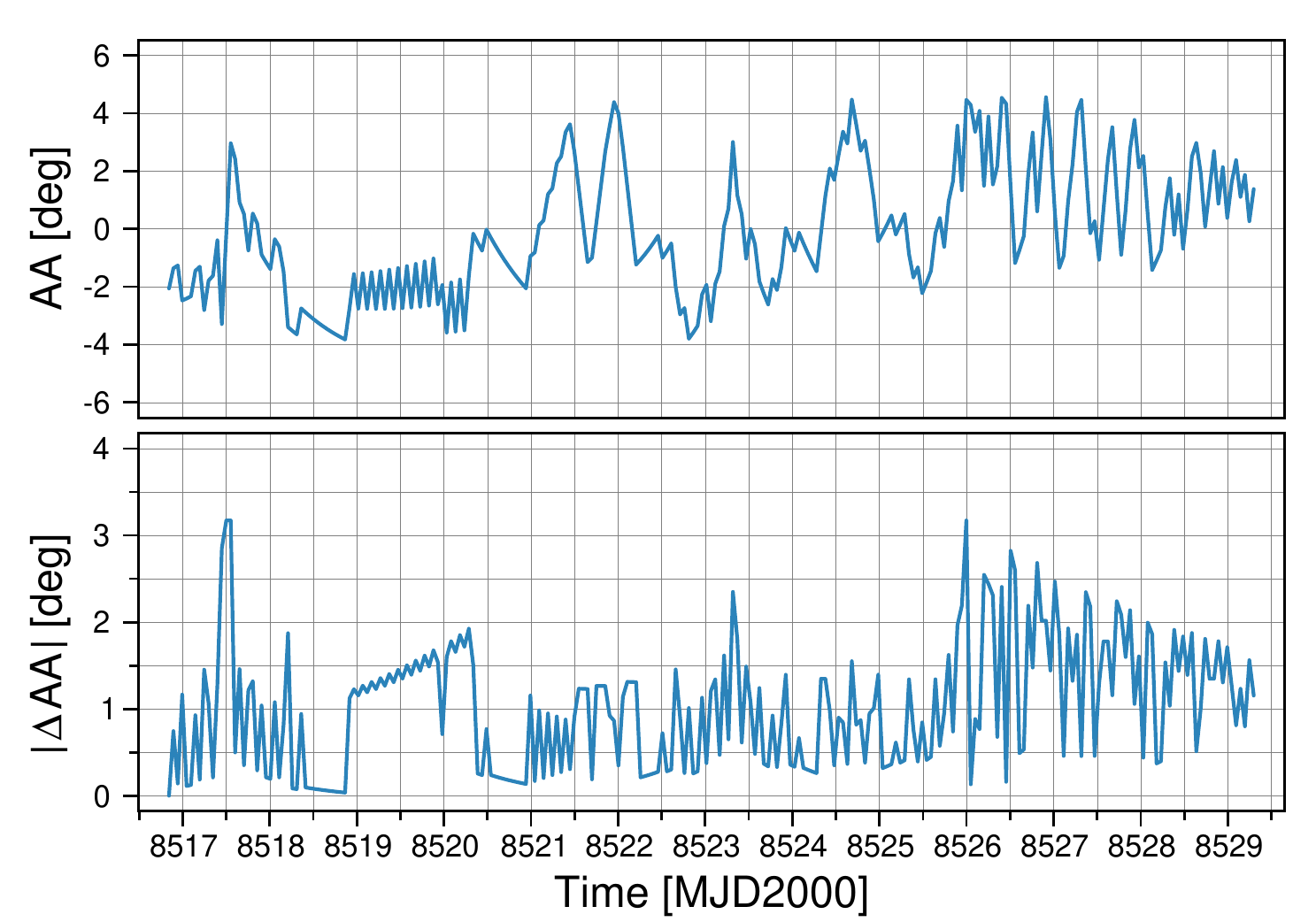}
\hfill
\includegraphics[scale=0.62,trim=0 0 8 0]{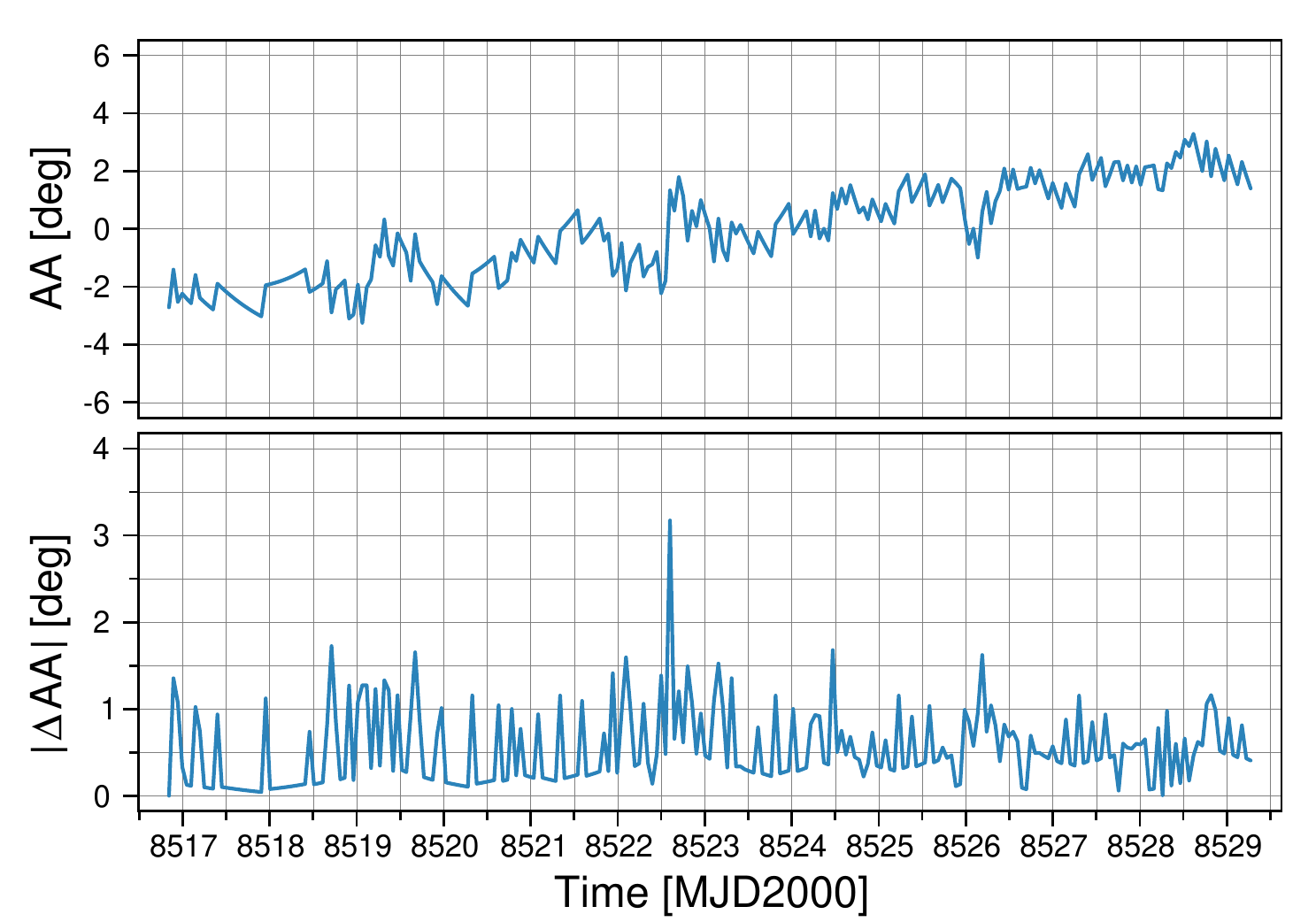}	
	\caption{Variation of AA when observing the same EWS patch with the two different algorithms: look-ahead (left panel) and  diffusion (right panel). The latter has comparatively a more monotonic progression with much smaller field to field AA variations and spikes. The bottom panels are the time derivative of the top panels.}
\label{Fig:delta_alpha}
\end{centering}
\end{figure*}

This process is illustrated in Fig.~\ref{Fig:diffusion_algorithm}. Figure~\ref{Fig:diffusion_algorithm_a} shows the first step, extending threads from the initial row (in blue) to the row immediately above. Given the difference in row length, one thread is held up. After a few iterations, Fig.~\ref{Fig:diffusion_algorithm_b} shows a case where a thread on the left-side jumps over four threads (approximately at longitude \ang{210;;}, latitude 60-\ang{68;;}), and a thread in the middle (approximately at longitude \ang{209;;}, latitude 66-\ang{68;;}) is kept straightly vertical by skipping a row. This is an example of the capability of the diffusion process to adjust to a varying row length, by ``squeezing'' more threads than the length of each individual row, promoting straighter threads. After a few more iterations, Fig.~\ref{Fig:diffusion_algorithm_c} shows two mergers of two neighbouring threads (rows at latitudes \ang{70;;} and \ang{71;;}). This reduces the number of threads from seven to five (which is enough to cover the tiles of the rows above). This mechanism ensures a monotonous decrease of the number of threads, keeping it close to the row's capacity. Figure~\ref{Fig:diffusion_algorithm_d} shows the threads fully extended upwards (with two pair of threads ended being merged in the process). In this case, there is no need to extend the threads downwards also, since the starting row is also the bottom row. Lastly,  Fig.~\ref{Fig:diffusion_algorithm_e} shows the final sequence, obtained after connecting adjacent threads (arrows show the temporal sequence of the covering).

The net effect of the above strategy is to grow threads upwards, striving to be as vertical as possible, and merging adjacent threads when necessary, to cope with the reduction of rows capacity (with latitude). In a way, this growth resembles a diffusion process, hence its name. It should be noticed that the number of threads crossing each row may be larger than the actual number of tiles at that row. This is desirable. It is the mechanism that enables the threads to accommodate to the slight irregularities of the patch-segment, while still be vertically aligned. In addition, this same mechanism allows the threads to skip tiles containing blinding stars (to be implemented in a future release).

The input patch-segment is bounded, by construction, to a lat-long rectangle. Therefore, in general, all the threads traverse the full extent of the patch-segment, from top to bottom (the exception being the cases truncated by the RoI boundary). So, the thread endpoints (top and bottom) are expected to be close to each other meridian-wise. The final step of the computation is to pairwise connect the threads from right to left. This last step creates larger moves in longitude, detrimental for PSF stability. However, these moves are limited to the number of threads, which is much lower than the length of the patch-segment). 

The diffusion algorithm is by far the most expensive part of the EWS computation, with the combinatorial exploration taking most of its cost. However, it has so far proven to be stable and robust, solving a large number of diverse patch-segment configurations while optimising thermal stability, thus yielding a much more stable PSF. Figure~\ref{Fig:delta_alpha} shows a comparison of the resulting AA field to field variation in the same patch, when scheduling it with the the look-ahead algorithm  versus diffusion algorithm. The time behaviour of AA in the latter case is much smoother and with fewer spikes than the previous results. This improvement translates into a better thermal stability and an overall decrease in time variations of the PSF, which therefore can be better modelled.

\subsubsection{Linking patch-segments into a patch}

\begin{figure}[tb]
\centering
\includegraphics[scale=.58]{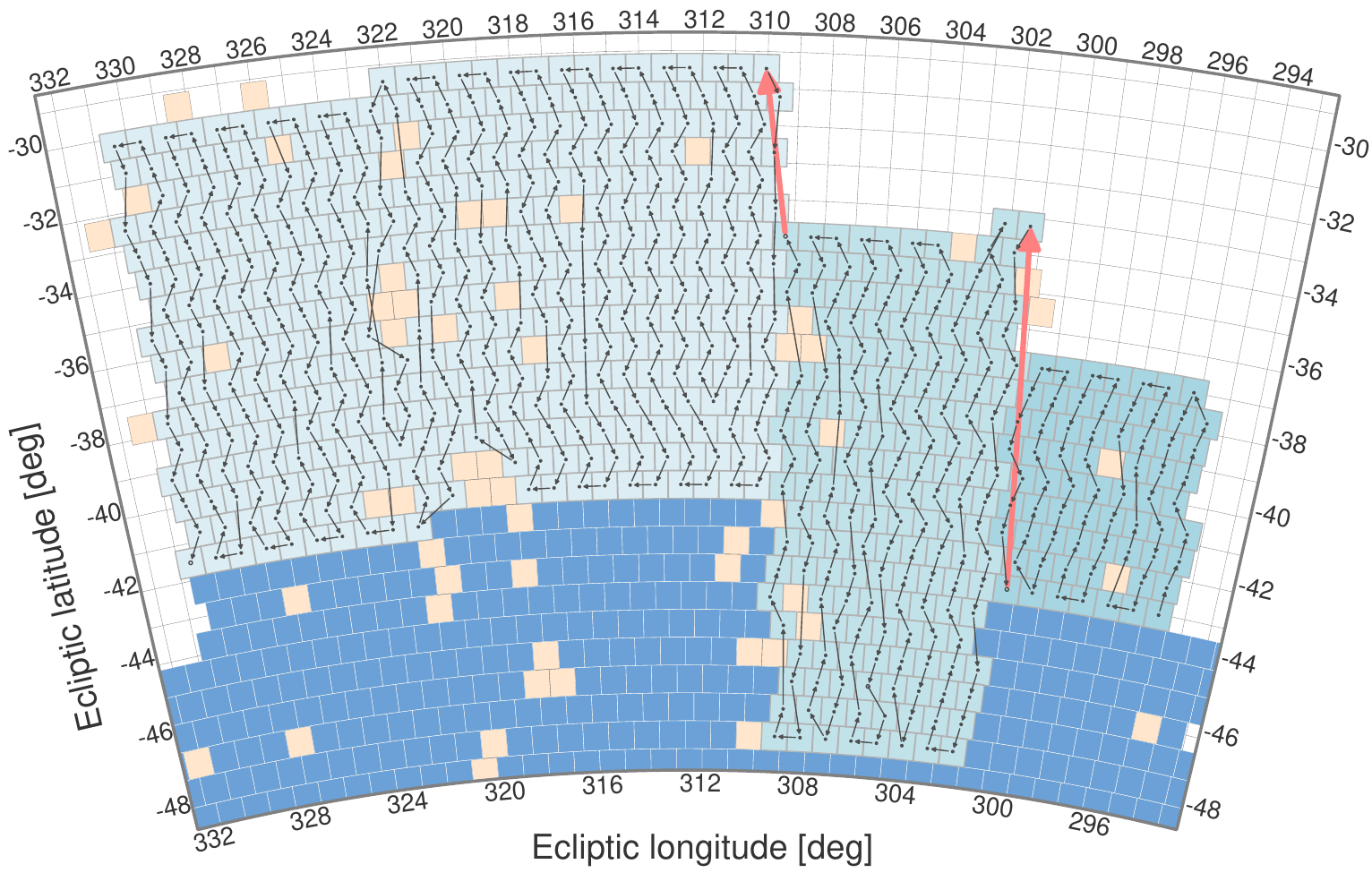}
\caption{Example of a three sibling patch-segments (cf. Fig.~\ref{Fig:source_segments}).  Patch-segments are displayed in alternating colours.  Previously observed tiles in dark blue.  Red arrows represent large-slews (moving from patch-segment to patch-segment).}
\label{Fig:multi_segment_patch}
\end{figure}

The diffusion algorithm proposes a sequence of observations for a pattern of tiles, which is a segment of a patch. This is the building block for computing a patch. First, the diffusion algorithm is applied to all patch-segments of a patch-source. Then, if all are successful, the resulting patch-segments are linked together in a single observation sequence, by assigning timestamps for observation, generating a single sequence that covers the time slot of a survey-window; i.e., a patch (see Fig.~\ref{Fig:multi_segment_patch}).

Timestamps are assigned adding observation and slew times along the order of the sequence, checking compliance with constraints. This is performed first forward in time, starting at the pointing (quaternion and timestamp) that defines the start of the survey-window. Next, the timestamp of the first observation is computed by adding the slew time from the start pointing. Then, the timestamp of the second time observation is computed by adding the ROS observation time (a fixed value) plus the slew time from the previous observation (a variable value). The process is repeated throughout the sequence, assigning timestamps sequentially. In this process, slew-times are computed according to the slew-time estimator. By construction, slews within a patch-segment are expected to be small slews. In contrast, jumps between patch-segments are considered to be large slews, adding to the large-slew budget. However, this number of large-slews is relatively small. Typically, survey-windows are filled with patches generated from one or two patch-sources, with each patch being split in a low number of patch-segments, at most.  Hence, the initial ${\sim}100$ survey-windows produce no more than a few hundreds of patch-segments in the end.

As explained above, the algorithm first attempts to assign timestamps from the beginning of the survey-window forward. If this succeeds, it creates a patch flushed backward in time, leaving some idle time at the end of the window. This is because in general, an integer number of observations with varying slew-times does not fit perfectly into the slot of time previously defined by a survey-window. If this succeeds, this is the preferable solution. If it fails, then a reverse assignment is attempted, assigning timestamps from the end of the survey-window backward (and reversing the computations of slew-time). If this succeeds, it may leave some idle-time at the beginning of the window.

Usually, there is a sufficient number of tiles distributed along the range of longitudes (covered by a patch-source) to generate a patch covering all the time slot of a survey-window. However, in some cases, such as when a patch-source intersects only a corner of a quadrant, it is not possible to have a path extending the full width of the respective survey-window. Trying to flush a patch to both end-sides of a survey-window enables patching those odd cases, promoting also the generation of patches that adapt to the boundary of the RoI.

In the process of assigning timestamps, an observation of a field with a given timestamp may fail to comply with the constraints of SAA and/or AA. If, at any point, a failure of compliance is encountered, the computation is terminated and all segments are discarded. The rationale for this strategy is simple: if a sequence, computed as parallel to meridians as possible, fails the timestamp assignment, then it might not be viable in the first place. Most likely, in those cases, the patch has some geometry feature that stretches the scheduling flexibility too far. The result of this stage is a patch that is flushed, time-wise, to the beginning or to the end of the interval of time allotted to it. The path is thus a composite structure, made by a string of segments.

\begin{figure*}
	\begin{centering}
	\includegraphics[width=\textwidth]{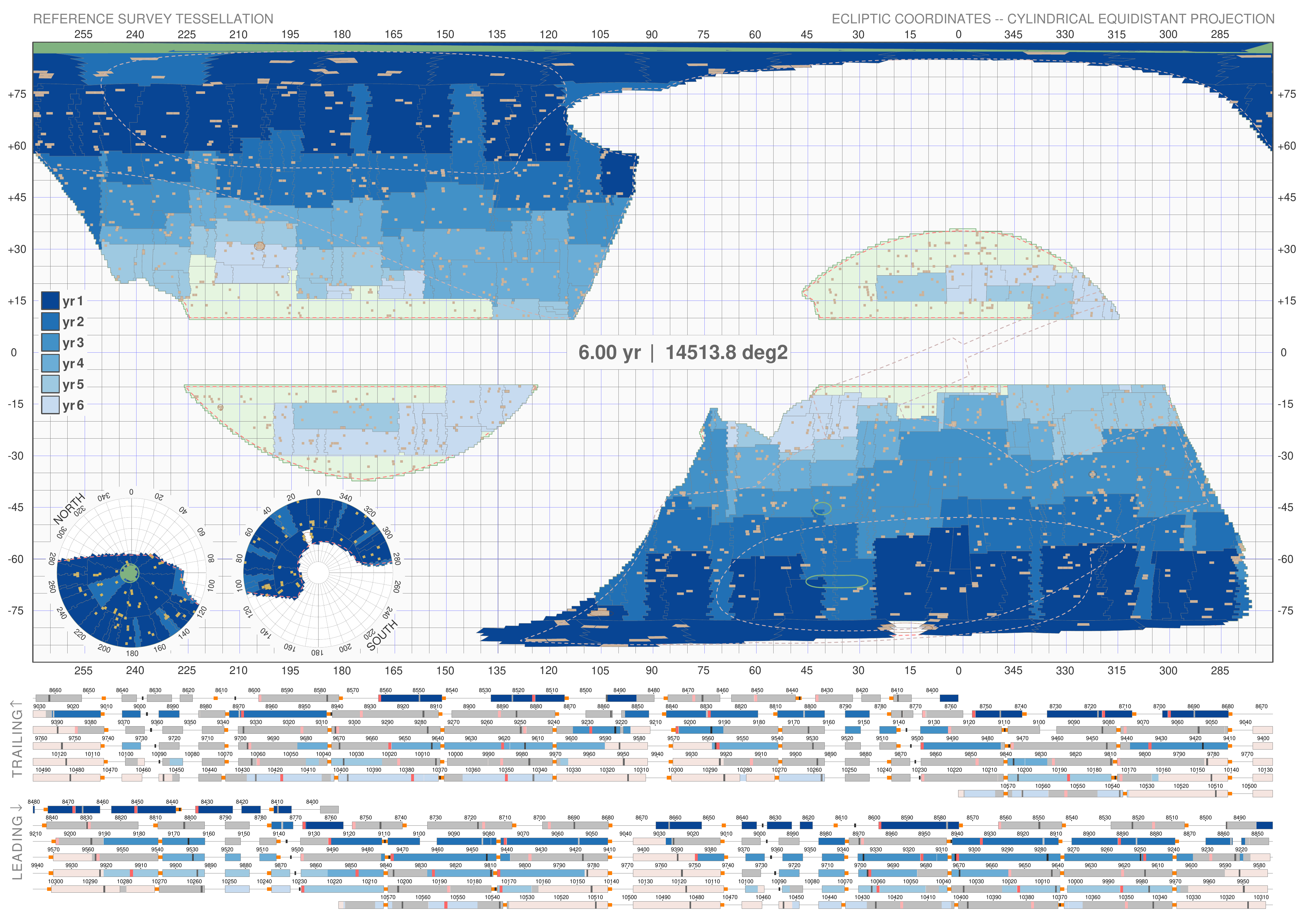}
		\caption{An example the \ECTile output, showing the 
		build-up of the coverage of the EWS in RSD\_2021A (cf. Sect.~\ref{sec:RSD}) over time in ecliptic coordinates. In the main panel the EWS patches are coloured according to different years, in correspondence to the time bars below. The part of the RoI which is not observed is in light green (we remind that the RoI covers a larger area than the one needed to be observed).  
		Under the main panel, the progression over the six years is unfolded, with two rows for each year indicating the trailing and leading directions of the telescope. Different colours represent the progression in time, from deep blue to light blue, whereas the thin line segments represent time intervals reserved for calibrations and Deep Fields. The light grey boxes for the leading or trailing directions mark when the telescope is actually observing in the opposite direction (for each grey box there is a corresponding coloured box). Towards the end of the mission some of the survey-windows are completely or partially light-red, showing that particular time slot has run out of sky areas to observe.} 
		\label{fig:ECTile_all}
	\end{centering}
\end{figure*}

\begin{figure*}[h!bt]
	\begin{centering}
		\resizebox{\twocolspan}{!}{ %
			\includegraphics{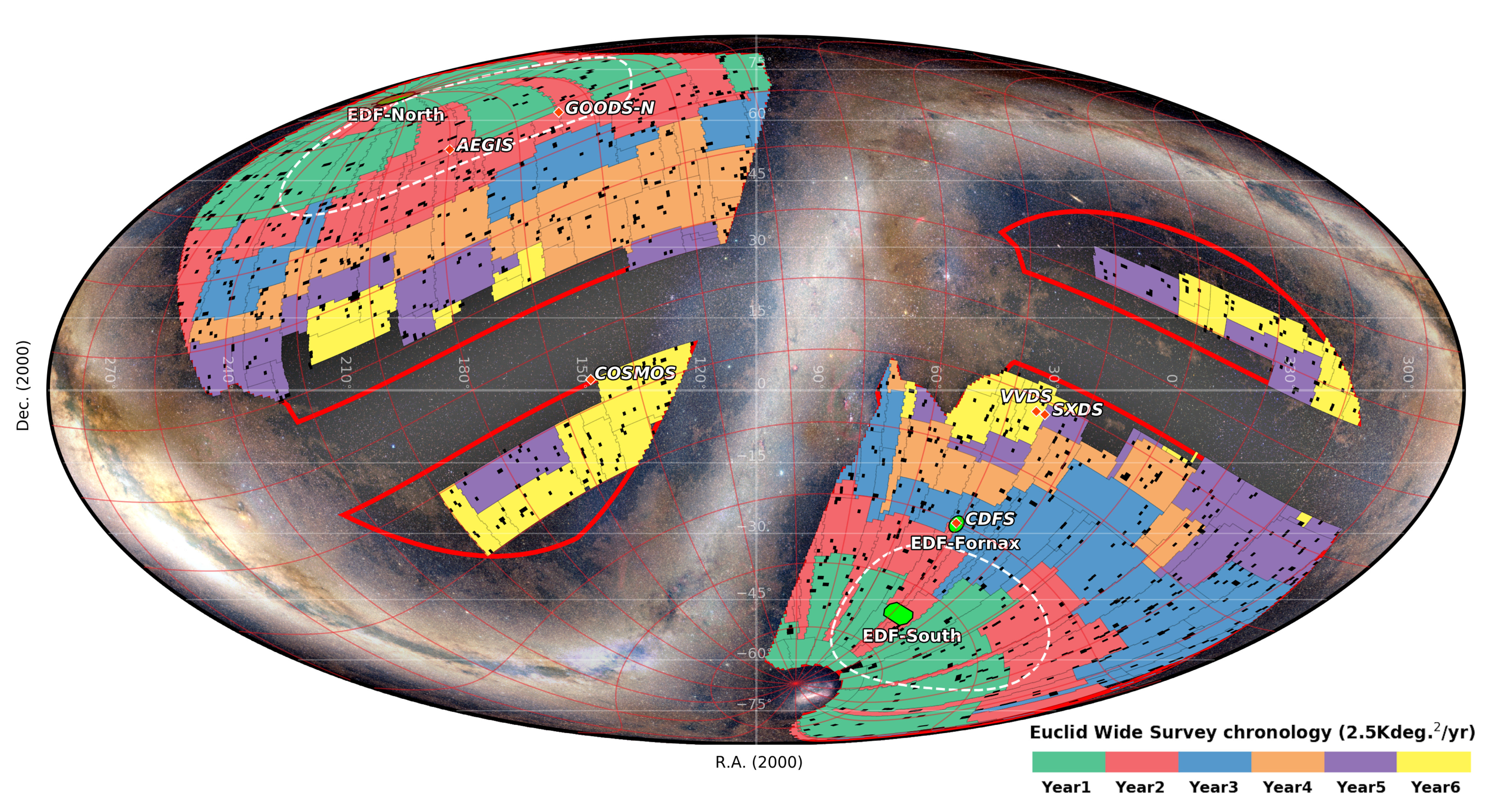} 
			}
		\caption{Reference Survey Definition 2021A (14\,514\,deg$^2$) chronology shown in  celestial coordinates. RoI boundaries are shown as solid red lines. Blinding stars cause 809 avoidance areas within the reference survey, with an average of $0.785\,\text{deg}^2$ per avoidance area, totalling  $635\, \deg^2$  Dashed lines (1300\,deg$^2$ in white per Galactic cap) delimit the highest SNR areas. The ecliptic referential is over-plotted in red. The three EDFs (bright green) and the six EAFs (red diamonds, not in size) are shown.}
		\label{Fig:coverage_by_year_moll}
	\end{centering}
\end{figure*}

\subsection{Placing patches}\label{sec:patch_puzzle}

The EWS is computed sequentially, covering the RoI by placing one patch at a time. In some way, it is like solving a  jigsaw puzzle on a surface of a sphere, but where the shapes of the pieces are not fixed from the start, instead being computed as the scheduling progresses. Two objectives guide the build-up of the EWS, which can be summarised as: ``observe as much as possible, as early as possible''. More specifically,

\begin{enumerate}
\item We want a compact footprint (per quadrant), so that the footprint of observations forms a single, continuous region. This means patches should match perfectly next to each other, with no holes in between;\footnote{The only planned ``holes'' are the tiles skipped because of the presence of blinding stars.}
      
\item We want to observe fields as early as possible, which means that all slots in survey-windows are assigned to observations, provided there is an unobserved part of the RoI within reach. After a few iterations, the RoI within reach of some survey-windows is typically scheduled (given the cyclic nature of the process).
\end{enumerate}

Eventually, either no area is left to schedule (within a range of latitudes of the RoI) or there is no time left (within the six years of the mission). When no suitable area can be scheduled for a given period, the schedule just leaves it unassigned, giving rise to unallocated time. This highlights that presently it is not possible to assign it to particular observations, but that it can be done at a later time.

The EWS is computed by filling the RoI one survey-window at the time, in chronological order. At each step, a survey-window is processed by assigning observations to its time slot. As explained below, this may require a few iterations. Only when all possible assignments are handled, the computation moves on to the next survey-window.

As discussed in Sect.~\ref{sec:patch_sources}, each survey-window generates one or more patch-sources, which in turn may generate one patch (made of a single patch-segment or of a string of several patch-segments), flushed to the beginning or to the end of its survey-window.  All the patches are continuously linked to a particular \stageone observation (e.g. a calibration, deep-field, or polar cap patch).  All these patches compete for the same slot of time, and we need to decide which patch to add to the survey.  In general, the choice is to build the EWS layer by layer of ecliptic latitude, striving to go from high to low SNR regions.  When this criteria is not decisive, the choice is to select the patch that better matches (horizontally, along ecliptic latitudes) some previously selected adjacent patch.

The last step in the processing of a survey-window is to place PSF calibrations and SOPs.  At most, one PSF calibrations is inserted per window. This is achieved by analysing the patches just selected, identifying the timestamps where a PSF calibration target is within reach.  A jump to a PSF calibration takes place at the end of some wide observation.  Typically, a single patch generates several such candidates.  In parallel, candidate timestamps for the required SOPs are identified.  Again, these are timestamps of the end of some wide observation.  There are one to two SOP candidates per window, at most.  Then, the two types of candidates are considered, inserting them in chronological order; one PSF calibration, if available, and one or two required SOPs.  Among all PSF candidates, preference is given to the ones occurring between patch-segments or between a preceding SOP and a following patch-segment, in order to save large-slews.  The insertion is performed by cutting a wide patch at the required timestamp, adding the PSF calibration or the SOP, and pushing the remaining of the patch forward in time.  This process may require also an adjustment of the following calibration block, pushing it forward in time (if now overlapped by the preceding patch).  The process of SOP insertion is also applied to the following calibration block before processing the next survey-window.

The islands are located at lower latitudes, and comprise less area. Consequently, most of the survey build up is shared between the two mainlands (\FancyRoman{1} and \FancyRoman{4}, see Fig.~\ref{Fig:roi_outline}) with the islands (\FancyRoman{2} and \FancyRoman{3}) becoming relevant in the last year of the survey. This can be seen in Fig.~\ref{fig:ECTile_all}, which shows an example of the build-up of EWS patches over the duration of the mission, coloured from deep blue to light blue as time progresses.  This strategy does not only forces the two mainlands to grow at an equal rate but, more importantly, it is crucial to guarantee a compact survey footprint.

The height of a patch is dictated by the speed of the orbit, which varies slightly with latitude. 
It roughly corresponds to the time the orbit takes to scan the width of a FoV divided by the nominal observation time. To some degree, the height (and width) may vary slightly from its natural size (approximately, $\pm 2$ rows).  Hence, in practice the height of a patch is more or less fixed, making the layer approach optimal. It guarantees a maximum of free RoI above previously computed patches, giving ample space (i.e., height) for the generation of each new patch. The sole exception is when the top layer of the survey reaches the boundary of the RoI.  If the available height is less then the minimum patch height, the patch generation fails. Due to the layering approach, this obstacle arises only for a small part of the RoI, namely the top layer of each quadrant. For the rest of the RoI, the stacking of patches ensures a compact filling.

\section{The Euclid reference survey definition}\label{sec:RSD}

In this section we present the latest version of the 
``\Euclid reference survey definition'' (RSD), RSD\_2021A.
This is the result of the \stageone and \stagetwo scheduling procedures (cf Sects.~\ref{sec:first_stage} and \ref{sec:wide}), and 
the corresponding \ECTile outputs are the ones shown in 
 Figs.~\ref{Fig:full_schedule} and \ref{fig:ECTile_all}. 

We recall that the RSD observations start three months after launch.  They are preceded by a one-month commissioning phase, followed by a two-month performance verification (PV) phase. During the PV phase the first survey-like data will be obtained, which are used to verify the data processing, and to validate and eventually tune or adjust the nominal sequence of operations (possible minor changes to the RSD can be implemented in less than one week time, producing a new version of the RSD; more complex changes of course require more time). Moreover, during PV some survey specific observations will be carried out. The latter will, for instance, provide direct measurements of the zodiacal light and stray light to verify and refine our models. 

\subsection{Survey performance} \label{sec:statistics}

Figure~\ref{Fig:coverage_by_year_moll} shows the footprint of RSD\_2021A.
The different colours indicate different observing epochs of the EWS. The three EDFs and the six EAFs (cf. Sect.~\ref{sec:deep_fields}) are tied into the EWS. In the two mainlands of the EWS, the observations start from the ecliptic poles and progress towards the equator. The best sky areas around both Galactic caps are covered within the first three years of the mission. The observations of the two EWS  islands only take place in the final two years of the mission.  Note that some of the worst parts of the RoI (totalling an area of ${\sim}2696\,\text{deg}^2$) are left unobserved. These correspond to the uncoloured areas in the islands and at low latitudes in the mainlands. The areas of the sky with longitudes between $150\degr$ and  $225\degr$,  and between $330\degr$ and $45\degr$, are observable at the same time since they are separated by $180\degr$. They contain much area within the RoI, and moreover the EDF-F and EDF-S are also located there. This means that there is not enough time for the EWS to observe all that area in the six years of the mission and hence their worst-quality regions are not observed.

\begin{figure*}[htb]
	\begin{centering}
		{\includegraphics[scale=.68]{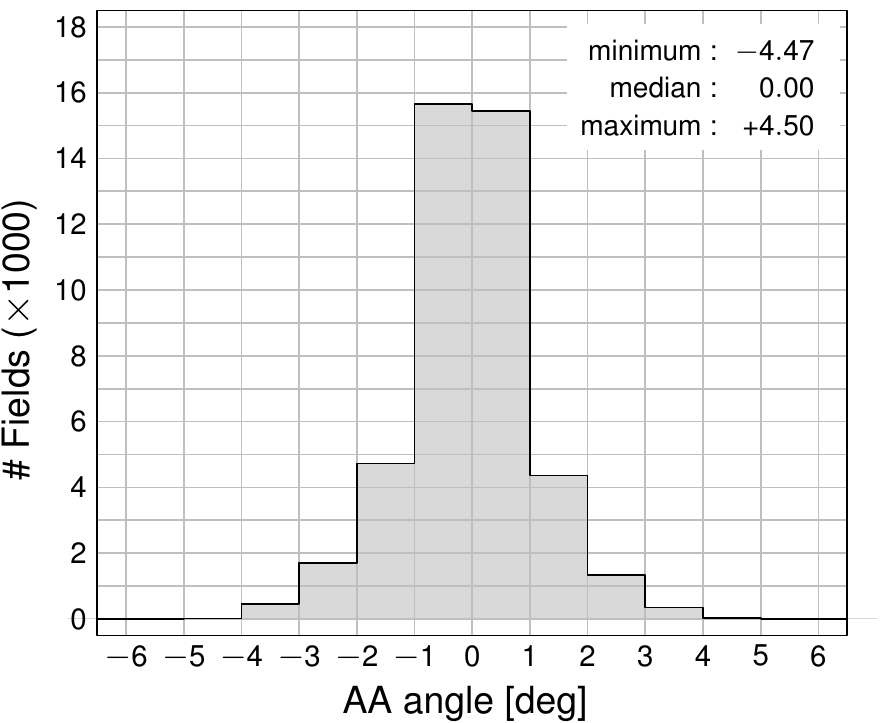}} 
		\hfill
		{\includegraphics[scale=.68]{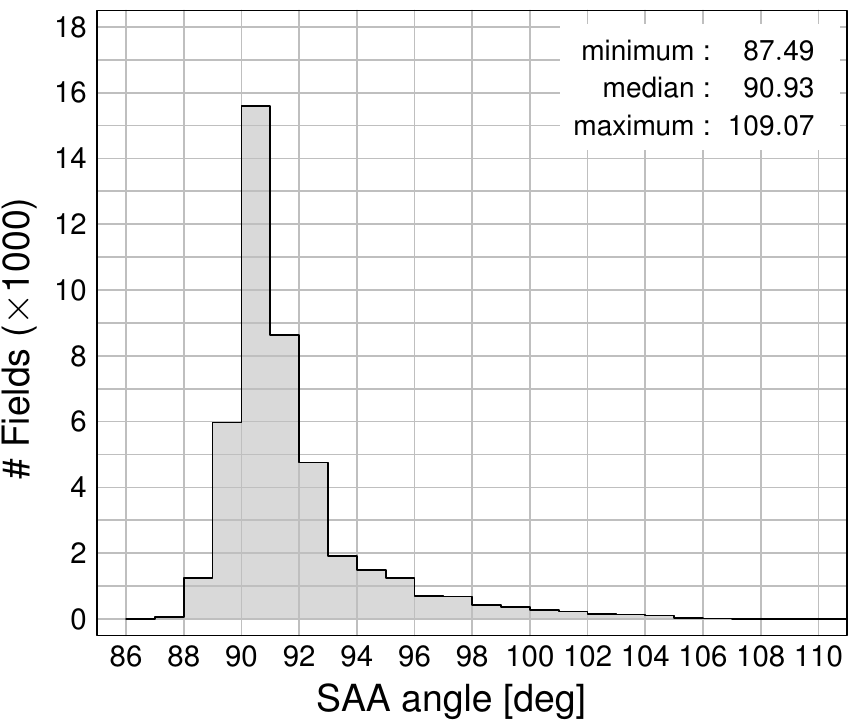}}
		\hfill
		{\includegraphics[scale=.68]{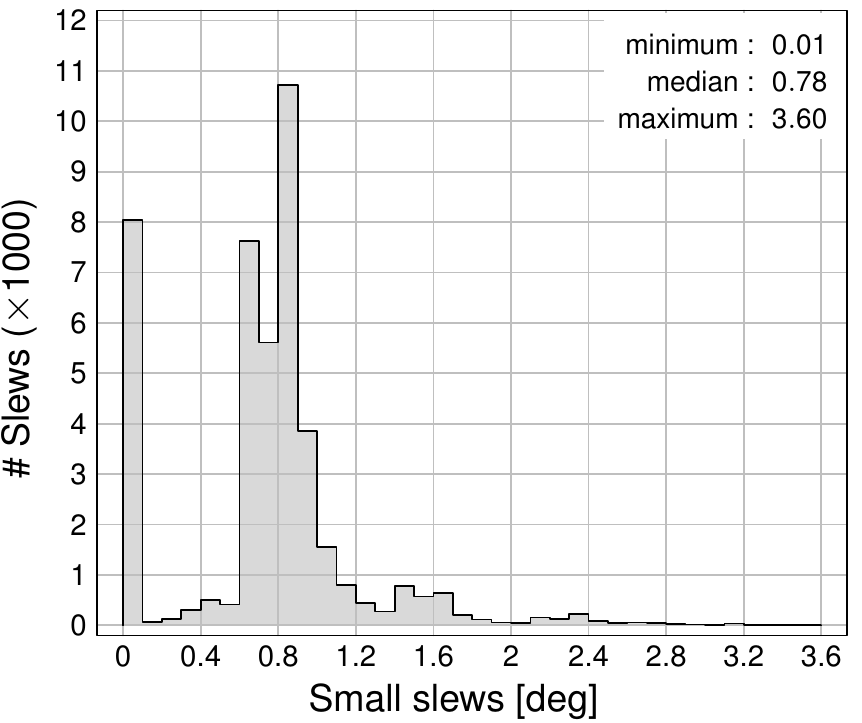}} 
		\caption{The distributions of angles and slews. \emph{Left panel:} AA angle.  \emph{Central panel:} SAA angle. \emph{Right panel:} field slews up to ${\ang{3.6;;}}$.  All quantities do comply with the allowed ranges.}
		\label{Fig:survey_histograms}
	\end{centering}
\end{figure*}

The RSD contains 44\,065 fields (28\,080 to build the EWS and 15\,985 for EDFs, EAFs and calibration targets observations). The EWS fields are contained in 256 patches (seen in Figs.~\ref{fig:ECTile_all} and \ref{Fig:coverage_by_year_moll}). The vast majority of the field slews, used to point the telescope, are below $1\fdg2$, as shown in the right panel of Fig.~\ref{Fig:survey_histograms}. This is the most efficient slew regime in terms of propellant usage.  As shown
in Fig.~\ref{Fig:survey_histograms}, all telescope rotations are done within the allowed SAA and AA limits. Most of the observations are done close to transit, with $90\%$ of the SAA values used between $88\degr$ and $94\degr$. The statistics of AA usage shows that  $71\%$ of the telescope rotations are done with $|AA| < 1\degr$. Even though SAA and AA values spread over the full range allowed, the field-to-field variations (between consecutive observations) of SAA and AA are very small throughout the survey: smaller than $1\degr$ in $97.4\%$ (SAA) and $98.6\%$ (AA) of the field-to-field transitions over the full mission. This feature is extremely important for the thermal stability, which ensures a stable PSF for WL shape measurements. It was possible to achieve this performance thanks to the implementation of the diffusion algorithm described in Sect.~\ref{sec:diffusion}.  

\subsection{Unallocated time}
\label{sec:unallocated_time}

\begin{figure}[!ht]
	\begin{centering}
		\resizebox{\onecolspan}{!}
		{\includegraphics[]{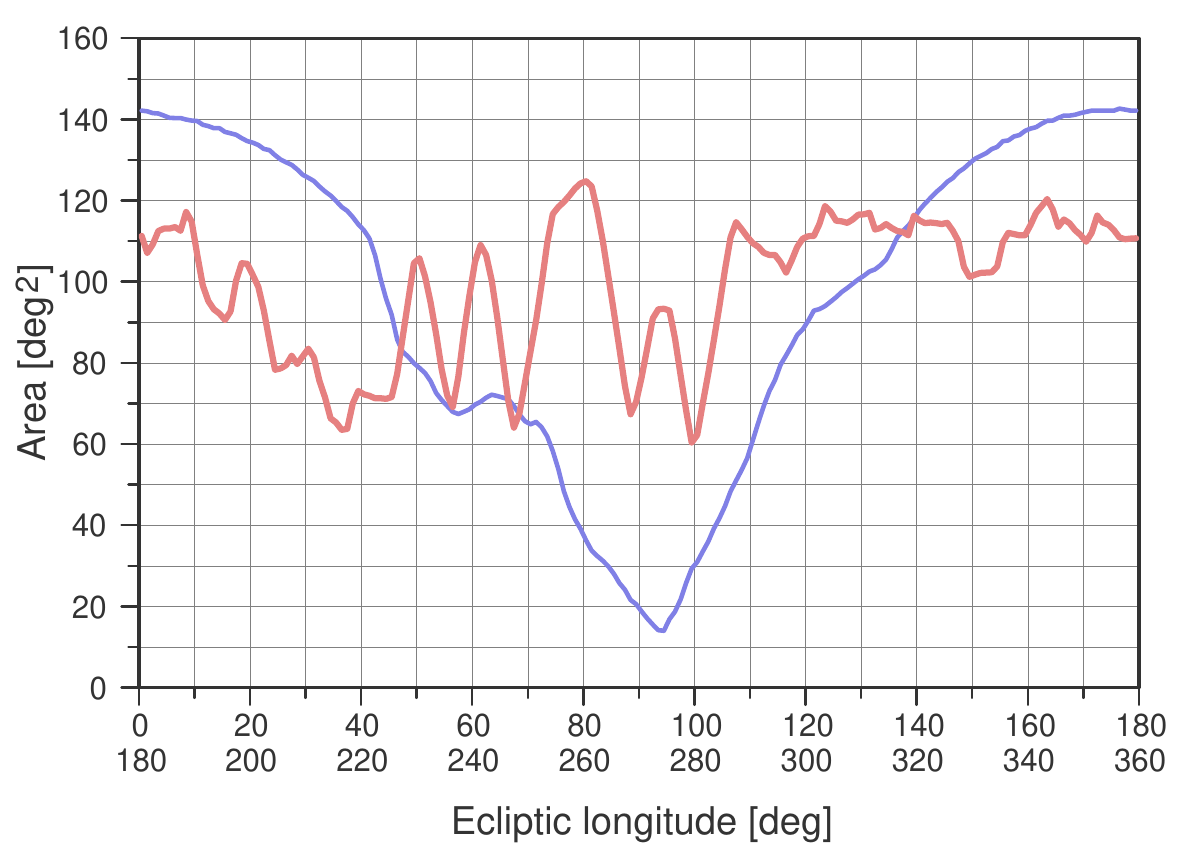}}
		\caption{Area of the RoI as function of (folded) longitude (blue). Equivalent area visible as function of (folded) longitude (red).  Area is the available RoI area at transit per degree of longitude.  The red curve assumes that the EWS observations are done at transit. Unallocated time arises when the red curve is above the blue curve. See text for details. }
		\label{Fig:area_vs_folded_longitude}
	\end{centering}
\end{figure}

The existence of a deficit of area on some longitudes (see Sect.~\ref{sec:roi_four_areas}), compared to the available observing time, is evident from Fig.~\ref{Fig:area_vs_folded_longitude}. The blue curve is the area available in the RoI at a given ecliptic longitude (in bins of 1\degr). The RoI areas in longitudes separated by $180\degr$ are added, since that pair of longitudes can be observed at the same time, from the trailing or the leading direction. Due to this six-month periodicity, the $x$--axis range only extends to 180\degr. The red curve denotes the cumulated number of days during which a given longitude is visible for EWS observations, assuming transit observations, and converted to equivalent area (1 day corresponding to $10\,\deg^2$). The available time is not uniform, it is determined after the \stageone schedule is defined (see Sect.~\ref{sec:first_stage}), which creates a strong variation along the year (i.e. wiggles in the red curve). For example, the absolute minimum corresponds to the highly booked longitudes of the EDFF and EDFS, where less time is left for EWS observations. 
 
 In longitudes where the red curve is above the blue curve, there is a deficit of area for the time available for EWS, leading to unallocated time. Conversely, in longitudes where the blue curve is above the red curve, there is an excess of area for the time available for EWS, leading to unobserved areas in the RoI. In Fig.~\ref{Fig:coverage_by_year_moll} this corresponds to the areas with no patches, which clearly are on the areas of the RoI of worst quality.

Note that the presence of unallocated time in the EWS schedule does not mean that there will be any idle time, because some areas of the EWS may be re-observed or new areas that do not qualify for the EWS, but have scientific value nonetheless, may be observed instead. In doing so, we can either consider fields that are observable within the thermal and pointing constraints enforced for the EWS, or we can operate outside these constraints, thus with the risk of perturbing the continuation of the EWS afterwards. Therefore one needs to have the real in-flight characteristics to get a solid picture of the possibilities and constraints. 

Some examples that maintain the strict survey limitations are: 

\begin{itemize}
\item possible multiple exposures ($\> 4$) on areas in the ecliptic plane which do not qualify for SNR with ROS single visit;
\item possible decontamination procedures;
\item repeat some suitable but lower-quality EWS regions to boost their SNR, or fill possible gaps due to unexpected events which might interrupt the basic scheduling;
\item increase the depth of the self-cal field, which could become a reference field for a dedicated supernova program in a possible extension of the mission;
\item build a medium deep field (EMDF) covering $200-300$\,deg$^2$
that is one magnitude deeper than the EWS (this would require five additional visits that would preferentially be done once every year);
\item observe suitable astronomical objects that are located outside the RoI that would benefit from localised repeats. Examples are observations on the ecliptic or Galactic plane, Galactic bulge (e.g. microlensing), specific low surface brightness objects, nearby galaxies, or clusters of galaxies;
\item use of the blue grism on targets during new or repeat visits on specific targets.
\end{itemize}

\noindent How to best use the time that cannot be used for single pass EWS will be decided at a later stage upon a consolidated scenario by the Euclid Science Team and ESA.

\begin{figure}[htb]
	\begin{centering}
		\resizebox{0.48\textwidth}{!}
		{\includegraphics{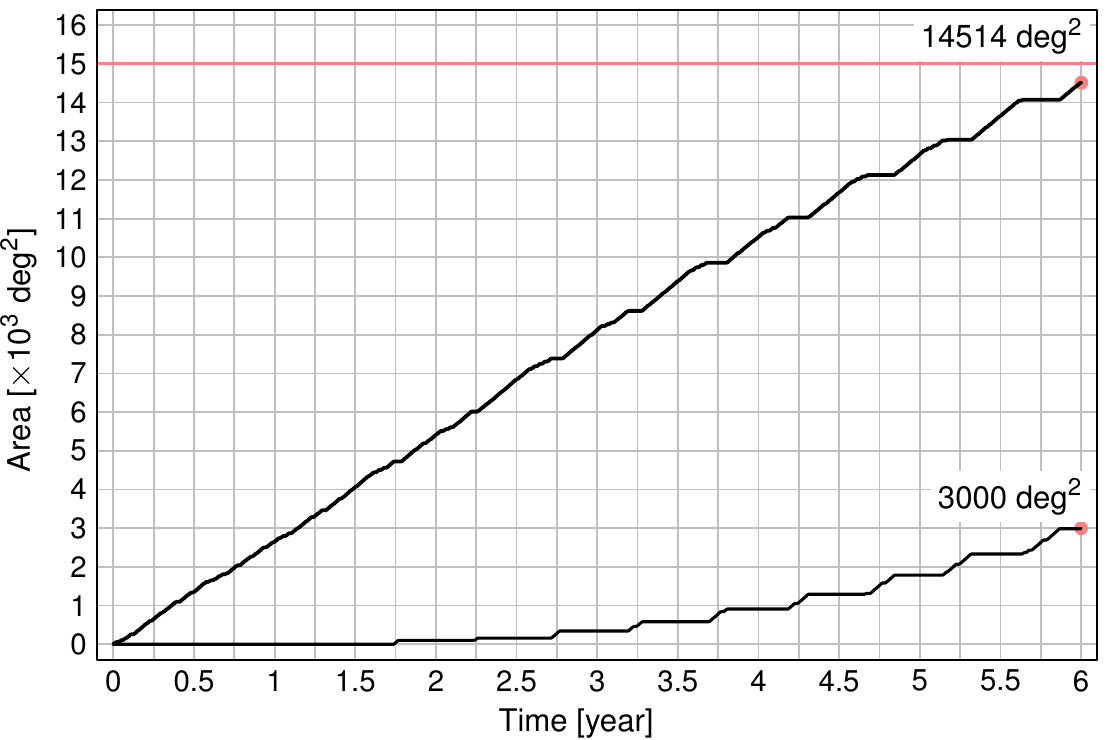}}
		\caption{Growth of the area surveyed by the RSD\_2021A EWS with time (upper curve). Growth of unallocated time with time (lower curve), here in units of EWS equivalent area (one month corresponds to about 300\,deg$^2$).}
		\label{Fig:deepn-20deg-area-time-plot}
	\end{centering}
\end{figure}

\begin{table}[hbt]
\caption{EWS area observed in RSD\_2021A per year and cumulatively at the end of each year. Last column shows the cumulative unallocated time (see text).} 
	\centering           
	\scalebox{0.9}{  \renewcommand{\arraystretch}{1.5} 
		\begin{tabular}{| c | c | c | c | c | c |} 
			\hline 
 \multirow{2}{*}{RSD\_2021A} & \multicolumn{2}{c|}{Area (per year)} & \multicolumn{2}{c|}{Cumulative area} & Unallocated time\\
 \cline{2-6}
year & [deg$^2$] & [$\%$] & [deg$^2$] & [$\%$] & [days]  \\
 \hline
 1 & 2\,656 & 18.3 & \phantom{0}2\,656 & \phantom{0}18.3 & \phantom{00}0 \\
 \hline
 2 & 2\,762 & 19.0 & \phantom{0}5\,418 & \phantom{0}37.3 & \phantom{0}10 \\
 \hline
 3 & 2\,708 & 18.7 & \phantom{0}8\,126 & \phantom{0}56.0 & \phantom{0}33 \\
 \hline
 4 & 2\,400 &  16.5 & 10\,526 & \phantom{0}72.5 & \phantom{0}89 \\
 \hline
 5 & 2\,147 & 14.8 & 12\,673 & \phantom{0}87.3 & 174 \\
 \hline
 6 & 1\,840 & 12.7 & 14\,514 & 100.0 & 292 \\
\hline
		\end{tabular}
		}
	\label{tab:EWSyear}  
\end{table}

 \subsection{Euclid footprint area} \label{sec:area_coverage}

The area of the RSD\_2021A \Euclid footprint is 14\,514 $\deg^2$. Table~\ref{tab:EWSyear} lists the observed EWS area at the end of each year of the mission, while the growth with time of the area covered by the EWS is shown in Fig.~\ref{Fig:deepn-20deg-area-time-plot}. 

During the first year many calibration observations are scheduled to support the first data releases. As a result, the EWS initially progresses slower than in the second and third year, but it still reaches an area in excess of 2500\,deg$^2$.
At the end of the third year, more periods occur when standard EWS observations cannot be made due to the increasing paucity of available unobserved areas within the RoI; the slope of the growth in time flattens there, causing the staircase-like pattern seen in Fig.~\ref{Fig:deepn-20deg-area-time-plot}.
These periods of `unallocated time' (see Sect.~\ref{sec:unallocated_time}) have a 6-month periodicity due to the intersection of the Galactic plane with the ecliptic plane, have an increasing duration and slow down the progression of the EWS.

\begin{figure}[h!tp]
	\begin{centering}
		\resizebox{\onecolspan}{!}{ %
		\includegraphics{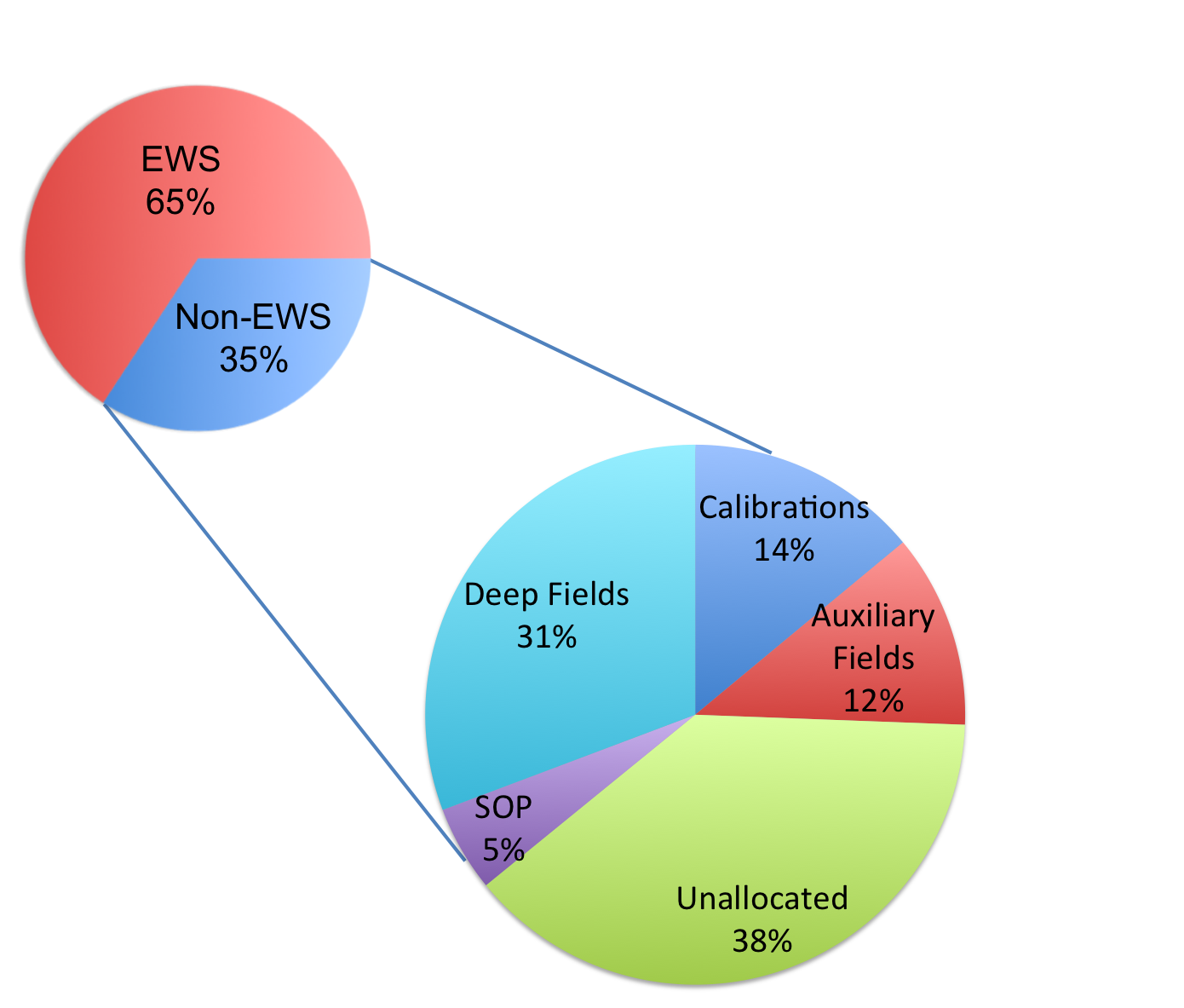} }
		\caption{The breakdown of the total time of the Euclid mission (2190 days) in EWS and non-EWS usage, for RSD\_2021A, corresponding to the schedule depicted in Fig.~\ref{Fig:full_schedule}. 
		} 
		\label{Fig:twopies}
	\end{centering}
\end{figure}

The time available to implement the EWS, once the other mandatory observations are carried out and the geometry of the RoI is taken into account, is an important input for the construction of the EWS and the final covered areas. In RSD\_2021A this time is $2190-427-39-292=1432$ days, where 427 days are used for calibrations, EDAs and EDFs observations, 39 days are reserved for SOP and there are 292 unallocated days (Fig.~\ref{Fig:twopies} depicts the time allocation breakdown).
Given the length of the ROS, close to 4400\,s per field (Sect.~\ref{sec:std_sequence}), and the FoV of 0.53\,deg$^2$ (Sect.~\ref{sec:common_fov}), the effective EWS area increases by 10.1\,deg$^2$ per day, (considering the effective average field overlap of 3\%). This means that, with the current calibration, EDFs, EAFs, and SOP required times, the EWS can  reach an area of 15\,000\,deg$^2$ only if the unallocated time is shorter than 240 days. 

\begin{figure}[htb]
	\begin{centering}
		\resizebox{\onecolspan}{!}
		{\includegraphics{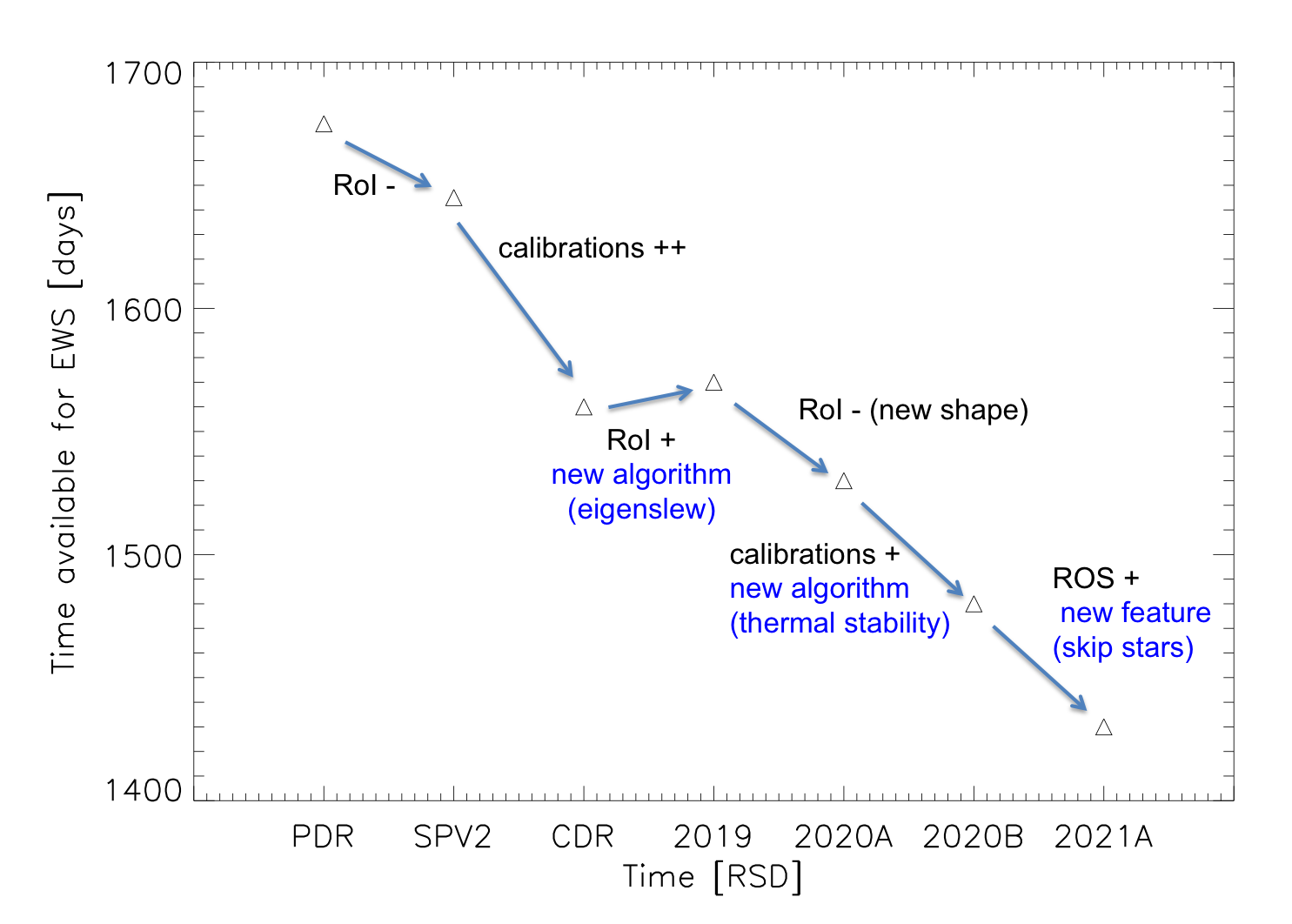}}
		\caption{Evolution of the time available for EWS observations in the RSDs produced from 2015 to 2021.  The changes are be due to increase or decrease (denoted by the symbols + or -, larger changes have double symbol) of calibrations, RoI area (directly impacting the unallocated time) and ROS time. See text for details.}
		\label{Fig:time_for_EWS}
	\end{centering}
\end{figure}

The time available for the EWS has generally decreased over the years, as the mission matured. Figure~\ref{Fig:time_for_EWS} depicts this evolution, indicating the driving changing factor for each transition (increse/decrease of calibrations, RoI or ROS). In earlier surveys, such as the RSD\_2015A prepared for PDR, the time needed for calibrations was small and the RoI was larger (driven by WL counts). In early 2018, RSD\_2018A used for SPV2 introduced a smaller RoI based on the more restrictive conditions for GC, which led to a larger distance from the Galactic plane (because of effects from star density on spectra), with a smaller area to be observed and an increase of unallocated time (cf. Sect.~\ref{sec:unallocated_time}). Shortly after, a major redefinition of the calibration plan, increased its allocated time by 100 days. At the same time the redefinition of the slew concept (see Sect.~\ref{sec:polar_caps}) made it no longer possible to schedule the EWS at high latitudes with the global tessellation, leading to a faster filling of the EWS and a faster build-up of unallocated time. These two factors led to a strong decrease of the EWS available time in RSD\_2018B for CDR \citep{Laureijs20}. In 2019, RSD\_2019A introduced a new algorithm to schedule the high-latitude regions. Given the fact that the throughputs were measured to be larger than specified, the RoI was reverted to the limits defined in \cite{RedBook}. The larger area of this RoI led to an increase in the available time for the EWS, and to a solution that covered 15\,000\,deg$^2$. The year 2020 introduced the new, better high-quality RoI, described in Sect.~\ref{sec:keyparams}, based on the latest estimates of SNR. This again decreased the available EWS time, now causing the RSD\_2020A to barely reach 15\,000\,deg$^2$. In RSD\_2020B the diffusion algorithm was introduced to increase thermal stability. This was achieved without decreasing the schedule efficiency and the available time for the EWS. Nevertheless the EWS time decreased due to a further increase of time needed for the calibrations. Finally, 
in RSD\_2021A we implemented the skipping of bright stars, together with the latest revised times for both ROS (significant increase of the dither steps duration) and SOP (halved).  

RSD\_2021A falls short of covering the desired 15\,000 deg$^2$ area for the EWS by $\sim 3\%$. The missing 486\,deg$^2$  are the equivalent of one and a half months of EWS observing time. In part this can be recovered by making local tessellations, to allow shifting the position of the centre of the tiles affected by stars,  such as to avoid the blinding stars falling on the detectors, instead of skipping the whole tile.
Moreover, it is expected that the EWS available time will further decrease sightly with the insertion of the complete set of PSF calibrations. In fact,
the expected cadence is not yet fully respected in RSD\_2021A, because the stringent stability constraints could not always be fulfilled, especially in the final years of the survey.
Study is underway to tackle this issue and, once this problem will be solved, the time needed amounts to $\approx$ 10 days (100 deg$^2$) to be taken off from the EWS present coverage.
It must be noticed, however, that a 3\% reduction in survey area translates into a comparable reduction in the final dark energy FoM (actually even less since the still uncovered areas of the RoI are of lower quality than the average). An area a few percent lower than originally expected  is therefore of no consequence for achieving the original goal of the mission, i.e. a FoM larger than 400 from the two main probes for a $w_0-w_a$~CDM model with no other prior information \citep{RedBook}. Indeed, 
for the present coverage, one preliminarily expects the FoM to scale down from 500 \citep{euclid_coll_istf_2020} to $\approx 480$.

Nevertheless, the final area of the Euclid footprint is not yet fixed and its value can change in either direction. On the one hand, possible further synergies between targets and optimisations will likely increase the area covered by the EWS. On the other hand,  modifications in the ROS, the calibrations plan, and in the amount of overlap between tiles might further reduce the final area.


\section{Summary}\label{sec:summary}

We have presented the status of the reference survey for the \Euclid mission at the beginning of year 2021. 
 The reference survey encompasses all the six years of the mission baseline, fulfilling all the constraints while combining observations of the EWS, all the slews, the calibrations, the deep and auxiliary field plus other calibration fields. We also discussed and gave models of the main backgrounds which affect the space observations. We also presented the complementary ground-based observations needed for photometric redshifts.  
 
 This is a non-trivial achievement, because of the complex nature of the mission: the \Euclid step-and-stare strategy is severely constrained by pointing limitations with respect to the Sun, whilst it also has to carry out a large number of dedicated observations for calibration purposes and sample characterisation. This leads to a complex interplay of timing observations, visibilities and manoeuvres. Although the two main instruments are operated in concert, they do require three separate calibration strategies, thus adding to the complexity. This makes \Euclid not only different from a typical observatory mission (e.g. \emph{Herschel}), but also from all-sky missions that have a single scanning strategy (e.g. \emph{Planck} and Gaia). 

Despite these challenges, we have found a highly optimised solution, that takes into account the main spacecraft and instrument characteristics and limitations, the current models for the expected background, the various calibrations, the dithering pattern and the methods to cover the wide area expected to be observed. 
The latest version of the \Euclid Reference Survey Definition (RSD\_2021A) fulfils practically all requirements, resulting in a EWS that covers  $\approx 14\,500 \deg^2$ of the extragalactic sky. This is 3\% short of initial target mainly because of the paucity of good sky for \Euclid and of the severe constraints on the pointing (if feasible, 10 months of presently unallocated time would allow in principle to add 3000 square degrees to EWS). A companion paper will detail the rationale and results for the Euclid Deep Fields that cover about 40 deg$^2$, but at much greater depth.

The definition of the reference survey is an ongoing process, and some of the results presented will continue to evolve because the Euclid Consortium has developed both the expertise and the specific tools that allow one to probe significant variations of any among the multiple boundary conditions which originate from either the knowledge of the spacecraft, the instruments or the astronomical environment.
The current solution is, however, a close proxy for the final survey. Crucially, it demonstrates the feasibility of the core mission within the numerous constraints. Future plans include:
 \begin{itemize}
\item further optimisations to reach the nominal 15\,000\,deg$^2$ (e.g. double exposures on the ecliptic plane to compensate the local high background);
\item implement refined simulations;
\item implement refined background models;
\item incorporate possible future changes in parameters (e.g. changes in ROS or in the calibration plan);
\item consider ``what if'' scenarios for non-recurrent operations (decontamination, phase diversity calibrations) and possible failures (electronics or other systems);
\item make revisions and updates based on real measurements and in-flight performance.
\end{itemize}

These further improvements will present challenges in their own right, but given the current level of maturity, there is little doubt that \Euclid will dramatically advance our understanding of the nature of dark matter and dark energy, whilst impacting many aspects of astronomy thanks to the tremendous legacy value of these unique data.

\begin{acknowledgements}
	\AckEC   
\end{acknowledgements}

%
%

\bibliographystyle{aa}

\bibliography{ecsurv}

\appendix 

\twocolumn

\section{Acronyms}

\tablefirsthead{
\hline
  \multicolumn{1}{|l}{Acronym} &
  \multicolumn{1}{l|}{Name} \\
\hline
}
\tabletail{\hline}
\tablehead{\hline
  \multicolumn{1}{|l}{Acronym} &
  \multicolumn{1}{l|}{Name} \\
\hline
}
\topcaption{Acronyms used in the paper.}
\noindent
{\footnotesize
\begin{supertabular} {| l p{6cm}|}
        1D & 1-Dimensional \\
        2D & 2-Dimensional \\
        AA & Alpha Angle \\
        AOCS & Attitude and Orbit Control System \\
        APE & Absolute Pointing Error \\
        BGS & ``Blue'' Grism  \\
        CCD(s) & Charge-Coupled Device(s) \\
        CDR & Critical Design Review \\
        CPC & Completeness Purity Calibration field \\
        CPC-N & CPC-North \\
        DE & Dark Energy \\
        DGL & Diffuse Galactic Light (cirrus) \\
        DM & Dark Matter \\
        DR\# & Data Release number \#\\
        EAFs & Euclid Auxiliary Fields \\
        EC & Euclid Consortium \\
        ECTile & EC Tiling program \\
        EDFs & Euclid Deep Fields \\
        EDFN & EDF-North \\
        EDFS & EDF-South \\
        EDFF & EDF-Fornax \\
        EDS & Euclid Deep Survey \\
        EMDF & Euclid Medium Deep Field \\
        EOL & End Of Life \\
        ERS & Euclid Reference Survey \\
        ESA & European Space Agency \\
        ESSPT & Euclid Sky Survey Planning Tool \\
        EWS & Euclid Wide Survey \\
        FGS & Fine Guidance Sensor \\
        FoM & Figure of Merit \\
        FoV & Field of View \\
        FPA & Focal Plane Array \\
        FWA & Filter Wheel Assembly \\ 
        FWHM & Full Width Half Maximum \\
        GC & Galaxy Clustering \\
        GWA & Grism Wheel Assembly \\
        LMC & Large Magellanic Cloud \\
        LoS & Line of Sight \\
        LSB & Low Surface Brightness \\
        $\Lambda$CDM & $\Lambda$ Cold Dark Matter \\
        M1 & Main mirror \\
        MC & Monte Carlo \\
        MCDR & Mission Critical Design Review \\
        NDI & Normalised Diffusion Irradiance profile \\
        NEO & Near Earth Object \\
        NEP & Northern Ecliptic Pole \\
        NIR & Near-InfraRed \\
        NISP & Near infrared Imager and SPectrometer \\
        NISP-S & NISP Spectroscopy \\
        NISP-P & NISP Photometry \\
        Q\# & Quick Data Release number \# \\
        PDR & Preliminary Design Review \\
        PLM & PayLoad Module \\
        PN & Planetary Nebula \\
        PSF & Point Spread Function \\
        PV & Performance Verification \\
        RGS\# & ``Red'' Grism \# \\
        RoI & Region of Interest \\
        ROS & Reference Observation Sequence \\
        RSD & Reference Survey Definition \\
        RSD\_\#  & Reference Survey Definition \_\#\\
        SAA & Solar Aspect Angle \\
        SEDs & Spectral Energy Distributions \\
        SEL2 & Sun-Earth Lagrangian point L2 \\
        SiC & Silicon Carbide \\
        SMC & Small Magellanic Cloud \\
        SNR & Signal to Noise Ratio \\
	SOP & Spacecraft Orbit and Platform maintenance \\ 
        SPSAA & Solar Panel Solar Aspect Angle \\
        SPV\# & Science Performance Verification \# \\
        STOP & Satellite Structural Thermal Optical Performance \\
        SVM &  SerVice Module \\
        VIS & VISible instrument \\
        WL & Weak gravitational Lensing \\
        \hline  
        \hline
        2MASS & Two Micron All-Sky Survey \\
        AEGIS & All-wavelength Extended Groth strip International Survey \\
        ASAP & Analysis of Stellar Atmospheres and Pulsation \\
        ATLAS & Asteroid Terrestrial-impact Last Alert System \\
        CANDELS & Cosmic Assembly Near-IR Deep Legacy Survey \\
        CDFS & Chandra Deep Field South  \\
        CFHT-MegaCam & Canada-France-Hawaii Telescope Mega Camera \\
        CFHT & Canada-France-Hawaii Telescope \\
        CFIS & Canada-France Imaging Survey \\
        COBE & COsmic Background Explorer \\
        COSMOS & Cosmic Evolution Survey \\
        DES & Dark Energy Survey \\
        DETF & Dark Energy Task Force \\
        DUNE & Dark UNiverse Explorer \\
        GOODS-N & Great Observatories Origins Deep Survey - North \\
        HSC & Subaru Hyper Suprime-Cam \\
        HST & \HST \\
        HUDF & HST Ultra Deep Field \\
        IPAC & Infrared Processing and Analysis Center \\
        IRAS & InfraRed Astronomical Satellite \\
        JEDIS-g & Javalambre-Euclid Deep Imaging Survey in g band \\
        JST & Javalambre Survey Telescope \\
        JWST & James Webb Space Telescope \\
        LSST & Legacy Survey of Space and Time \\
        Pan-STARRS & Panoramic Survey Telescope \& Rapid Response System \\
        SDSS & Sloan Digital Sky Survey \\
        SMEI & Solar Mass Ejection Imager \\
        SNAP & SuperNova / Acceleration Probe  \\
        SXDS & Subaru/XMM-Newton Deep Survey \\
        UDF & Ultra Deep Field \\
        UNIONS &  Ultraviolet Near-Infrared Optical  Northern Survey \\
        VVDS & VIsible Multi Object Spectrograph (VIMOS) Very Large Telescope (VLT) Deep Survey \\   
        VVDS-Deep & VIsible Multi Object Spectrograph (VIMOS) Very Large Telescope (VLT) Deep Survey -- Deep \\  
        WISHES & Wide Imaging with Subaru HSC of the \Euclid Sky \\
        \hline 
        \hline 
        AKM\# & Arendt, Kashlinski and Mosley \# \\
        Sc21 & Scaramella {\it et al.}, in prep. (Euclid Deep Survey) \\
\hline
\end{supertabular}
} 

\end{document}